\pdfoutput=1
\documentclass{JHEP3}
\usepackage{amsmath}
\usepackage{graphics}
\usepackage{epsfig}

\def\({\left(}
\def\){\right)}
\def\[{\left[}
\def\]{\right]}
\def\<{\langle}
\def\>{\rangle}
\def\sg{\sigma}

\newcommand{\be}{\begin{equation}}
\newcommand{\ee}{\end{equation}}
\newcommand{\bal}{\begin{aligned}}
\newcommand{\eal}{\end{aligned}}

\newcommand{\labell}[1]{\label{#1}}

\title{4-particle Amplituhedronics for 3-5 Loops}

\author{Junjie Rao$^{a}$\footnote{Email: jrao@aei.mpg.de}\\
{$^a$Max Planck Institute for Gravitational Physics (Albert Einstein Institute), 14476 Potsdam, Germany}}

\abstract{Following the direction of 1712.09990 and 1712.09994, this article continues to excavate more interesting
aspects of the 4-particle amplituhedron for a better understanding of the 4-particle integrand
of planar $\mathcal{N}\!=\!4$ SYM to all loop orders, from the perspective of positive geometry.
At 3-loop order, we introduce a much more refined dissection of the amplituhedron to understand its essential structure
and maximally simplify its direct calculation, by fully utilizing its symmetry as well as the efficient Mondrian way
for reorganizing all contributing pieces. Although significantly improved, this approach immediately encounters its
technical bottleneck at 4-loop. Still, we manage to alleviate this difficulty by imitating the traditional (generalized)
unitarity cuts, which is to use the so-called positive cuts. Given a basis of dual conformally invariant (DCI)
loop integrals, we can figure out the coefficient of each DCI topology using its $d\log$ form via
positivity conditions. Explicit examples include all 2+5 non-rung-rule topologies at 4- and 5-loop respectively.
These results remarkably agree with previous knowledge, which confirms the validity of amplituhedron up to 5-loop
and develops a new approach of determining the coefficient of each distinct DCI loop integral.}


\begin{document}
\maketitle

\section{Introduction and the 3-loop Amplituhedron Revisited}

The amplituhedron proposal for 4-particle all-loop integrand of planar $\mathcal{N}\!=\!4$ SYM
\cite{Arkani-Hamed:2013jha,Arkani-Hamed:2013kca} is a novel reformulation
which only uses positivity conditions for all physical poles to construct the integrand.
At $L$-loop order, for any two sets of loop variables labelled by $i,j\!=\!1,\ldots,L$
we have the mutual positivity condition
\be
D_{ij}=(x_j-x_i)(z_i-z_j)+(y_j-y_i)(w_i-w_j)>0,
\ee
where $x_i\!=\!\<A_iB_i\,14\>$, $y_i\!=\!\<A_iB_i\,34\>$, $z_i\!=\!\<A_iB_i\,23\>$, $w_i\!=\!\<A_iB_i\,12\>$
and $D_{ij}\!=\!\<A_iB_i\,A_jB_j\>$ are all possible physical poles in terms of momentum twistor contractions,
and $x_i,y_i,z_i,w_i$ are trivially set to be positive.
A simplest nontrivial case is the 2-loop integrand given in \cite{Arkani-Hamed:2013kca}.
Though the dominating principle is simple and symmetric up to all loops,
as the loop order increases, its calculational complexity grows explosively due to the highly nontrivial
intertwining of all $L(L\!-\!1)/2$ positivity conditions.

So far the 4-particle amplituhedron has been fully understood up to 3-loop \cite{Rao:2017fqc},
from which we have incidentally found an intriguing pattern valid at all loop orders for a special subset of
dual conformally invariant (DCI) loop integrals: the Mondrian diagrammatics \cite{An:2017tbf}.
Even though there still remain unknown characteristics of the connection between this neat formalism and
down-to-earth physics, to say the very least, it offers us a much more efficient way for reorganizing
the 3-loop results via a direct calculation, by extensively using the properties of ordered subspaces
which further refine the space spanned by $x,y,z,w$.

This work continues the exploration of 4-particle amplituhedron at higher loop orders, which mainly includes two parts:
a more refined understanding of the 3-loop case, and the motivation and application of positive cuts at 4- and 5-loop.
We will see that even the maximally refined recipe can hardly handle the 4-loop case, hence we are forced to verify
the amplituhedron proposal in a somehow compromised way but even this concession is very interesting and nontrivial,
and most importantly, it is consistent with known results via the traditional approach.

Let's first briefly summarize some notions with relevant notations introduced in
\cite{Rao:2017fqc,An:2017tbf} which are frequently used in this work.

For the 3-loop amplituhedron as an example, given positive variables $x_1,x_2,x_3$, an \textit{ordered subspace}
$X(abc)$ denotes the region in which $x_a\!<\!x_b\!<\!x_c$. There are $3!\!=\!6$ such subspaces and they together
make up the space spanned by $x_1,x_2,x_3$. We also use $X(abc)$ as its corresponding $d\log$ form, namely
\be
X(abc)=\frac{1}{x_a(x_b-x_a)(x_c-x_b)}\equiv\frac{1}{x_ax_{ba}x_{cb}},
\ee
note that we have omitted the measure factor, following the convention of \cite{Rao:2017fqc,An:2017tbf}.
Originally, the full $d\log$ form is defined as
\be
d\log x=\frac{dx}{x},
\ee
where $x$ must be positive, and it becomes singular when $x\!\to\!0$. For $x\!>\!a$, the $d\log$ form is then
\be
\frac{da}{a}\frac{d(x-a)}{x-a}=\frac{da}{a}\frac{dx}{x-a},
\ee
since the measure factor remains the same, we can safely omit such universal factors for convenience when
triangulating positive regions. Back to $X(abc)$, obviously there is a completeness relation
\be
X(123)+X(132)+X(213)+X(231)+X(312)+X(321)=\frac{1}{x_1x_2x_3}.
\ee
The same notion applies for loop variables $x,y,z,w$, for example, $X(123)Z(321)Y(123)W(123)$ is simply
a direct product of these four subspaces, and the overall $d\log$ form is the product of their corresponding $d\log$ forms.

Each subspace admits some \textit{Mondrian seed diagrams} \cite{Rao:2017fqc}, for example, $X(123)Z(321)Y(123)W(123)$
admits the ladder diagram in figure \ref{fig-5}, which can be characterized by a
Mondrian factor $X_{12}X_{23}D_{13}$, with $X_{ij}\!=\!(x_j\!-\!x_i)(z_i\!-\!z_j)$,
$Y_{ij}\!=\!(y_j\!-\!y_i)(w_i\!-\!w_j)$ and $D_{ij}\!=\!X_{ij}\!+\!Y_{ij}$. This factor is determined by
the contact rules between any two loops defined in \cite{Rao:2017fqc,An:2017tbf} as
\be
\bal
\textrm{horizontal contact: }&X_{ij}\\
\textrm{vertical contact: }&Y_{ij}\\
\textrm{no contact: }&D_{ij}~(\textrm{always taking }i\!<\!j\textrm{ for }D_{ij})
\eal
\ee
For a particular subspace we can derive its $d\log$ form by demanding $D_{12},D_{13},D_{23}\!>\!0$.
Then multiplying its form by all positive denominators gives its \textit{proper numerator},
and the \textit{dimensionless ratio} between this numerator and $D_{12}D_{13}D_{23}$
encodes the positivity constraints, which becomes 1 if the positivity is trivial.
For example, the $d\log$ form of $X(123)Z(321)Y(123)W(123)$ takes the form
\be
\frac{1}{x_1x_{21}x_{32}}\frac{1}{z_3z_{23}z_{12}}\frac{1}{y_1y_{21}y_{32}}\frac{1}{w_1w_{21}w_{32}}
\frac{N}{D_{12}D_{23}D_{13}},
\ee
then $N$ is its proper numerator and $N/(D_{12}D_{23}D_{13})$ is the dimensionless ratio. In contrast,
the $d\log$ form of $X(123)Z(321)Y(123)W(321)$ simply reads
\be
\frac{1}{x_1x_{21}x_{32}}\frac{1}{z_3z_{23}z_{12}}\frac{1}{y_1y_{21}y_{32}}\frac{1}{w_3w_{23}w_{12}}
\ee
since $D_{12},D_{13},D_{23}$ are trivially positive, then the proper numerator is $D_{12}D_{23}D_{13}$
and the dimensionless ratio is simply 1.

The difference between the proper numerator and all admitted Mondrian factors (or the \textit{contributing part})
of a particular subspace is called the \textit{spurious part}. The spurious parts sum to zero
(over all ordered subspaces) at the end as their name implies.

For a DCI topology as those given in figures \ref{fig-7}, \ref{fig-10} and \ref{fig-11}, which can be Mondrian or
non-Mondrian, to enumerate all relevant DCI loop integrals, one must consider all its \textit{orientations} and
\textit{configurations of loop numbers}. For each topology by dihedral symmetry there can be 8, 4, 2, or 1 orientations,
depending on the additional symmetries it may have \cite{An:2017tbf}, and for each orientation there are $L!$
configurations of loop numbers. This finishes the summary.

Now we would like to improve all these techniques to extract the essential structure of the 4-particle amplituhedron
by fully utilizing the symmetry of (mutual) positivity conditions. Before this, let's briefly review the standard
calculation for the 2-loop case as a simplest nontrivial example below. For its single positivity condition
\be
D_{12}=(x_2-x_1)(z_1-z_2)+(y_2-y_1)(w_1-w_2)>0,
\ee
without loss of generality, we can fix the ordered subspace as $X(12)$ in which $x_1\!<\!x_2$, so it becomes
\be
z_1-z_2+\frac{(y_2-y_1)(w_1-w_2)}{x_{21}}>0,
\ee
where $x_{21}\!=\!x_2\!-\!x_1$ is a positive variable. Then depending on the choice of ordered subspaces of $y,w$,
there are 4 combinations to be considered, while the $z$-space is used for imposing $D_{12}\!>\!0$.
After that, we sum the result over all permutations of loop numbers, which are just $1,2$ in the 2-loop case
\cite{Arkani-Hamed:2013kca}. This has been used for the 3-loop case as well \cite{Rao:2017fqc}, while for the
latter we have to deal with three intertwining conditions $D_{12},D_{23},D_{13}\!>\!0$.
Though such a straightforward approach successfully works for the first two nontrivial cases, it inevitably
gets complicated by the tension between the simplicity of each contributing piece of a corresponding ordered subspace,
and the number and variety of such building blocks. That is to say, the more refined each piece is, naturally, the simpler
it looks, but there are more situations to be considered and hence their sum will be more involved, as one has to
carefully ensure that all spurious poles brought by the subspace division must be wiped off after the summation.
This disadvantage is due to overlooking the symmetry of positivity conditions. In the following, instead of picking
subspace $X(123)$ at 3-loop, we will treat all $x,y,z,w$ variables on the same footing.

To classify all possible positive configurations in a totally symmetric way, let's first explicitly write
\be
D_{12}=X_{12}+Y_{12},~~D_{23}=X_{23}+Y_{23},~~D_{13}=X_{13}+Y_{13},
\ee
with $X_{ij}\!=\!(x_j\!-\!x_i)(z_i\!-\!z_j)$ and $Y_{ij}\!=\!(y_j\!-\!y_i)(w_i\!-\!w_j)$ as introduced before.
For each $D_{ij}$, there are three possible configurations: $X_{ij}$ is positive while $Y_{ij}$ is negative and
the other way around, as well as both $X_{ij}$ and $Y_{ij}$ are positive. It goes without saying, the configuration
of which both $X_{ij}$ and $Y_{ij}$ are negative must be excluded. We can use a convenient notation to precisely
characterize each configuration, such as
\be
\{(+-)_{12},(+-)_{23},(+-)_{13}\}, \labell{eq-1}
\ee
which means $X_{12},X_{23},X_{13}$ are positive and $Y_{12},Y_{23},Y_{13}$ are negative.
Since the positivity conditions are symmetric in combinations $12,23,13$, the counting of all possible configurations is
given by a ``generating function'' which does not distinguish $12,23,13$, namely
\be
(D+X+Y)^3=D^3+3\,D^2(X+Y)+3\,D\(X^2+Y^2\)+6\,DXY+\(X^3+Y^3\)+3\(X^2Y+XY^2\), \labell{eq-2}
\ee
where $D,X,Y$ stand for both $X$ and $Y$ are positive, only $X$ is positive and only $Y$ is positive respectively.
Essentially there are only 6 distinct configurations, as we also treat $X$ and $Y$ on the same footing, which leads to
switching $x,z\!\leftrightarrow\!y,w$. We see the coefficient 1, 3 or 6 above precisely represents the number of combinations
within each distinct configuration. For example, for the 2nd term in the RHS above $3D^2X$ tells that $X$ can be chosen
to be $X_{12}$, $X_{23}$ or $X_{13}$, and also for the 4th term there are $3!\!=\!6$ combinations of $12,23,13$
for $D,X,Y$. Moreover, we can count the number of ordered subspaces for each configuration and sum them as
\be
36+24\times6+24\times6+16\times6+36\times2+16\times6=588, \labell{eq-3}
\ee
where each number in the sum will be explained in a detailed analysis of its corresponding configuration. On the other hand,
the total number of ordered subspaces of $x,y,z,w$ is $(3!)^4\!=\!1296$, so we see that the contributing pieces take up
$49/108$ of all subspaces. By this more refined dissection, we immediately get rid of more than half of all
subspaces which do not contribute, since they violate positivity conditions. In contrast, the standard way used in
\cite{Rao:2017fqc} has implicitly taken all non-contributing subspaces into account so it naturally looks more involved
and contains more repetitive calculation. Using notations of \eqref{eq-1}, we select one representative
for each of the 6 distinct configurations above for further calculation, as summarized in the following list:
\be
\bal
\{(++)_{12},(++)_{23},(++)_{13}\},~~\{(++)_{12},(++)_{23},(+-)_{13}\},~~\{(++)_{12},(+-)_{23},(+-)_{13}\},\\
\{(++)_{12},(+-)_{23},(-+)_{13}\},~~\{(+-)_{12},(+-)_{23},(+-)_{13}\},~~\{(+-)_{12},(+-)_{23},(-+)_{13}\}.
\eal
\ee
Note that after we obtain the $d\log$ forms of these 6 configurations,
the multiplicity in \eqref{eq-2} must be taken into account for correctly
summing all relevant terms. Now we start to analyze them one by one.

\subsection{Configuration $\{(++)_{12},(++)_{23},(++)_{13}\}$}

\begin{figure}
\begin{center}
\includegraphics[width=0.75\textwidth]{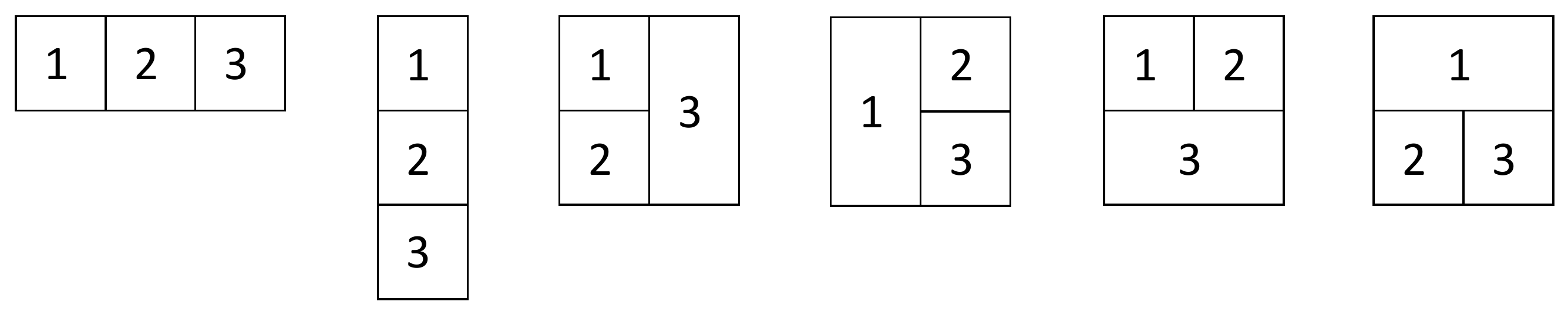}
\caption{Mondrian seed diagrams in subspace $X(123)Z(321)\otimes Y(123)W(321)$.} \label{fig-1}
\end{center}
\end{figure}

For the simplest configuration $\{(++)_{12},(++)_{23},(++)_{13}\}$, since it is totally positive for all $X_{ij}$'s
and $Y_{ij}$'s, there is no multiplicity as its coefficient in \eqref{eq-2} is simply 1. This corresponds to
the collection of ordered subspaces (here $\otimes$ is used for separating $X,Z$ and $Y,W$ only, it is equivalent to
the ordinary product)
\be
X(\sg_1\sg_2\sg_3)\,Z(\sg_3\sg_2\sg_1)\otimes Y(\tau_1\tau_2\tau_3)\,W(\tau_3\tau_2\tau_1), \labell{eq-5}
\ee
which means the orderings of $x_1,x_2,x_3$ are always opposite to those of $z_1,z_2,z_3$ and the same for $y_1,y_2,y_3$
and $w_1,w_2,w_3$. For $x$- and $z$-space there are $3!\!=\!6$ combinations, so there are in total 36 ordered subspaces
in this collection, which explains the counting in \eqref{eq-3}.
Since for each $D_{ij}$, both $X_{ij}$ and $Y_{ij}$ are positive, the positivity of
$D_{ij}$ is trivial, which leads to the proper numerator
\be
N=D_{12}D_{23}D_{13}
\ee
in the $d\log$ form (of any subspace in this collection)
\be
\frac{1}{x_{\sg_1}x_{\sg_2\sg_1}x_{\sg_3\sg_2}}\frac{1}{z_{\sg_3}z_{\sg_2\sg_3}z_{\sg_1\sg_2}}
\frac{1}{y_{\tau_1}y_{\tau_2\tau_1}y_{\tau_3\tau_2}}\frac{1}{w_{\tau_3}w_{\tau_2\tau_3}w_{\tau_1\tau_2}}
\frac{N}{D_{12}D_{23}D_{13}}.
\ee
To make use of the Mondrian diagrammatics, we pick an explicit subspace $X(123)Z(321)\!\otimes\!Y(123)W(321)$ as a
representative to separate its contributing and spurious parts. As extensively discussed
in \cite{Rao:2017fqc,An:2017tbf}, the identity
\be
D_{12}D_{23}D_{13}=X_{12}X_{23}D_{13}+Y_{12}Y_{23}D_{13}
+X_{13}X_{23}Y_{12}+X_{12}X_{13}Y_{23}+X_{12}Y_{13}Y_{23}+Y_{12}Y_{13}X_{23}
\ee
results in a vanishing spurious part, denoted by $S\!=\!0$. The relevant Mondrian seed diagrams are given
in figure \ref{fig-1}, corresponding to the six terms in the RHS above. This separation has significantly simplified
the summation as we only need to check whether the final sum of all spurious parts vanishes.

\subsection{Configuration $\{(++)_{12},(++)_{23},(+-)_{13}\}$}

\begin{figure}
\begin{center}
\includegraphics[width=0.55\textwidth]{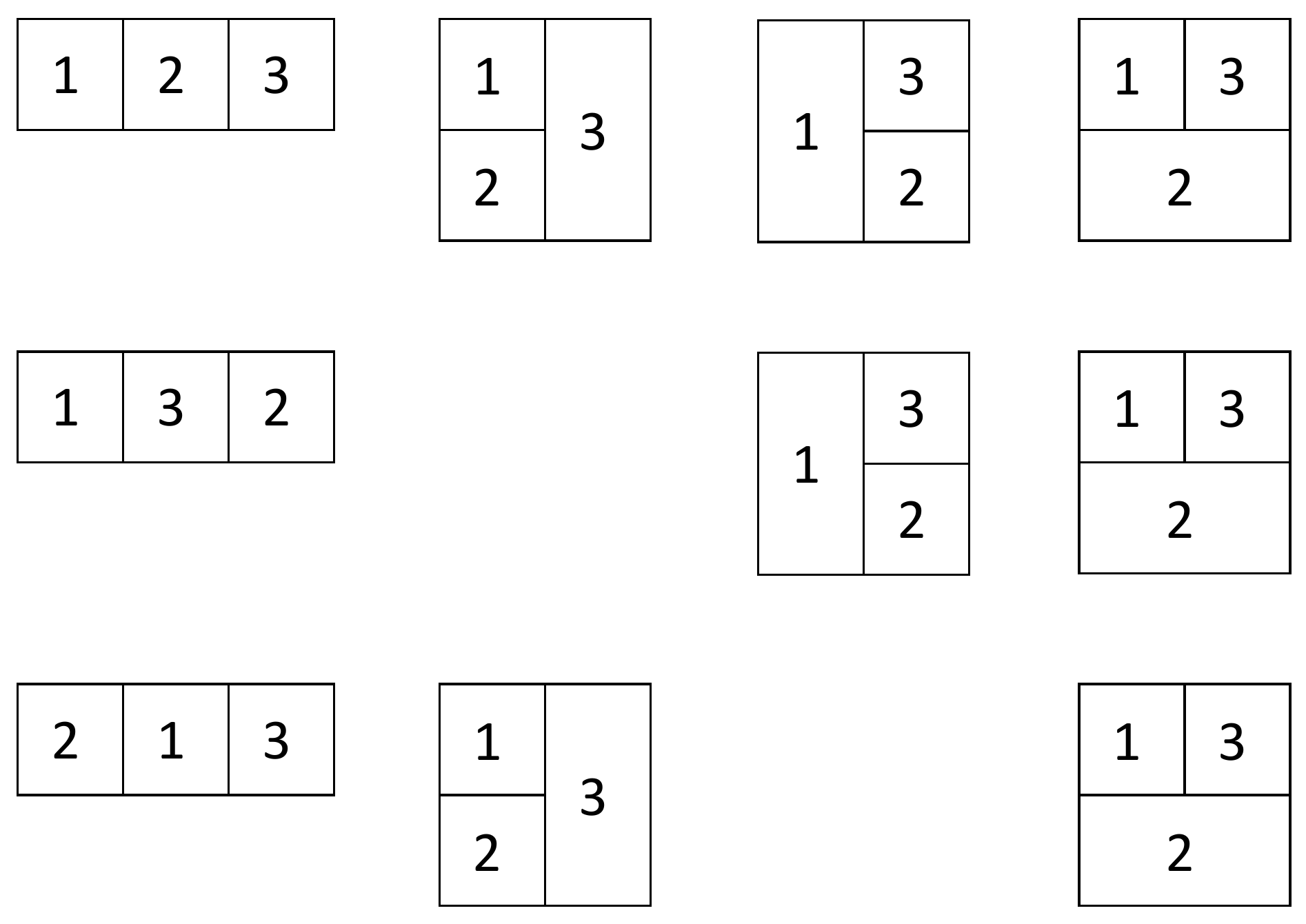}
\caption{Mondrian seed diagrams in subspaces $X(123)Z(321)\otimes Y(132)W(213)$, $X(132)Z(231)\otimes Y(132)W(213)$
and $X(213)Z(312)\otimes Y(132)W(213)$. Each row corresponds to one subspace respectively.} \label{fig-2}
\end{center}
\end{figure}

If we flip one plus into minus in the former case, we obtain the configuration $\{(++)_{12},(++)_{23},(+-)_{13}\}$.
Here $Y_{13}$ is chosen to be negative but of course, the negative quantity can be
$Y_{12}$, $Y_{23}$, $X_{12}$, $X_{23}$ or $X_{13}$ as well,
which explains the multiplicity of $3D^2(X\!+\!Y)$ in \eqref{eq-2}.
This corresponds to the collection of ordered subspaces
\be
X(\sg_1\sg_2\sg_3)\,Z(\sg_3\sg_2\sg_1)\otimes Y(\cdot\cdot2)W(2\cdot\cdot),
\ee
where
\be
\bal
Y(\cdot\cdot2)W(2\cdot\cdot)&\equiv Y(132)W(213)+Y(231)W(312)+(Y\!\leftrightarrow\!W)\\
&=Y(132)W(213)+Y(231)W(312)+Y(213)W(132)+Y(312)W(231) \labell{eq-4}
\eal
\ee
is the part satisfying $Y_{12},Y_{23}\!>\!0$ and $Y_{13}\!<\!0$. It is clear that there are in total $6\!\times\!4\!=\!24$
ordered subspaces in this collection. With the extra multiplicity $3\!\times\!2$, this explains the counting $24\!\times\!6$
in \eqref{eq-3}. To calculate the proper numerator, we observe that since only $Y_{13}$ is negative, the 2-loop analysis
for loop numbers 1,3 already suffices. Therefore we have
\be
N=D_{12}D_{23}X_{13}.
\ee
Then as usual, we pick some explicit representative subspaces to separate their contributing and spurious parts,
which include $X(123)Z(321)$, $X(132)Z(231)$ and $X(213)Z(312)$ among $X(\sg_1\sg_2\sg_3)Z(\sg_3\sg_2\sg_1)$ as we can
get the rest three by reversing the orderings of loop numbers in all parentheses or switching $X\!\leftrightarrow\!Z$,
and similarly $Y(132)W(213)$ among $Y(\cdot\cdot2)W(2\cdot\cdot)$. The relevant Mondrian seed diagrams of these three
subspaces are given in figure \ref{fig-2}.

Among these three cases, the only one with a nonzero spurious part is $X(123)Z(321)\!\otimes\!Y(132)W(213)$ with
(recall that it is the difference between the proper numerator and Mondrian factors)
\be
S=D_{12}D_{23}X_{13}-X_{12}X_{23}D_{13}-X_{13}X_{23}Y_{12}-X_{13}X_{12}Y_{23}-X_{13}Y_{12}Y_{23}=-\,X_{12}X_{23}Y_{13}.
\ee
To collect all spurious parts of this configuration, we need to permutate $13,23,12$ and switch
$x,z\!\leftrightarrow\!y,w$. For compactness, we can consider those associated with $X(123)$ only \cite{Rao:2017fqc},
so the relevant terms are
\be
X(123)Z(321)\otimes Y(\cdot\cdot2)W(2\cdot\cdot):~~-X_{12}X_{23}Y_{13},
\ee
as well as
\be
\bal
\[Y(132)W(231)+Y(231)W(132)\]\otimes X(123)Z(312):~~-Y_{13}Y_{23}X_{12},\\
\[Y(213)W(312)+Y(312)W(213)\]\otimes X(123)Z(231):~~-Y_{12}Y_{13}X_{23}.
\eal
\ee
These results will be summed over the forms of corresponding ordered subspaces for proving
all spurious parts finally cancel.

\subsection{Configuration $\{(++)_{12},(+-)_{23},(+-)_{13}\}$}

If we flip one more plus into minus at the same side in the former case, we get $\{(++)_{12},(+-)_{23},(+-)_{13}\}$.
Its multiplicity is similar to that of $\{(++)_{12},(++)_{23},(+-)_{13}\}$ as can be seen in \eqref{eq-2}.
This corresponds to the collection of ordered subspaces
\be
X(\sg_1\sg_2\sg_3)\,Z(\sg_3\sg_2\sg_1)\otimes Y(\cdot\cdot3)W(\cdot\cdot3),
\ee
where
\be
\bal
Y(\cdot\cdot3)W(\cdot\cdot3)&\equiv Y(123)W(213)+Y(321)W(312)+(Y\!\leftrightarrow\!W)\\
&=Y(123)W(213)+Y(321)W(312)+Y(213)W(123)+Y(312)W(321)
\eal
\ee
is the part satisfying $Y_{12}\!>\!0$ and $Y_{23},Y_{13}\!<\!0$. Similarly, there are in total $6\!\times\!4\!=\!24$
ordered subspaces in this collection. This explains the counting $24\!\times\!6$ in \eqref{eq-3} with the
extra multiplicity $3\!\times\!2$. In this case, to calculate the proper numerator is nontrivial and we can again pick
some explicit representative subspaces to analyze, which similarly include $X(123)Z(321)$, $X(132)Z(231)$,
$X(213)Z(312)$ and also $Y(123)W(213)$. Note that $X(213)Z(312)\!\otimes\!Y(123)W(213)$ is identical to
$X(123)Z(321)\!\otimes\!Y(123)W(213)$ if we switch $1\!\leftrightarrow\!2$ and $Y\!\leftrightarrow\!W$,
so there are only two distinct cases under consideration.

\begin{figure}
\begin{center}
\includegraphics[width=0.27\textwidth]{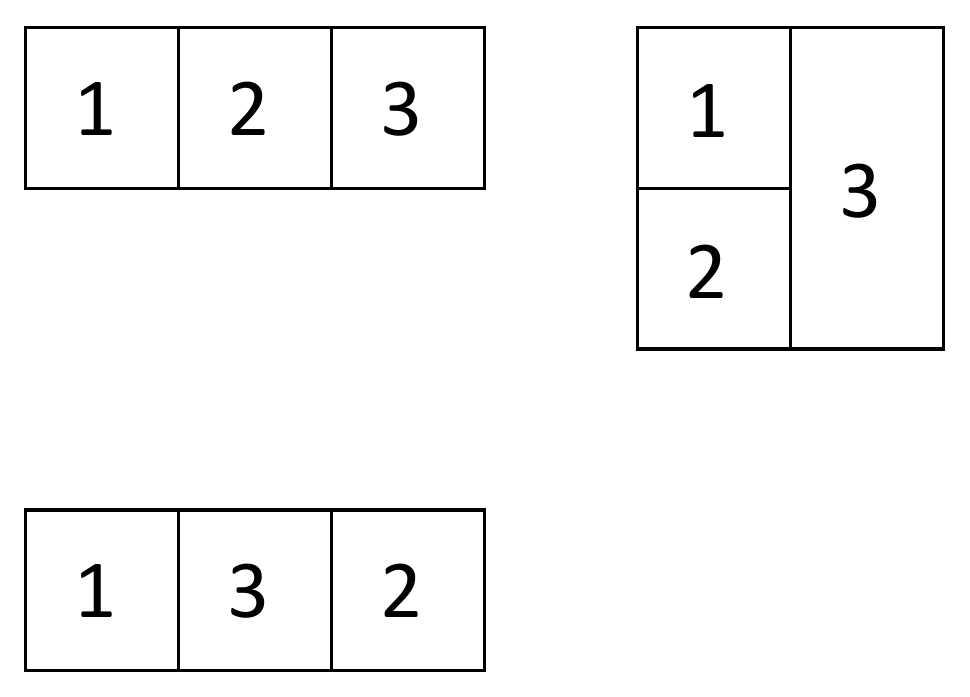}
\caption{Mondrian seed diagrams in subspaces $X(123)Z(321)\otimes Y(123)W(213)$
and $X(132)Z(231)\otimes Y(123)W(213)$.} \label{fig-3}
\end{center}
\end{figure}

For $X(123)Z(321)\!\otimes\!Y(123)W(213)$, $D_{12}$ is trivially positive, so we need to impose
\be
D_{23}=x_{32}z_{23}-y_{32}(w_{31}+w_{12})>0,~~D_{13}=(x_{32}+x_{21})(z_{12}+z_{23})-(y_{32}+y_{21})w_{31}>0.
\ee
For $D_{23}$ let's define
\be
z'_{23}\equiv z_{23}-\frac{y_{32}(w_{31}+w_{12})}{x_{32}}>0,
\ee
and its $d\log$ form is simply (for later convenience we multiply it by $z_{23}$ to make a dimensionless ratio)
\be
\frac{z_{23}}{z'_{23}}=\frac{X_{23}}{D_{23}}.
\ee
Next, for $D_{13}$ we have
\be
\bal
z_{12}+z_{23}-\frac{(y_{32}+y_{21})w_{31}}{x_{32}+x_{21}}
&=z_{12}+z'_{23}+\frac{y_{32}(w_{31}+w_{12})}{x_{32}}-\frac{(y_{32}+y_{21})w_{31}}{x_{32}+x_{21}}\\
&=z_{12}+z'_{23}+\frac{y_{32}}{x_{32}}\(w_{12}+w_{31}\frac{x_{21}}{x_{32}+x_{21}}\)
-\frac{y_{21}w_{31}}{x_{32}+x_{21}}>0,
\eal
\ee
we can focus on $z_{12}$, $z'_{23}$ and $y_{32}$, so its $d\log$ form is simply
(omitting $z_{12}$, $z'_{23}$ and $y_{32}$ in the denominator to make a dimensionless ratio,
and the form of $x_1\!+\!\ldots\!+\!x_n\!>\!a$ can be referred in \cite{Rao:2017fqc})
\be
\bal
&\[z_{12}+z'_{23}+\frac{y_{32}}{x_{32}}\(w_{12}+w_{31}\frac{x_{21}}{x_{32}+x_{21}}\)\]\bigg/
\[z_{12}+z'_{23}+\frac{y_{32}}{x_{32}}\(w_{12}+w_{31}\frac{x_{21}}{x_{32}+x_{21}}\)-\frac{y_{21}w_{31}}{x_{32}+x_{21}}\]\\
=\,&\,\frac{D_{13}+y_{21}w_{31}}{D_{13}}.
\eal
\ee
Collecting all three dimensionless ratios from the $d\log$ forms gives
\be
\frac{D_{12}}{D_{12}}\frac{X_{23}}{D_{23}}\frac{D_{13}+y_{21}w_{31}}{D_{13}},
\ee
the proper numerator is then $N\!=\!D_{12}X_{23}(D_{13}\!+\!y_{21}w_{31})$.
The relevant Mondrian seed diagrams of this subspace are given in the 1st row of figure \ref{fig-3},
and its spurious part is given by
\be
S=D_{12}X_{23}(D_{13}+y_{21}w_{31})-X_{12}X_{23}D_{13}-X_{13}X_{23}Y_{12}=X_{23}(Y_{12}Y_{13}+D_{12}\,y_{21}w_{31}).
\ee

For $X(132)Z(231)\!\otimes\!Y(123)W(213)$, similarly we need to impose
\be
D_{23}=x_{23}z_{32}-y_{32}(w_{31}+w_{12})>0,~~D_{13}=x_{31}z_{13}-(y_{32}+y_{21})w_{31}>0.
\ee
If we focus on $x_{23}$ and $x_{31}$, we find these two conditions in fact ``decouple''. Then the
dimensionless ratios (as a product) are simply
\be
\frac{D_{12}}{D_{12}}\frac{X_{23}}{D_{23}}\frac{X_{13}}{D_{13}},
\ee
with the proper numerator $N\!=\!D_{12}X_{23}X_{13}$. The relevant Mondrian seed diagram is given in the 2nd row
of figure \ref{fig-3}, and its spurious part is obviously $S\!=\!0$.

To collect all spurious parts of this configuration, we again permutate $13,23,12$ and switch
$x,z\!\leftrightarrow\!y,w$ for $X(123)Z(321)\!\otimes\!Y(123)W(213)$ and its derivative subspaces via
reversing the orderings of loop numbers and/or switching $Y\!\leftrightarrow\!W$. Fixing $X(123)$, the relevant terms are
\be
\bal
X(123)Z(321)&\otimes Y(123)W(213):~~X_{23}(Y_{12}Y_{13}+D_{12}\,y_{21}w_{31}),\\
\ldots\,&\otimes Y(321)W(312):~~X_{23}(Y_{12}Y_{13}+D_{12}\,y_{12}w_{13}),\\
\ldots\,&\otimes Y(213)W(123):~~X_{23}(Y_{12}Y_{13}+D_{12}\,w_{21}y_{31}),\\
\ldots\,&\otimes Y(312)W(321):~~X_{23}(Y_{12}Y_{13}+D_{12}\,w_{12}y_{13}),
\eal
\ee
\be
\bal
X(123)Z(321)&\otimes Y(321)W(231):~~X_{12}(Y_{23}Y_{13}+D_{23}\,y_{23}w_{13}),\\
\ldots\,&\otimes Y(123)W(132):~~X_{12}(Y_{23}Y_{13}+D_{23}\,y_{32}w_{31}),\\
\ldots\,&\otimes Y(231)W(321):~~X_{12}(Y_{23}Y_{13}+D_{23}\,w_{23}y_{13}),\\
\ldots\,&\otimes Y(132)W(123):~~X_{12}(Y_{23}Y_{13}+D_{23}\,w_{32}y_{31}),
\eal
\ee
where $\ldots$ stands for the repetitive subspace (and similar below), as well as
\be
\bal
&\[Y(123)W(321)+Y(321)W(123)\]\otimes X(123)Z(213)\!\!\!\!&:&~~~Y_{23}(X_{12}X_{13}+D_{12}\,x_{21}z_{31}),\\
&\[Y(213)W(312)+Y(312)W(213)\]\otimes\ldots\!\!\!\!&:&~~~Y_{13}(X_{12}X_{23}+D_{12}\,z_{12}x_{32}),
\eal
\ee
\be
\bal
&\[Y(321)W(123)+Y(123)W(321)\]\otimes X(123)Z(132)\!\!\!\!&:&~~~Y_{12}(X_{23}X_{13}+D_{23}\,x_{32}z_{31}),\\
&\[Y(231)W(132)+Y(132)W(231)\]\otimes\ldots\!\!\!\!&:&~~~Y_{13}(X_{23}X_{12}+D_{23}\,z_{23}x_{21}).
\eal
\ee
These results will be used for proving all spurious parts finally cancel.

\subsection{Configuration $\{(++)_{12},(+-)_{23},(-+)_{13}\}$}

If we replace $(+-)_{13}$ by $(-+)_{13}$ in the former case, we get $\{(++)_{12},(+-)_{23},(-+)_{13}\}$.
Now its multiplicity becomes 6 as can be seen in \eqref{eq-2}. This corresponds to the collection of ordered subspaces
\be
X(\cdot\cdot2)Z(2\cdot\cdot)\otimes Y(\cdot\cdot1)W(1\cdot\cdot),
\ee
where $X(\cdot\cdot2)Z(2\cdot\cdot)$ and $Y(\cdot\cdot1)W(1\cdot\cdot)$ are similarly defined by \eqref{eq-4}.
There are in total $4^2\!=\!16$ ordered subspaces in this collection, which explains the counting
$16\!\times\!6$ in \eqref{eq-3}. To get the proper numerator, we again pick a representative subspace
$X(132)Z(213)\!\otimes\!Y(231)W(123)$ to analyze.

\begin{figure}
\begin{center}
\includegraphics[width=0.23\textwidth]{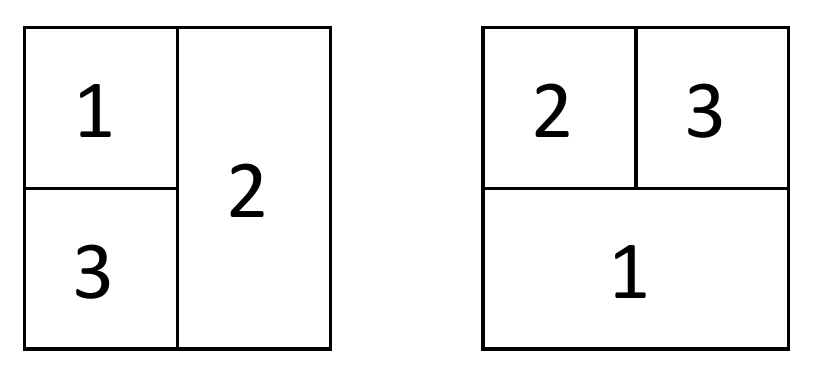}
\caption{Mondrian seed diagrams in subspace $X(132)Z(213)\otimes Y(231)W(123)$.} \label{fig-4}
\end{center}
\end{figure}

Since $D_{12}$ is trivially positive, we need to impose
\be
D_{23}=x_{23}(z_{31}+z_{12})-y_{32}w_{32}>0,~~D_{13}=-\,x_{31}z_{31}+y_{13}(w_{32}+w_{21})>0.
\ee
Focusing on $x_{23}$ and $x_{31}$, we find these two conditions decouple. Then the
dimensionless ratios are
\be
\frac{D_{12}}{D_{12}}\frac{X_{23}}{D_{23}}\frac{Y_{13}}{D_{13}},
\ee
with the proper numerator $N\!=\!D_{12}X_{23}Y_{13}$. The relevant Mondrian seed diagrams are given in
figure \ref{fig-4}, and its spurious part is obviously $S\!=\!0$. Therefore, similar to configuration
$\{(++)_{12},(++)_{23},(++)_{13}\}$, in this case there is no spurious part to be collected.

\subsection{Configuration $\{(+-)_{12},(+-)_{23},(+-)_{13}\}$}

For this configuration, we have three minus signs at the same side.
Its multiplicity is 2, due to switching $X\!\leftrightarrow Y$ in \eqref{eq-2}.
This corresponds to the collection of ordered subspaces
\be
X(\sg_1\sg_2\sg_3)\,Z(\sg_3\sg_2\sg_1)\otimes Y(\tau_1\tau_2\tau_3)\,W(\tau_1\tau_2\tau_3).
\ee
Similar to \eqref{eq-5}, there are in total 36 ordered subspaces in this collection,
which explains the counting $36\!\times\!2$ in \eqref{eq-3}. We again pick some representative subspaces to analyze,
in fact there are only two distinct cases: $X(123)Z(321)\!\otimes\!Y(123)W(123)$ and
$X(123)Z(321)\!\otimes\!Y(132)W(132)$.

\begin{figure}
\begin{center}
\includegraphics[width=0.15\textwidth]{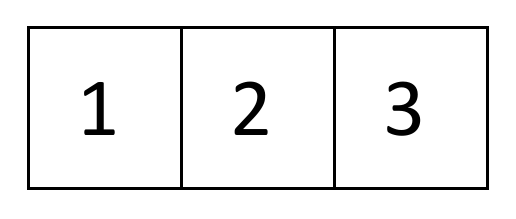}
\caption{Mondrian seed diagram in subspaces $X(123)Z(321)\otimes Y(123)W(123)$ and
$X(123)Z(321)\otimes Y(132)W(132)$.} \label{fig-5}
\end{center}
\end{figure}

For $X(123)Z(321)\!\otimes\!Y(123)W(123)$, we need to impose
\be
\bal
&D_{12}=x_{21}z_{12}-y_{21}w_{21}>0,~~D_{23}=x_{32}z_{23}-y_{32}w_{32}>0,\\
&D_{13}=(x_{32}+x_{21})(z_{12}+z_{23})-(y_{32}+y_{21})(w_{32}+w_{21})>0.
\eal
\ee
For $D_{12}$ and $D_{23}$ let's define
\be
z'_{12}\equiv z_{12}-\frac{y_{21}w_{21}}{x_{21}}>0,~~z'_{23}\equiv z_{23}-\frac{y_{32}w_{32}}{x_{32}}>0,
\ee
next for $D_{13}$ we have
\be
\bal
&\,z'_{12}+z'_{23}-\(\frac{(y_{32}+y_{21})(w_{32}+w_{21})}{x_{32}+x_{21}}
-\frac{y_{21}w_{21}}{x_{21}}-\frac{y_{32}w_{32}}{x_{32}}\)\\
=\,&\,z'_{12}+z'_{23}-\frac{x_{21}}{x_{32}(x_{32}+x_{21})}
\(y_{32}-y_{21}\,\frac{x_{32}}{x_{21}}\)\(\frac{x_{32}}{x_{21}}\,w_{21}-w_{32}\)>0,
\eal
\ee
this condition is only nontrivial when
\be
a\equiv\frac{x_{21}}{x_{32}(x_{32}+x_{21})}
\(y_{32}-y_{21}\,\frac{x_{32}}{x_{21}}\)\(\frac{x_{32}}{x_{21}}\,w_{21}-w_{32}\)>0,
\ee
so its $d\log$ form is (omitting $z'_{12}$ and $z'_{23}$ in the denominator as usual)
\be
\bal
\!\!\!\!\!\!\!\!\!
&\[\frac{1}{y_{32}-y_{21}x_{32}/x_{21}}\(\frac{1}{w_{32}}-\frac{1}{w_{32}-w_{21}x_{32}/x_{21}}\)
+\(\frac{1}{y_{32}}-\frac{1}{y_{32}-y_{21}x_{32}/x_{21}}\)\frac{1}{w_{32}-w_{21}x_{32}/x_{21}}\]
\frac{z'_{12}+z'_{23}}{z'_{12}+z'_{23}-a}\\
\!\!\!\!\!\!\!\!\!
&+\[\frac{1}{y_{32}-y_{21}x_{32}/x_{21}}\frac{1}{w_{32}-w_{21}x_{32}/x_{21}}
+\(\frac{1}{y_{32}}-\frac{1}{y_{32}-y_{21}x_{32}/x_{21}}\)\(\frac{1}{w_{32}}-\frac{1}{w_{32}-w_{21}x_{32}/x_{21}}\)\]\\
\!\!\!\!\!\!\!\!\!
=\,&\,\frac{D_{13}+y_{32}w_{21}+y_{21}w_{32}}{y_{32}w_{32}D_{13}}.
\eal
\ee
Collecting all three dimensionless ratios gives
\be
\frac{X_{12}}{D_{12}}\frac{X_{23}}{D_{23}}\frac{D_{13}+y_{32}w_{21}+y_{21}w_{32}}{D_{13}},
\ee
with the proper numerator $N\!=\!X_{12}X_{23}(D_{13}\!+\!y_{32}w_{21}\!+\!y_{21}w_{32})$.
The relevant Mondrian seed diagram is given in figure \ref{fig-5}, and its spurious part is obviously
$S\!=\!X_{12}X_{23}(y_{32}w_{21}\!+\!y_{21}w_{32})$.

For $X(123)Z(321)\!\otimes\!Y(132)W(132)$, similarly we need to impose
\be
\bal
&D_{12}=x_{21}z_{12}-(y_{23}+y_{31})(w_{23}+w_{31})>0,~~D_{23}=x_{32}z_{23}-y_{23}w_{23}>0,\\
&D_{13}=(x_{32}+x_{21})(z_{12}+z_{23})-y_{31}w_{31}>0.
\eal
\ee
Focusing on $z_{12}$ and $z_{23}$, we find $D_{12}\!>\!0$ and $D_{23}\!>\!0$ decouple, and $D_{12}\!>\!0$
can trivialize $D_{13}\!>\!0$. Then the dimensionless ratios are
\be
\frac{X_{12}}{D_{12}}\frac{X_{23}}{D_{23}}\frac{D_{13}}{D_{13}},
\ee
with the proper numerator $N\!=\!X_{12}X_{23}D_{13}$. The relevant Mondrian seed diagram is identical to
that of $X(123)Z(321)\!\otimes\!Y(123)W(123)$ given in figure \ref{fig-5}, and its spurious part is obviously $S\!=\!0$.

To collect all spurious parts of this configuration, we again permutate $13,23,12$ and switch
$x,z\!\leftrightarrow\!y,w$ for $X(123)Z(321)\!\otimes\!Y(123)W(123)$ and its derivative subspaces.
Fixing $X(123)$, the relevant terms are
\be
\bal
X(123)Z(321)&\otimes Y(123)W(123):~~X_{12}X_{23}(y_{32}w_{21}+y_{21}w_{32}),\\
\ldots\,&\otimes Y(321)W(321):~~X_{12}X_{23}(y_{23}w_{12}+y_{12}w_{23}),
\eal
\ee
as well as
\be
\bal
Y(123)W(321)&\otimes X(123)Z(123)\!\!\!\!&:&~~~\,Y_{12}Y_{23}(x_{32}z_{21}+x_{21}z_{32}),\\
Y(321)W(123)&\otimes\ldots\!\!\!\!&:&~~~\,Y_{12}Y_{23}(x_{32}z_{21}+x_{21}z_{32}).
\eal
\ee
These results will be used for proving all spurious parts finally cancel.

\subsection{Configuration $\{(+-)_{12},(+-)_{23},(-+)_{13}\}$}

\begin{figure}
\begin{center}
\includegraphics[width=0.1\textwidth]{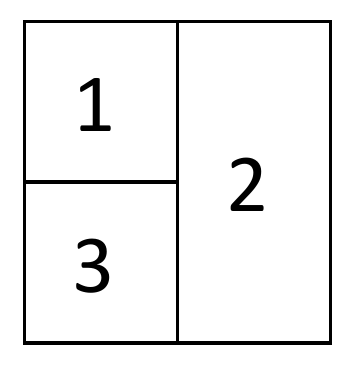}
\caption{Mondrian seed diagram in subspace $X(132)Z(213)\otimes Y(132)W(312)$.} \label{fig-6}
\end{center}
\end{figure}

If we replace $(+-)_{13}$ by $(-+)_{13}$ in the former case, we get $\{(+-)_{12},(+-)_{23},(-+)_{13}\}$.
Its multiplicity is $3\!\times\!2$, due to choosing one of $12,23,13$ to assign $(-+)$ and
switching $X\!\leftrightarrow Y$ in \eqref{eq-2}. This corresponds to the collection of ordered subspaces
\be
X(\cdot\cdot2)Z(2\cdot\cdot)\otimes Y(\cdot\cdot2)W(\cdot\cdot2).
\ee
There are in total $4^2\!=\!16$ ordered subspaces in this collection, which explains the counting
$16\!\times\!6$ in \eqref{eq-3}. To get the proper numerator,
we again pick a representative subspace $X(132)Z(213)\!\otimes\!Y(132)W(312)$ to analyze, for which we need to impose
\be
\bal
&D_{12}=(x_{23}+x_{31})z_{12}-(y_{23}+y_{31})w_{21}\equiv(x_{23}+x_{31})z'_{12}>0,\\
&D_{23}=x_{23}(z_{31}+z_{12})-y_{23}(w_{21}+w_{13})>0,\\
&D_{13}=-\,x_{31}z_{31}+y_{31}w_{13}\equiv y_{31}w'_{13}>0,
\eal
\ee
where similarly $z'_{12}$ and $w'_{13}$ are positive variables, so that for $D_{23}$ we have
\be
\(1-\frac{y_{23}}{x_{23}}\frac{x_{31}}{y_{31}}\)z_{31}+z'_{12}
+\(\frac{y_{23}+y_{31}}{x_{23}+x_{31}}-\frac{y_{23}}{x_{23}}\)w_{21}-\frac{y_{23}}{x_{23}}\,w'_{13}>0,
\ee
note that
\be
\frac{y_{23}}{x_{23}}\lessgtr\frac{y_{31}}{x_{31}}\Longrightarrow
\frac{y_{23}}{x_{23}}\lessgtr\frac{y_{23}+y_{31}}{x_{23}+x_{31}}\lessgtr\frac{y_{31}}{x_{31}},
\ee
which determines signs of the factors of $z_{31}$ and $w_{21}$,
so its $d\log$ form is (omitting $z_{31}$, $z'_{12}$ and $w_{21}$ in the denominator)
\be
\bal
&\frac{1}{y_{31}-y_{23}\,x_{31}/x_{23}}\[\(1-\frac{y_{23}}{x_{23}}\frac{x_{31}}{y_{31}}\)z_{31}+z'_{12}
+\(\frac{y_{23}+y_{31}}{x_{23}+x_{31}}-\frac{y_{23}}{x_{23}}\)w_{21}\]\frac{x_{23}}{D_{23}}\\
&+\(\frac{1}{y_{31}}-\frac{1}{y_{31}-y_{23}\,x_{31}/x_{23}}\)\frac{z'_{12}\,x_{23}}{D_{23}}\\
=\,&\,\frac{1}{y_{31}D_{23}}\(x_{23}(z_{31}+z_{12})-\frac{x_{23}}{x_{23}+x_{31}}\,y_{23}w_{21}\).
\eal
\ee
Collecting all three dimensionless ratios gives
\be
\frac{X_{12}}{D_{12}}\frac{Y_{13}}{D_{13}}\frac{1}{D_{23}}\(X_{23}-\frac{x_{23}}{x_{23}+x_{31}}\,y_{23}w_{21}\),
\ee
with the proper numerator $N\!=\!X_{12}Y_{13}(X_{23}\!-\!y_{23}w_{21}x_{23}/(x_{23}\!+\!x_{31}))$.
The relevant Mondrian seed diagram is given in figure \ref{fig-6}, and its spurious part is obviously
$S\!=\!X_{12}Y_{13}(-y_{23}w_{21}x_{23}/(x_{23}\!+\!x_{31}))$.

To collect all spurious parts of this configuration, we again permutate $13,23,12$ and switch
$x,z\!\leftrightarrow\!y,w$ for $X(132)Z(213)\!\otimes\!Y(132)W(312)$ and its derivative subspaces.
Fixing $X(123)$, the relevant terms are
\be
\bal
X(123)Z(312)&\otimes Y(123)W(213):~~X_{13}Y_{12}\(-\frac{x_{32}}{x_{32}+x_{21}}\,y_{32}w_{31}\),\\
\ldots\,&\otimes Y(321)W(312):~~X_{13}Y_{12}\(-\frac{x_{32}}{x_{32}+x_{21}}\,y_{23}w_{13}\),\\
\ldots\,&\otimes Y(213)W(123):~~X_{13}Y_{12}\(-\frac{x_{32}}{x_{32}+x_{21}}\,w_{32}y_{31}\),\\
\ldots\,&\otimes Y(312)W(321):~~X_{13}Y_{12}\(-\frac{x_{32}}{x_{32}+x_{21}}\,w_{23}y_{13}\),
\eal
\ee
\be
\bal
X(123)Z(231)&\otimes Y(231)W(321):~~X_{12}Y_{23}\(-\frac{z_{13}}{z_{13}+z_{32}}\,y_{13}w_{12}\),\\
\ldots\,&\otimes Y(132)W(123):~~X_{12}Y_{23}\(-\frac{z_{13}}{z_{13}+z_{32}}\,y_{31}w_{21}\),\\
\ldots\,&\otimes Y(321)W(231):~~X_{12}Y_{23}\(-\frac{z_{13}}{z_{13}+z_{32}}\,w_{13}y_{12}\),\\
\ldots\,&\otimes Y(123)W(132):~~X_{12}Y_{23}\(-\frac{z_{13}}{z_{13}+z_{32}}\,w_{31}y_{21}\),
\eal
\ee
\\ \\ \\ \\ \\ \\
as well as
\be
\bal
Y(123)W(312)&\otimes X(123)Z(213)\!\!\!\!&:&~~~Y_{13}X_{12}\(-\frac{y_{32}}{y_{32}+y_{21}}\,x_{32}z_{31}\),\\
Y(321)W(213)&\otimes\ldots\!\!\!\!&:&~~~Y_{13}X_{12}\(-\frac{y_{23}}{y_{12}+y_{23}}\,x_{32}z_{31}\),\\
Y(312)W(123)&\otimes\ldots\!\!\!\!&:&~~~Y_{13}X_{12}\(-\frac{w_{32}}{w_{32}+w_{21}}\,x_{32}z_{31}\),\\
Y(213)W(321)&\otimes\ldots\!\!\!\!&:&~~~Y_{13}X_{12}\(-\frac{w_{23}}{w_{12}+w_{23}}\,x_{32}z_{31}\),
\eal
\ee
\be
\bal
Y(231)W(123)&\otimes X(123)Z(132)\!\!\!\!&:&~~~Y_{12}X_{23}\(-\frac{y_{13}}{y_{13}+y_{32}}\,z_{31}x_{21}\),\\
Y(132)W(321)&\otimes\ldots\!\!\!\!&:&~~~Y_{12}X_{23}\(-\frac{y_{31}}{y_{23}+y_{31}}\,z_{31}x_{21}\),\\
Y(123)W(231)&\otimes\ldots\!\!\!\!&:&~~~Y_{12}X_{23}\(-\frac{w_{13}}{w_{13}+w_{32}}\,z_{31}x_{21}\),\\
Y(321)W(132)&\otimes\ldots\!\!\!\!&:&~~~Y_{12}X_{23}\(-\frac{w_{31}}{w_{23}+w_{31}}\,z_{31}x_{21}\).
\eal
\ee
These results will be used for proving all spurious parts finally cancel.

\subsection{Final sum of all spurious parts}

One might notice that, even though we treat all $x,y,z,w$ variables on the same footing and preserve the
symmetry in combinations $12,23,13$, we can still consider terms associated with $X(123)$ only because
we would like to confirm the sum of all spurious parts in subspace $X(123)$ matches the result in \cite{Rao:2017fqc}.

Explicitly, we collect those nonzero spurious parts in configurations $\{(++)_{12},(++)_{23},(+-)_{13}\}$,\\
$\{(++)_{12},(+-)_{23},(+-)_{13}\}$, $\{(+-)_{12},(+-)_{23},(+-)_{13}\}$ and $\{(+-)_{12},(+-)_{23},(-+)_{13}\}$
then sum them over the forms of corresponding ordered subspaces, which gives the proper numerator
\be
S_{123}=x_{21}(-2\,z_1y_2y_3w_2w_3-z_1y_1w_1(y_2w_3+y_3w_2)+z_2y_3w_3(y_1w_2+y_2w_1)+z_3y_2w_2(y_1w_3+y_3w_1)),
\ee
and hence the final sum over permutations of loop numbers
\be
S_{123}X(123)+(5\textrm{ permutations of 1,2,3})=0.
\ee
In fact, this vanishing result can be further refined as $S_{123}X(123)\!+\!S_{132}X(132)\!=\!0$,
which has not been noticed in \cite{Rao:2017fqc}.
\\ \\ \\ \\

\subsection{Technical bottleneck at 4-loop}

Completing the 3-loop proof, it is appealing to continue this approach at 4-loop. We can have a glance at the variety of
its positive configurations via the generating function, as a generalization of \eqref{eq-2}:
\be
\bal
\!\!\!\!\!\!\!\!\!\!\!\!\!\!\!\!\!\!\!\!\!\!\!\!
(D\!+\!X\!+\!Y)^6\!=\,&D^6\!+\!6\,D^5(X\!+\!Y)\!+\!15\,D^4\(X^2\!+\!Y^2\)\!+\!30\,D^4XY
\!+\!20\,D^3\(X^3\!+\!Y^3\)\!+\!60\,D^3\(X^2Y\!+\!XY^2\)\\
&\!+\!15\,D^2\(X^4\!+\!Y^4\)\!+\!60\,D^2\(X^3Y\!+\!XY^3\)\!+\!90\,D^2X^2Y^2
\!+\!6\,D\(X^5\!+\!Y^5\)\!+\!30\,D\(X^4Y\!+\!XY^4\)\\
&\!+\!60\,D\(X^3Y^2\!+\!X^2Y^3\)\!+\!\(X^6\!+\!Y^6\)\!+\!6\(X^5Y\!+\!XY^5\)\!+\!15\(X^4Y^2\!+\!X^2Y^4\)\!+\!20\,X^3Y^3,
\eal
\ee
so there are 16 distinct configurations. Taking $X^6$ as one of the most nontrivial examples, or equivalently,
the configuration in terms of plus and minus signs
\be
\{(+-)_{12},(+-)_{23},(+-)_{34},(+-)_{13},(+-)_{24},(+-)_{14}\},
\ee
we can pick the representative subspace $X(1234)Z(4321)\!\otimes\!Y(1234)W(1234)$ to analyze,
for which we need to impose
\be
\bal
&D_{12}=x_{21}z_{12}-y_{21}w_{21}>0,~~D_{23}=x_{32}z_{23}-y_{32}w_{32}>0,~~D_{34}=x_{43}z_{34}-y_{43}w_{43}>0,\\
&D_{13}=(x_{32}+x_{21})(z_{12}+z_{23})-(y_{32}+y_{21})(w_{32}+w_{21})>0,\\
&D_{24}=(x_{43}+x_{32})(z_{23}+z_{34})-(y_{43}+y_{32})(w_{43}+w_{32})>0,\\
&D_{14}=(x_{43}+x_{32}+x_{21})(z_{12}+z_{23}+z_{34})-(y_{43}+y_{32}+y_{21})(w_{43}+w_{32}+w_{21})>0.
\eal
\ee
For $D_{12}$, $D_{23}$ and $D_{34}$ let's define
\be
z'_{12}\equiv z_{12}-\frac{y_{21}w_{21}}{x_{21}}>0,~~z'_{23}\equiv z_{23}-\frac{y_{32}w_{32}}{x_{32}}>0,~~
z'_{34}\equiv z_{34}-\frac{y_{43}w_{43}}{x_{43}}>0,
\ee
then for $D_{13}$, $D_{24}$ and $D_{14}$ we have
\be
\bal
(x_{32}+x_{21})(z'_{12}+z'_{23})
&-x_{32}\,x_{21}\(\frac{y_{32}}{x_{32}}-\frac{y_{21}}{x_{21}}\)\(\frac{w_{21}}{x_{21}}-\frac{w_{32}}{x_{32}}\)>0,\\
(x_{43}+x_{32})(z'_{23}+z'_{34})
&-x_{43}\,x_{32}\(\frac{y_{43}}{x_{43}}-\frac{y_{32}}{x_{32}}\)\(\frac{w_{32}}{x_{32}}-\frac{w_{43}}{x_{43}}\)>0,\\
(x_{43}+x_{32}+x_{21})(z'_{12}+z'_{23}+z'_{34})
&-x_{32}\,x_{21}\(\frac{y_{32}}{x_{32}}-\frac{y_{21}}{x_{21}}\)\(\frac{w_{21}}{x_{21}}-\frac{w_{32}}{x_{32}}\)\\
-\,x_{43}\,x_{32}\(\frac{y_{43}}{x_{43}}-\frac{y_{32}}{x_{32}}\)\(\frac{w_{32}}{x_{32}}-\frac{w_{43}}{x_{43}}\)
&-x_{43}\,x_{21}\(\frac{y_{43}}{x_{43}}-\frac{y_{21}}{x_{21}}\)\(\frac{w_{21}}{x_{21}}-\frac{w_{43}}{x_{43}}\)>0.
\labell{eq-8}
\eal
\ee
Note that this smallest sector of the 4-loop amplituhedron almost has the complexity of the entire 3-loop case
already! As the loop order increases, the calculational complexity grows explosively. This advises us to stop at
4-loop even though we have a maximally refined recipe to dissect the iceberg of amplituhedron.

\subsection{Motivation of positive cuts}

Before moving on to the 4-loop amplituhedron using a different approach, it is pedagogical to manipulate the known
3-loop case first to see how it works. Naturally, we would like to impose traditional cuts on the amplituhedron and check
the validity of positivity conditions in this simplified situation.

Back to the two distinct 3-loop topologies, namely the diagrams given in figures \ref{fig-5} and \ref{fig-6}
without loss of generality, we can tentatively cut all of their external propagators and evaluate the $d\log$ forms
of the remaining variables. Explicitly, for figure \ref{fig-5} the corresponding integrand is
\be
\frac{1}{x_1z_3\,y_1y_2y_3w_1w_2w_3D_{12}D_{23}},
\ee
cutting all external propagators as $x_1\!=\!z_3\!=\!y_1\!=\!y_2\!=\!y_3\!=\!w_1\!=\!w_2\!=\!w_3\!=\!0$ gives
\be
D_{12}=x_2(z_1-z_2),~~D_{23}=z_2(x_3-x_2),~~D_{13}=x_3z_1.
\ee
The remaining variables are $x_2,x_3,z_1,z_2$, and we need to further impose $z_1\!>\!z_2$ and $x_3\!>\!x_2$ to ensure
the positivity of $D_{12}$ and $D_{23}$, while $D_{13}$ is trivially positive. The residue of this integrand is
\be
\frac{1}{D_{12}D_{23}}=\frac{1}{x_2(x_3-x_2)z_2(z_1-z_2)},
\ee
and the RHS above is clearly the $d\log$ form of remaining variables $x_2,x_3,z_1,z_2$, consistent with positivity.
Then for figure \ref{fig-6} with the integrand (numerator $x_2$ below is the rung rule
factor \cite{Bern:2006ew,Bern:2007ct})
\be
\frac{x_2}{x_1x_3z_2\,y_1y_2w_2w_3D_{12}D_{23}D_{13}},
\ee
similarly the cuts $x_1\!=\!x_3\!=\!z_2\!=\!y_1\!=\!y_2\!=\!w_2\!=\!w_3\!=\!0$ lead to
\be
D_{12}=x_2z_1,~~D_{23}=x_2z_3,~~D_{13}=y_3w_1.
\ee
The remaining variables are $x_2,z_1,z_3,y_3,w_1$, and since $D_{12},D_{23},D_{13}$ are all trivially positive,
there is no further positivity condition to be imposed. The residue of this integrand is
\be
\frac{x_2}{D_{12}D_{23}D_{13}}=\frac{1}{x_2z_1z_3\,y_3w_1},
\ee
and the RHS above is trivially the $d\log$ form of $x_2,z_1,z_3,y_3,w_1$.

From these simple examples we see the traditional cuts work in an even easier way in the context of amplituhedron,
which inspires us to apply these techniques at higher loop orders, and it is interesting to check the
consistency between amplituhedron and the known results obtained via cuts.

In fact, in the first case of figure \ref{fig-5} above, we can even further cut internal
propagators $D_{12}$ and $D_{23}$ by setting $z_1\!=\!z_2$ and $x_3\!=\!x_2$, which are the
positive cuts that we will introduce immediately. Compared to the straightforward approach,
calculation of amplituhedron with positive cuts is much simpler, but we need the ansatz of a basis of
DCI loop integrals as explained in the next section.

\section{Positive Cuts at 4-loop}

For the 4-loop case besides continuing a direct derivation, we will also alleviate the calculational difficulty
by imitating the traditional (generalized) unitarity cuts, which is to use the positive cuts.
In this way, we can peel off the unnecessary flesh of the amplituhedron and concentrate on its essential skeleton -- the
pole structure. Given a basis of DCI loop integrals, we can first assign each DCI topology with an
undetermined coefficient. Then after imposing as many positive cuts as possible for various pole structures, in general
we obtain a set of equations by equating each resulting $d\log$ form via positivity conditions,
and the deformed integrand as a sum of all non-vanishing DCI diagrams under the corresponding cuts.
These equations will be complete for determining all coefficients.

\begin{figure}
\begin{center}
\includegraphics[width=0.6\textwidth]{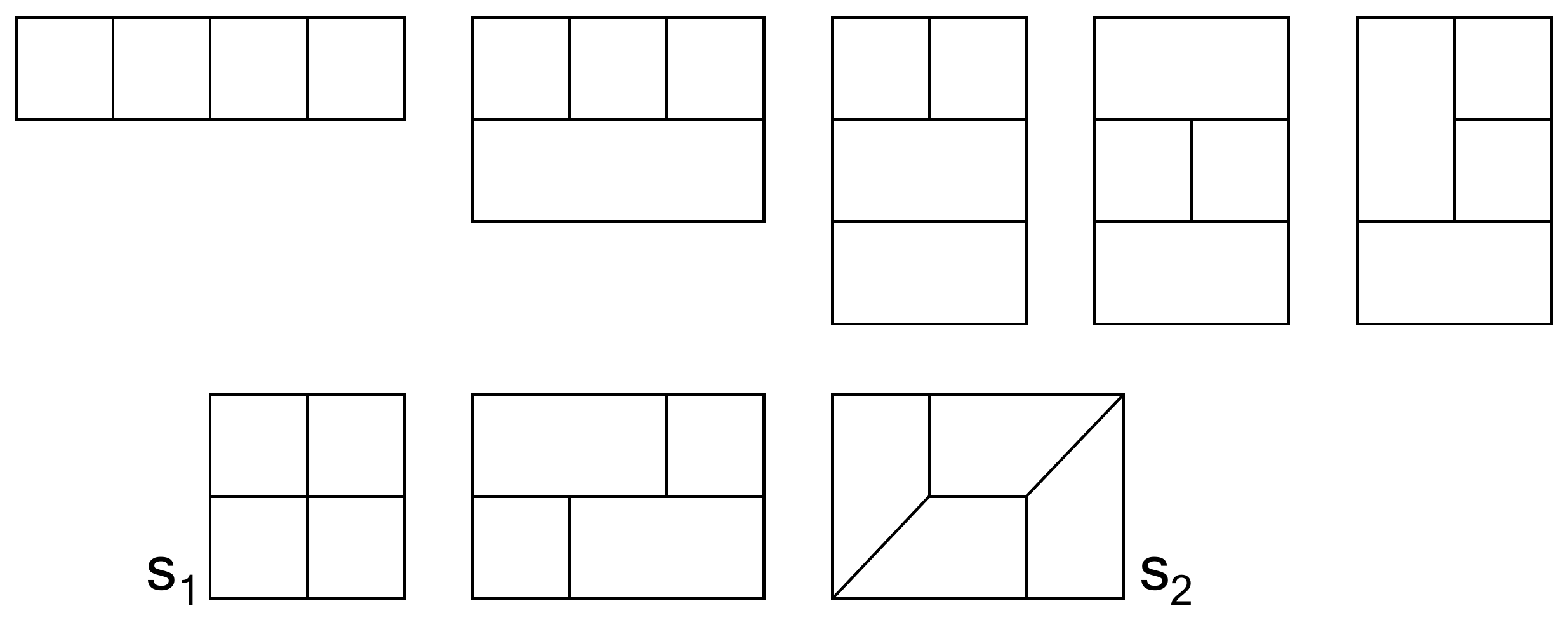}
\caption{All eight distinct DCI topologies at 4-loop. $s_1$ and $s_2$ are coefficients
of two non-rung-rule topologies.} \label{fig-7}
\end{center}
\end{figure}

\begin{figure}
\begin{center}
\includegraphics[width=0.165\textwidth]{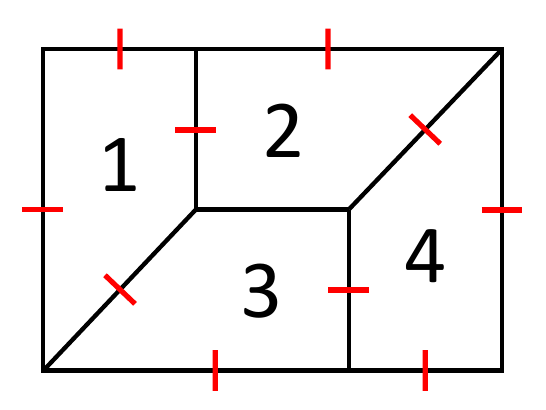}
\caption{A particular diagram of the non-Mondrian topology at 4-loop with 6 external and 4 internal cuts.} \label{fig-8}
\end{center}
\end{figure}

However, as a simplified demonstration, below we will focus on the non-rung-rule topologies at 4-loop
(of course, it is an interesting and challenging problem to prove the rung rule preserves coefficients of DCI topologies
while increasing the number of loops, using the amplituhedron approach). First, we enumerate all eight distinct
DCI topologies at 4-loop in figure \ref{fig-7}, among which the cross and the only non-Mondrian topology
are of the non-rung-rule type, while the other six rung-rule (and also Mondrian) topologies
are all associated with the coefficient $+1$. It is important to recall that, the term `DCI topology' includes the
numerator part as this matters for dual conformal invariance \cite{An:2017tbf}, but for convenience we will not
draw the extra numerators explicitly as they can be inferred from the rung rule, as long as there is no ambiguity
in the choices of DCI numerator. Then we assign the cross and non-Mondrian topologies with
coefficients $s_1$ and $s_2$ respectively, and consider a particular diagram of the latter type
given in figure \ref{fig-8}.

For this diagram, we can first maximally impose all 6 available external cuts, as indicated by the red segments
around its rim. Following the convention of external face variables in \cite{Rao:2017fqc,An:2017tbf}, these 6 cuts
result in $x_1\!=\!y_1\!=\!y_2\!=\!z_4\!=\!w_4\!=\!w_3\!=\!0$, which can simplify the six $D$'s as
\be
\bal
&D_{12}=x_2(z_1-z_2),\\
&D_{34}=z_3(x_4-x_3),\\
&D_{13}=x_3(z_1-z_3)+y_3w_1,\\
&D_{24}=z_2(x_4-x_2)+y_4w_2,\\
&D_{23}=(x_3-x_2)(z_2-z_3)+y_3w_2,\\
&D_{14}=x_4z_1+y_4w_1.\\
\eal
\ee
Now for part of these $D$'s as internal propagators, we can either cut them or impose their positivity.
Note that there is no way to further cut $D_{14}$ by fixing one variable, as discussed in \cite{Arkani-Hamed:2013kca},
but since it is manifestly positive already, there is no positivity condition to be imposed. By tentatively setting
\be
z_1=z_2,~~x_4=x_3,~~z_3=z_2+\frac{y_3w_1}{x_3}\equiv\hat{z}_3,~~x_2=x_3+\frac{y_4w_2}{z_2}\equiv\hat{x}_2, \labell{eq-6}
\ee
we can turn off $D_{12},D_{34},D_{13},D_{24}$, and incidentally we have
\be
D_{23}=y_3w_2\(1+\frac{y_4w_1}{x_3z_2}\),
\ee
which is also manifestly positive, therefore we are done with this further simplification. Note the solutions
of $D_{12}\!=\!D_{34}\!=\!D_{13}\!=\!D_{24}\!=\!0$, namely \eqref{eq-6}, are also manifestly positive. In contrast,
solutions that involve relative minus signs, such as $z_3\!=\!z_2\!-\!y_3w_1/x_3$, are clearly not, since we also
have to impose $z_2\!>\!y_3w_1/x_3$. Such a category of manifestly positive solutions will be
named as the \textit{positive cuts}.

\begin{figure}
\begin{center}
\includegraphics[width=0.56\textwidth]{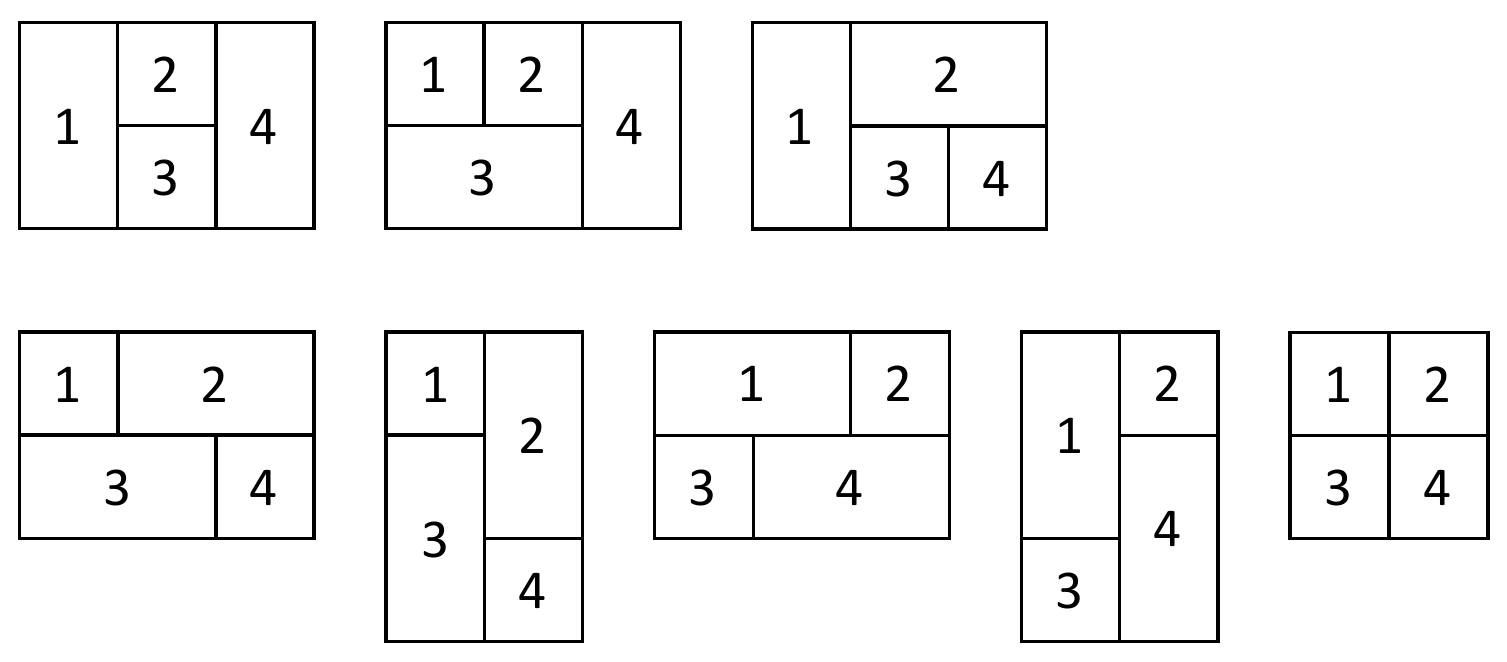}
\caption{All other 8 diagrams that survive the 10 cuts
$x_1\!=\!y_1\!=\!y_2\!=\!z_4\!=\!w_4\!=\!w_3\!=\!D_{12}\!=\!D_{34}\!=\!D_{13}\!=\!D_{24}\!=\!0$.} \label{fig-9}
\end{center}
\end{figure}

The further 4 internal cuts are also drawn in figure \ref{fig-8},
and besides this diagram, other diagrams of all topologies, orientations and configurations of loop numbers
at 4-loop that survive these 10 cuts, are given in figure \ref{fig-9},
as can be enumerated from the topologies in figure \ref{fig-7} then picked out by identifying all 10 poles
$x_1,y_1,y_2,z_4,w_4,w_3,D_{12},D_{34},D_{13},D_{24}$. Let's define the sum of these 9 surviving diagrams
as a function of $x,y,z,w$ (we only sum their proper numerators as usual)
\be
\bal
&S\,(x_1,y_1,z_1,w_1,x_2,y_2,z_2,w_2,x_3,y_3,z_3,w_3,x_4,y_4,z_4,w_4)\\
=\,&\,x_2x_3x_4z_1z_2z_3\,y_3w_2\,D_{14}(s_2\,y_4w_1+D_{14})+x_2x_4z_1z_3\,y_3w_2\,D_{14}(x_4z_2\,y_3w_1+x_3z_1\,y_4w_2)\\
&+x_2x_4z_1z_3\,y_3y_4w_1w_2\,(y_3w_2D_{14}+x_2z_3D_{14}+y_4w_1D_{23}+x_4z_1D_{23}+s_1D_{14}D_{23}), \labell{eq-7}
\eal
\ee
where $s_1$ and $s_2$ are coefficients to be determined. Since the cross diagram in figure \ref{fig-9} can survive
these 10 cuts like the non-Mondrian one in figure \ref{fig-7}, we can fix both $s_1$ and $s_2$ in only one equation.
In contrast, if we impose all 8 external cuts available for the cross diagram, the non-Mondrian one cannot
survive these cuts and hence $s_2$ will disappear in this equation, then one more equation that involves $s_2$
is needed. This explains why to determine $s_1$ and $s_2$ in one equation, we choose a set of external cuts
in the non-Mondrian diagram which has less available external cuts than the cross diagram, as it is a general trick
to minimize the number of equations needed for determining all coefficients.

On the other hand, from the positivity conditions of the amplituhedron we have the following dimensionless ratios
with respect to $D_{12},D_{34},D_{13},D_{24}$:
\be
\bal
&\frac{z_1}{z_1-z_2}=\frac{x_2z_1}{D_{12}}\to\frac{\hat{x}_2z_2}{D_{12}},\\
&\frac{x_4}{x_4-x_3}=\frac{z_3x_4}{D_{34}}\to\frac{\hat{z}_3x_3}{D_{34}},\\
&z_3\(\frac{1}{z_3}-\frac{1}{z_3-\hat{z}_3}\)=\frac{x_3\hat{z}_3}{D_{13}},\\
&x_2\(\frac{1}{x_2}-\frac{1}{x_2-\hat{x}_2}\)=\frac{z_2\hat{x}_2}{D_{24}},
\eal
\ee
where $\hat{x}_2$ and $\hat{z}_3$ are defined in \eqref{eq-6}, and $\to$ denotes some variables are replaced
by the solutions of cuts. Since $D_{14}$ and $D_{23}$ are trivially positive, we get the proper numerator
\be
(\hat{x}_2x_3z_2\hat{z}_3)^2D_{14}D_{23}=\[\(x_3+\frac{y_4w_2}{z_2}\)\(z_2+\frac{y_3w_1}{x_3}\)x_3z_2\]^2
y_3w_2\(1+\frac{y_4w_1}{x_3z_2}\)(x_3z_2+y_4w_1).
\ee
Now the critical step is to equate the deformed $S$ defined in \eqref{eq-7} on the 10 cuts and the quantity above,
or consider their difference
\be
\bal
&\,S\(0\,,\,0\,,z_2,w_1,x_3\!+\!\frac{y_4w_2}{z_2},\,0\,,z_2,w_2,x_3,y_3,z_2\!+\!\frac{y_3w_1}{x_3},
\,0\,,x_3,y_4,\,0\,,\,0\)\\
&-\[\(x_3+\frac{y_4w_2}{z_2}\)\(z_2+\frac{y_3w_1}{x_3}\)x_3z_2\]^2
y_3w_2\(1+\frac{y_4w_1}{x_3z_2}\)(x_3z_2+y_4w_1)\\
=\,&\,y_3y_4w_1w_2\(1+\frac{y_4w_1}{x_3z_2}\)(x_3z_2+y_3w_1)(x_3z_2+y_4w_2)
\[(1+s_1)y_3w_2(x_3z_2+y_4w_1)+(1+s_2)x_3^2z_2^2\],
\eal
\ee
then it is clear that to make this difference vanish, we must take $s_1\!=\!s_2\!=\!-1$,
which agrees with \cite{Bern:2006ew}. For this 4-loop case, we see the analysis and calculation are very simple,
due to there is in fact no positivity condition to be imposed -- all $D$'s are either cut
or manifestly positive. But in general this simplicity does not always occur, as immediately at 5-loop
we will encounter some quite nontrivial and hence much more complicated examples. Still, with the aid of positive cuts,
our calculational capability is greatly enhanced so that unlike the hopeless case study of \eqref{eq-8},
we manage to tackle all 5-loop examples.

\section{Positive Cuts at 5-loop}

\begin{figure}
\begin{center}
\includegraphics[width=0.95\textwidth]{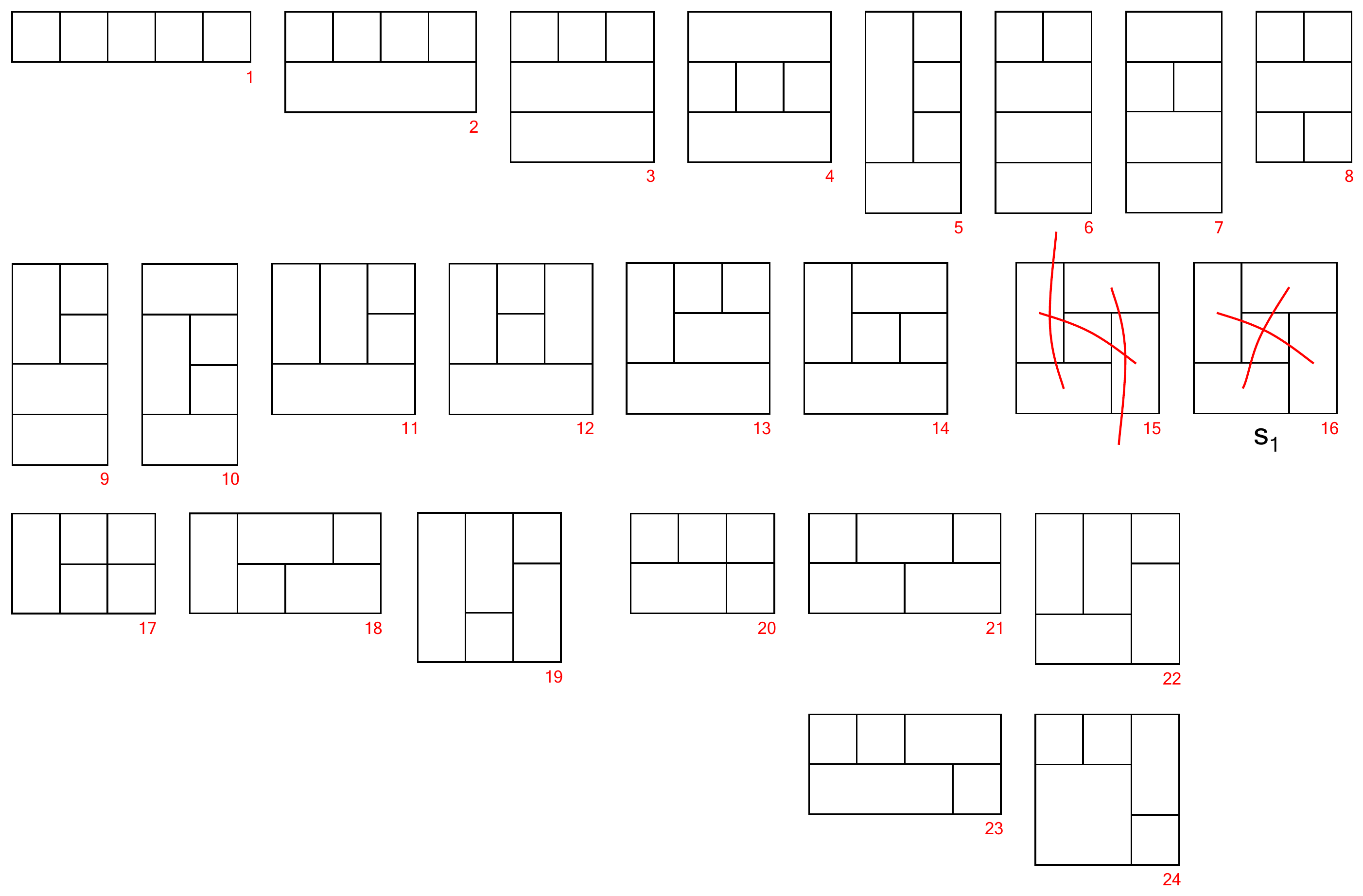}
\caption{Mondrian DCI topologies $T_1,\ldots,T_{24}$ at 5-loop. $T_{16}$ assigned with $s_1$
is a non-rung-rule topology (it is generated by the substitution rule).} \label{fig-10}
\end{center}
\end{figure}

\begin{figure}
\begin{center}
\includegraphics[width=0.665\textwidth]{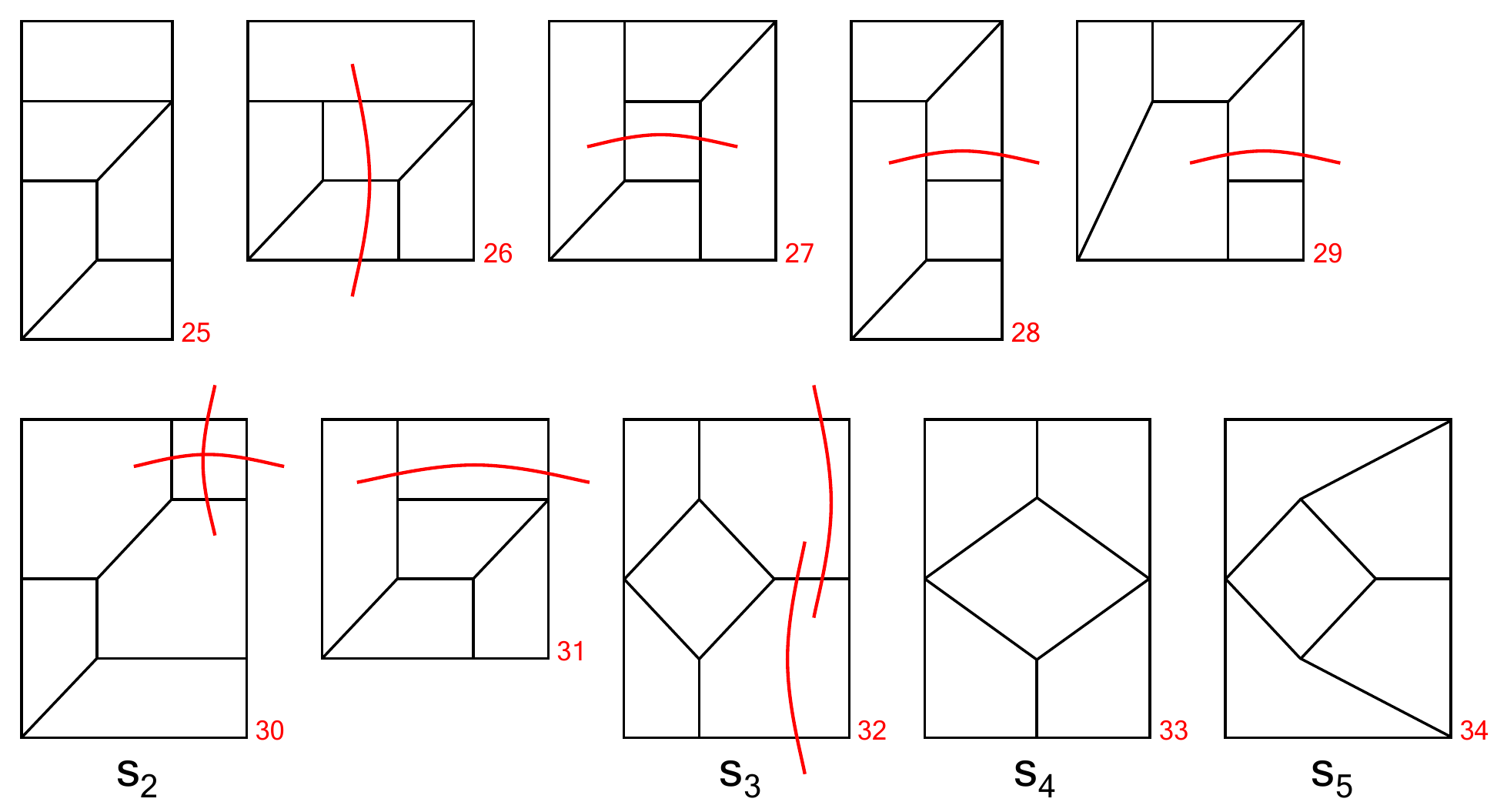}
\caption{Non-Mondrian DCI topologies $T_{25},\ldots,T_{34}$ at 5-loop. $T_{30},T_{32},T_{33},T_{34}$ assigned with
$s_2,s_3,s_4,s_5$ respectively are non-rung-rule topologies ($T_{30}$ is generated by the substitution rule
while $T_{32},T_{33},T_{34}$ are neither generated by the rung nor substitution rule).} \label{fig-11}
\end{center}
\end{figure}

For the 5-loop application of positive cuts, there is nothing new in its principle but we will see much more complexity
in various techniques, as well as its miraculous agreement with previous knowledge.
As usual, we first enumerate all 34 distinct DCI topologies at 5-loop:
figure \ref{fig-10} lists all 24 Mondrian DCI topologies labelled by $T_1,\ldots,T_{24}$, as indicated by the red
subscripts, and figure \ref{fig-11} all 10 non-Mondrian ones labelled by $T_{25},\ldots,T_{34}$ similarly.

Note that there exist two distinct choices of DCI numerator for the pinwheel's pole structure, namely $T_{15}$ and $T_{16}$
given in figure \ref{fig-10}, so we must explicitly draw their numerators while suppressing those of the rest Mondrian
topologies as they can be uniquely inferred from the rung rule. And for non-Mondrian ones in figure \ref{fig-11},
we draw all numerators explicitly since the rung rule cannot account for all of them. Among all these 34 topologies,
$T_{16},T_{30}$ are generated by applying the substitution rule to the 4-loop counterparts in figure \ref{fig-7},
which also preserves coefficients \cite{Bern:2007ct}, while the rules for $T_{32},T_{33},T_{34}$ are unknown,
and the rest are generated by the rung rule. As a simplified demonstration, we focus on non-rung-rule topologies only, so
$T_{16},T_{30},T_{32},T_{33},T_{34}$ assigned with coefficients $s_1,s_2,s_3,s_4,s_5$ respectively are of our concern.
Let's now determine these coefficients one by one using the amplituhedron approach.

\begin{figure}
\begin{center}
\includegraphics[width=0.16\textwidth]{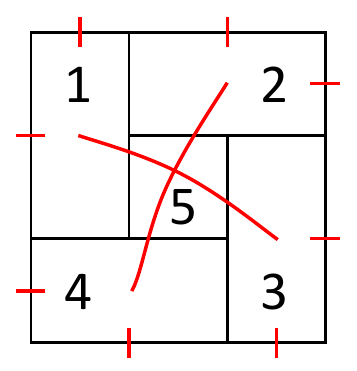}
\caption{A particular diagram of $T_{16}$ at 5-loop with 8 external cuts.} \label{fig-12}
\end{center}
\end{figure}

\subsection{Determination of $s_1$}

To determine $s_1$, let's consider a particular diagram of DCI topology $T_{16}$ given in figure \ref{fig-12}.
As usual, we can maximally impose all 8 available external cuts, as indicated by the red segments.
These 8 cuts result in $x_1\!=\!y_1\!=\!y_2\!=\!z_2\!=\!z_3\!=\!w_3\!=\!w_4\!=\!x_4\!=\!0$,
which can simplify the ten $D$'s as
\be
\bal
&D_{12}=x_2z_1,~~D_{23}=y_3w_2,~~D_{34}=x_3z_4,~~D_{14}=y_4w_1,\\
&D_{13}=x_3z_1+y_3w_1,~~D_{24}=x_2z_4+y_4w_2,
\eal
\ee
as well as
\be
\bal
&D_{15}=x_5z_1+y_5w_1-x_5z_5-y_5w_5,\\
&D_{25}=z_5x_2+y_5w_2-x_5z_5-y_5w_5,\\
&D_{35}=z_5x_3+w_5y_3-x_5z_5-y_5w_5,\\
&D_{45}=x_5z_4+w_5y_4-x_5z_5-y_5w_5.
\eal
\ee
Since $D_{12},D_{23},D_{34},D_{14},D_{13},D_{24}$ are manifestly positive, we only need to either cut
$D_{15},D_{25},D_{35},D_{45}$ or impose their positivity. However, there is no straightforward positive cut for
positivity conditions of the form $x\!+\!y\!>\!a$ in this case -- the discussion can be rather complicated.
Therefore let's keep their positivity and see what happens next, in fact, $D_{15},D_{25},D_{35},D_{45}$ totally
decouple partly due to the symmetry of the 8 external cuts in figure \ref{fig-12}, so that we can impose the positivity
for each $D_{i5}$ individually. This leads to the simple proper numerator
\be
\bal
N=\,&\,(x_5z_1+y_5w_1)(z_5x_2+y_5w_2)(z_5x_3+w_5y_3)(x_5z_4+w_5y_4)D_{12}D_{23}D_{34}D_{14}D_{13}D_{24}\\
=\,&\,(x_5z_1+y_5w_1)(z_5x_2+y_5w_2)(z_5x_3+w_5y_3)(x_5z_4+w_5y_4)
\,x_2x_3z_1z_4\,y_3y_4w_1w_2\,(x_3z_1+y_3w_1)(x_2z_4+y_4w_2).
\eal
\ee
On the other hand, diagrams of all topologies, orientations and configurations of loop numbers at 5-loop
that survive these 8 cuts are summarized below:
\be
\begin{array}{cccccccccc}
~T_8~ & ~T_{15}~ & ~T_{16}~ & ~T_{20}~ & ~T_{21}~ & ~T_{22}~ & ~T_{23}~ & ~T_{24}~ & ~T_{32}~ & ~T_{33}~ \\
\hline
2 & 4 & 1 & 8 & 4 & 8 & 8 & 8 & 4 & 2
\end{array}
\ee
where all orientations generated by dihedral symmetry of these topologies contribute and each orientation exactly
contributes one configuration of loop numbers, as given by the numbers of contributing diagrams of each $T_i$ above.
It is easy to enumerate all of them, and the sum of their proper numerators is
\be
\bal
&\,S\,(x_1,y_1,z_1,w_1,x_2,y_2,z_2,w_2,x_3,y_3,z_3,w_3,x_4,y_4,z_4,w_4,x_5,y_5,z_5,w_5)\\
=\,&\,x_2x_3x_5z_1z_4z_5\,y_3y_4y_5w_1w_2w_5\,(S_8+S_{15-16}+S_{20}+S_{21}+S_{22}+S_{23}+S_{24}+S_{32}+S_{33}),
\labell{eq-10}
\eal
\ee
where for compactness we have factored out a common factor, and each piece in the sum is given by
\be
S_8=\frac{y_5w_5}{x_5z_5}D_{13}D_{14}D_{23}D_{24}+\frac{x_5z_5}{y_5w_5}D_{12}D_{13}D_{24}D_{34},
~~~~~~~~~~~~~~~~~~~~~~~~~~~~~~~~~~~~~~~~~~~~~~~~~~~~
\ee
\be
S_{15-16}=D_{13}D_{24}(x_3z_1D_{24}+y_3w_1D_{24}+x_2z_4D_{13}+y_4w_2D_{13}+s_1D_{13}D_{24}),
~~~~~~~~~~~~~~~~~~~~~~~~~~~~~~~
\ee
\be
\bal
S_{20}=&-\frac{y_4}{y_5}D_{12}D_{13}D_{24}D_{35}-\frac{y_3}{y_5}D_{12}D_{13}D_{24}D_{45}
-\frac{w_1}{w_5}D_{13}D_{24}D_{25}D_{34}-\frac{w_2}{w_5}D_{13}D_{15}D_{24}D_{34}\\
&-\frac{z_4}{z_5}D_{13}D_{15}D_{23}D_{24}-\frac{x_3}{x_5}D_{13}D_{14}D_{24}D_{25}
-\frac{z_1}{z_5}D_{13}D_{23}D_{24}D_{45}-\frac{x_2}{x_5}D_{13}D_{14}D_{24}D_{35},~~~\,
\eal
\ee
\be
S_{21}=\frac{y_3y_4w_5}{y_5}D_{12}D_{13}D_{24}+\frac{y_5w_1w_2}{w_5}D_{13}D_{24}D_{34}
+\frac{x_2x_3z_5}{x_5}D_{13}D_{14}D_{24}+\frac{x_5z_1z_4}{z_5}D_{13}D_{23}D_{24},~\,
\ee
\be
\bal
S_{22}=\,&\,\frac{x_3z_5y_4}{y_5}D_{12}D_{13}D_{24}+\frac{x_5z_4y_3}{y_5}D_{12}D_{13}D_{24}
+\frac{x_2z_5w_1}{w_5}D_{13}D_{24}D_{34}+\frac{x_5z_1w_2}{w_5}D_{13}D_{24}D_{34}\\
+&\,\frac{x_2y_3w_5}{x_5}D_{13}D_{14}D_{24}+\frac{z_1y_4w_5}{z_5}D_{13}D_{23}D_{24}
+\frac{x_3y_5w_2}{x_5}D_{13}D_{14}D_{24}+\frac{z_4y_5w_1}{z_5}D_{13}D_{23}D_{24},
\eal
\ee
\be
\bal
S_{23}=\,&\,\frac{y_4^2w_2}{y_5}D_{12}D_{13}D_{35}+\frac{y_4w_2^2}{w_5}D_{13}D_{15}D_{34}
+\frac{y_3^2w_1}{y_5}D_{12}D_{24}D_{45}+\frac{y_3w_1^2}{w_5}D_{24}D_{25}D_{34}\\
+&\,\frac{x_2^2z_4}{x_5}D_{13}D_{14}D_{35}+\frac{x_2z_4^2}{z_5}D_{13}D_{15}D_{23}
+\frac{x_3z_1^2}{z_5}D_{23}D_{24}D_{45}+\frac{x_3^2z_1}{x_5}D_{14}D_{24}D_{25},~~~~~~~~~~~~\,
\eal
\ee
\be
\bal
S_{24}=\,&\,\frac{x_2z_4y_4}{y_5}D_{12}D_{13}D_{35}+\frac{x_3z_1y_3}{y_5}D_{12}D_{24}D_{45}
+\frac{x_3z_1w_1}{w_5}D_{24}D_{25}D_{34}+\frac{x_2z_4w_2}{w_5}D_{13}D_{15}D_{34}\\
+&\,\frac{x_2y_4w_2}{x_5}D_{13}D_{14}D_{35}+\frac{z_1y_3w_1}{z_5}D_{23}D_{24}D_{45}
+\frac{x_3y_3w_1}{x_5}D_{14}D_{24}D_{25}+\frac{z_4y_4w_2}{z_5}D_{13}D_{15}D_{23},
\eal
\ee
\be
S_{32}=s_3(y_3w_2D_{13}D_{14}D_{24}+y_4w_1D_{13}D_{23}D_{24}+x_3z_4D_{12}D_{13}D_{24}+x_2z_1D_{13}D_{24}D_{34}),~~~~~~~~~
\ee
\be
S_{33}=s_4(D_{13}D_{14}D_{23}D_{24}+D_{12}D_{13}D_{24}D_{34}).
~~~~~~~~~~~~~~~~~~~~~~~~~~~~~~~~~~~~~~~~~~~~~~~~~~~~~~~~~~~~\,
\ee
The difference between the deformed $S$ on the 8 cuts and the proper numerator from positivity conditions is then
\be
\bal
&\,S\,(0\,,\,0\,,z_1,w_1,x_2,\,0\,,\,0\,,w_2,x_3,y_3,\,0\,,\,0\,,\,0\,,y_4,z_4,\,0\,,x_5,y_5,z_5,w_5)\\
&-(x_5z_1+y_5w_1)(z_5x_2+y_5w_2)(z_5x_3+w_5y_3)(x_5z_4+w_5y_4)
\,x_2x_3z_1z_4\,y_3y_4w_1w_2\,(x_3z_1+y_3w_1)(x_2z_4+y_4w_2)\\
=\,&\,x_2x_3x_5z_1z_4z_5\,y_3y_4y_5w_1w_2w_5\,(x_3z_1+y_3w_1)(x_2z_4+y_4w_2)\\
&\times[(1+s_1)(x_3z_1\,y_4w_2+x_2z_4\,y_3w_1)+(2+s_1+2s_3+s_4)(x_3x_2z_1z_4+y_3y_4w_1w_2)],
\eal
\ee
to make this difference vanish we must take $s_1\!=\!-1$ which agrees with \cite{Bern:2006ew},
and $1\!+\!2s_3\!+\!s_4\!=\!0$. Even though $s_3$ and $s_4$ cannot be determined by these 8 external cuts yet,
we can determine one with the aid of further cuts then get the other via relation $1\!+\!2s_3\!+\!s_4\!=\!0$.

\subsection{Determination of $s_2,s_3,s_4$}

\begin{figure}
\begin{center}
\includegraphics[width=0.161\textwidth]{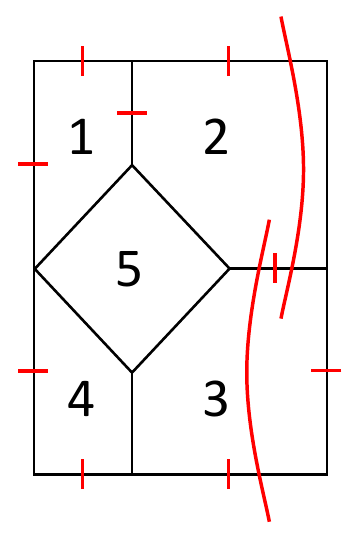}
\caption{A particular diagram of $T_{32}$ at 5-loop with 7 external and 2 internal cuts.
The external cut $z_2\!=\!0$ is traded for two internal cuts $D_{12}\!=\!D_{23}\!=\!0$
which are free of the subtlety of composite residues.} \label{fig-13}
\end{center}
\end{figure}

To figure out $s_3$ or $s_4$, we have to disentangle $T_{32}$ and $T_{33}$, otherwise combination $(1\!+\!2s_3\!+\!s_4)$
will always obstruct our intention. Since $T_{32}$ has one internal propagator more than $T_{33}$ while their other
topological features are identical, it is feasible to impose further internal cuts to kill $T_{33}$ but let $T_{32}$ survive
so that $s_3$ can be isolated then determined. If we consider a particular diagram of $T_{32}$ given in
figure \ref{fig-13}, a simplest choice is to impose $D_{12}\!=\!D_{23}\!=\!0$, as one can easily check that none of the
diagrams of $T_{33}$ can survive it regardless of orientations and number configurations (we also maintain
the 8 external cuts in figure \ref{fig-12}).

However, since $D_{12}\!=\!x_2z_1$ and $D_{23}\!=\!y_3w_2$, setting $D_{12}\!=\!D_{23}\!=\!0$ will force two external
propagators which do not belong to the diagram in figure \ref{fig-12} to vanish. This involves a technical subtlety
of composite residues, although there is no problem in this way after some clarification, we prefer to avoid this subtlety
for the moment. Therefore, a simplest alternative is to relax one external cut, which is chosen to be $z_2$.

In summary, upon the 7 external cuts $x_1\!=\!y_1\!=\!y_2\!=\!z_3\!=\!w_3\!=\!w_4\!=\!x_4\!=\!0$, we can further impose
\be
z_1=z_2,~~x_2=x_3+\frac{y_3w_2}{z_2}\equiv\hat{x}_2, \labell{eq-9}
\ee
so these $7\!+\!2$ cuts can simplify the ten $D$'s as
\be
D_{12}=D_{23}=0,~~D_{34}=x_3z_4,~~D_{14}=y_4w_1,~~D_{13}=x_3z_2+y_3w_1
\ee
which are either zero or manifestly positive, as well as
\be
\bal
&D_{15}=x_5z_2+y_5w_1-x_5z_5-y_5w_5,\\
&D_{45}=x_5z_4+w_5y_4-x_5z_5-y_5w_5,\\
&D_{35}=z_5x_3+w_5y_3-x_5z_5-y_5w_5,\\
&D_{24}=(x_3z_2+y_3w_2)\(\frac{z_4}{z_2}+\frac{y_4}{y_3+x_3z_2/w_2}-1\),\\
&D_{25}=(z_5-z_2)\(x_3+y_3\frac{w_2}{z_2}-x_5\)+y_5(w_2-w_5),
\eal
\ee
again there is no straightforward positive cut for any of these five positivity conditions, so it is better to keep
their positivity. In this case, $D_{15},D_{45},D_{35},D_{24},D_{25}$ do not trivially decouple, as we can see it
more clearly after the following reorganization:
\be
\bal
&\frac{z_2}{z_5+y_5w_5/x_5}+\frac{w_1}{w_5+x_5z_5/y_5}>1,\\
&\frac{x_3}{x_5+y_5w_5/z_5}+\frac{y_3}{y_5+x_5z_5/w_5}>1,~~~(z_5-z_2)\(x_3+y_3\frac{w_2}{z_2}-x_5\)+y_5(w_2-w_5)>0,\\
&\frac{z_4}{z_5+y_5w_5/x_5}+\frac{y_4}{y_5+x_5z_5/w_5}>1,~~~\frac{z_4}{z_2}+\frac{y_4}{y_3+x_3z_2/w_2}>1.
\eal
\ee
In the first line we focus on $z_2,w_1$, in the second $x_3,y_3$ and in the third $z_4,y_4$.
For the latter two lines, the discussion of imposing positivity is nontrivial, since we need to choose one condition
(or both) as the relations among several variables vary. Explicitly, the second line's discussion depends on how
$z_2$ varies in the first line, and the third line's discussion depends on how
$x_3,y_3$ vary in the second line. Its technical details are elaborated in appendix \ref{app1},
and below we just present the resulting $d\log$ form after analyzing all possible situations of variables
$z_2,w_1,w_2,y_3,x_3,z_4,y_4$:
\be
\frac{M}{z_2^3w_1w_2y_3x_3z_4y_4\,D_{15}D_{35}D_{25}D_{45}D_{24}}\equiv\frac{R}{z_2w_1w_2y_3x_3z_4y_4}, \labell{eq-19}
\ee
where the expression of $M$ is given below, as the result simplified by \textsc{Mathematica},
and $R$ is the desired dimensionless ratio.\\

$M = w_1 y_5 (w_5 x_5 y_3 z_2^2 (w_5 (y_4 - y_5) + x_5 z_4) (w_2 y_4 z_2 + w_2 y_3 z_4 +
          x_3 z_2 z_4) + (w_2 w_5 y_4 z_2 (w_2 w_5 y_3^2 (y_4 - y_5) +
            x_3 (w_5 y_3 y_4 + w_2 (-y_3 + y_4) y_5 -
               w_5 (y_3 + y_4) y_5) z_2 - (x_5^2 y_3 +
               x_3^2 y_5) z_2^2) + (w_2^2 w_5^2 y_3^3 y_4 +
            w_2 w_5 y_3 (w_5 x_3 y_4 (2 y_3 - y_5) +
               w_2 (x_5 y_3 (-y_3 + y_4) +
                  x_3 y_4 y_5)) z_2 + (w_5 y_3 (-w_2 x_5 (2 x_3 + x_5) y_3 +
                  x_3 (w_5 x_3 + w_2 x_5) y_4) +
               x_3 (w_2 w_5 x_5 y_3 + (w_2 - w_5) (w_5 x_3 +
                    w_2 x_5) y_4) y_5) z_2^2 -
            w_5 x_3 x_5 ((x_3 + x_5) y_3 - x_3 y_5) z_2^3) z_4 +
         x_5 (w_2 y_3 + x_3 z_2) (w_2 w_5 y_3^2 +
            x_3 (w_5 (y_3 - y_5) +
               w_2 y_5) z_2) z_4^2) z_5 + (w_2 w_5 y_4 z_2 (w_2 y_3 (-x_5 y_3 +
               x_3 y_4) + x_3 (x_3 y_4 - x_5 (y_3 + y_4)) z_2) +
         x_3 (w_2 y_3 + (x_3 - x_5) z_2) (x_3 z_2 (w_5 y_4 - x_5 z_2) +
            w_2 (w_5 y_3 y_4 + x_5 (-y_3 + y_4) z_2)) z_4 +
         x_3 x_5 (w_2 y_3 + x_3 z_2) (w_2 y_3 + (x_3 - x_5) z_2) z_4^2) z_5^2) +
   x_5 z_2^2 (w_2^2 x_3 y_5 (w_5 y_4 (-y_3 + y_4) z_2 + w_5 y_3 (y_4 - y_5) z_4 +
         x_5 z_4 (y_4 z_2 + y_3 (z_4 - z_5))) z_5 -
      w_5 x_3 y_3 z_2 z_4 (w_5 (y_4 - y_5) + x_5 (z_4 - z_5)) (w_5 y_5 +
         x_5 (-z_2 + z_5)) +
      w_2 (-w_5^3 y_3 (y_4 - y_5) y_5 (y_4 z_2 + y_3 z_4) +
         w_5^2 x_5 y_3 (y_4 z_2 + y_3 z_4) (-y_5 (z_2 + z_4 - 2 z_5) +
            y_4 (z_2 - z_5)) + x_3^2 x_5 y_5 z_2 z_4 (z_4 - z_5) z_5 +
         w_5 (x_5^2 y_3 (y_4 z_2 + y_3 z_4) (z_2 - z_5) (z_4 - z_5) -
            x_3^2 y_5 z_2 (y_4 z_2 - y_4 z_4 + y_5 z_4) z_5)));$\\

To get the overall dimensionless ratio, we also need
\be
\bal
&\frac{z_1}{z_1-z_2}=\frac{x_2z_1}{D_{12}}\to\frac{\hat{x}_2z_2}{D_{12}},\\
&x_2\(\frac{1}{x_2}-\frac{1}{x_2-\hat{x}_2}\)=\frac{z_2\hat{x}_2}{D_{23}},
\eal
\ee
where $\hat{x}_2$ is defined in \eqref{eq-9}, and since the positivity of $D_{34},D_{14},D_{13}$ is trivial,
we finally obtain
\be
\frac{\hat{x}_2z_2}{D_{12}}\,\frac{z_2\hat{x}_2}{D_{23}}\,\frac{D_{34}D_{14}D_{13}}{D_{34}D_{14}D_{13}}\,R
=\frac{(\hat{x}_2z_2)^2D_{34}D_{14}D_{13}}{D_{12}D_{23}D_{34}D_{14}D_{13}}
\,\frac{1}{D_{15}D_{35}D_{25}D_{45}D_{24}}\,\frac{M}{z_2^2},
\ee
therefore the proper numerator is
\be
N=\hat{x}_2^2\,D_{34}D_{14}D_{13}\,M=\(x_3+\frac{y_3w_2}{z_2}\)^2x_3z_4\,y_4w_1\,(x_3z_2+y_3w_1)\,M.
\ee
On the other hand, diagrams of all topologies, orientations and configurations of loop numbers at 5-loop
that survive these $7\!+\!2$ cuts are summarized below:
\be
\begin{array}{ccccccccccccccccccc}
T_3~ & T_8~ & T_9~ & T_{11}~ & T_{13}~ & T_{14}~ & T_{15}~ & T_{16}~ & T_{17}~ & T_{18}~ & T_{19}~ &
T_{20}~ & T_{21}~ & T_{22}~ & T_{23}~ & T_{24}~ & T_{30}~ & T_{31}~ & T_{32} \\
\hline
{} & {} & {} & {} & {} & {} & 4 & 1 & {} & {} & {} & 4~ & 2~ & 4~ & 4~ & 4~ & {} & {} & 2~ \\
1~ & 1~ & 1~ & 1~ & 1~ & 1~ & {} & {} & 1~ & 2~ & 2~ & 1~ & 1~ & 2~ & 1~ & 1~ & 1~ & 1~ & {}
\end{array}
\ee
where the first line denotes a subset of diagrams among \eqref{eq-10}, and the second line the additional
surviving contribution due to relaxing $z_2\!=\!0$.
Again, each orientation of $T_i$ can at most contribute one configuration of loop numbers.
The sum of their proper numerators is
\be
\bal
&\,S\,(x_1,y_1,z_1,w_1,x_2,y_2,z_2,w_2,x_3,y_3,z_3,w_3,x_4,y_4,z_4,w_4,x_5,y_5,z_5,w_5)\\
=\,&\,x_2x_3x_5z_1z_4z_5\,y_3y_4y_5w_1w_2w_5\,(S_{15-16}+S_{20}+S_{21}+S_{22}+S_{23}+S_{24}+S_{32})\\
&+S_3+S_8+S_9+S_{11}+S_{13}+S_{14}+S_{17-19}+S_{20-24}+S_{30}+S_{31}, \labell{eq-16}
\eal
\ee
where each piece in the sum is given by
\be
\!\!\!\!
S_{15-16}=D_{13}D_{24}(x_3z_1D_{24}+y_3w_1D_{24}+x_2z_4D_{13}+y_4w_2D_{13}+s_1D_{13}D_{24}),~~~~~~~~~~~~~
\ee
\be
~~S_{20}=-\,0-0-\frac{w_1}{w_5}D_{13}D_{24}D_{25}D_{34}-\frac{w_2}{w_5}D_{13}D_{15}D_{24}D_{34}
-0-\frac{x_3}{x_5}D_{13}D_{14}D_{24}D_{25}-0-\frac{x_2}{x_5}D_{13}D_{14}D_{24}D_{35},
\ee
\be
S_{21}=0+\frac{y_5w_1w_2}{w_5}D_{13}D_{24}D_{34}+\frac{x_2x_3z_5}{x_5}D_{13}D_{14}D_{24}+0,
~~~~~~~~~~~~~~~~~~~~~~~~~~~~~~~~~~~~~~~~~~~~~~~~~~~~~~\,
\ee
\be
~~S_{22}=0+0+\frac{x_2z_5w_1}{w_5}D_{13}D_{24}D_{34}+\frac{x_5z_1w_2}{w_5}D_{13}D_{24}D_{34}
+\frac{x_2y_3w_5}{x_5}D_{13}D_{14}D_{24}+0+\frac{x_3y_5w_2}{x_5}D_{13}D_{14}D_{24}+0,
\ee
\be
S_{23}=0+\frac{y_4w_2^2}{w_5}D_{13}D_{15}D_{34}+0+\frac{y_3w_1^2}{w_5}D_{24}D_{25}D_{34}
+\frac{x_2^2z_4}{x_5}D_{13}D_{14}D_{35}+0+0+\frac{x_3^2z_1}{x_5}D_{14}D_{24}D_{25},\,
\ee
\be
~~S_{24}=0+0+\frac{x_3z_1w_1}{w_5}D_{24}D_{25}D_{34}+\frac{x_2z_4w_2}{w_5}D_{13}D_{15}D_{34}
+\frac{x_2y_4w_2}{x_5}D_{13}D_{14}D_{35}+0+\frac{x_3y_3w_1}{x_5}D_{14}D_{24}D_{25}+0,
\ee
\be
S_{32}=s_3(y_3w_2D_{13}D_{14}D_{24}+0+0+x_2z_1D_{13}D_{24}D_{34})
~~~~~~~~~~~~~~~~~~~~~~~~~~~~~~~~~~~~~~~~~~~~~~~~~~~~~~~~~~
\ee
for the subset among \eqref{eq-10} (the zeros denote diagrams killed by $D_{12}\!=\!D_{23}\!=\!0$), as well as
\be
S_3=x_2^3x_3z_1z_2z_4z_5\,y_4y_5w_1w_5D_{13}D_{14}D_{34}D_{35},~~~~~~~~~~~~~~~~~~~~~~~~~~~~~~~~~~~~
\ee
\be
S_8=x_2^2x_3x_5z_1z_2^2z_4\,y_3y_4w_1w_5D_{13}D_{15}D_{34}D_{45},~~~~~~~~~~~~~~~~~~~~~~~~~~~~~~~~~~~\,
\ee
\be
S_9=x_2^2x_3x_5z_1z_2z_4z_5\,y_4y_5w_1^2D_{13}D_{24}D_{34}D_{35},~~~~~~~~~~~~~~~~~~~~~~~~~~~~~~~~~~~~
\ee
\be
S_{11}=x_2x_3^3z_1z_2z_4z_5\,y_4y_5^2w_1w_2w_5D_{13}D_{14}D_{24},~~~~~~~~~~~~~~~~~~~~~~~~~~~~~~~~~~~~~~
\ee
\be
S_{13}=x_2^2x_3^2z_1z_2z_4z_5\,y_4^2y_5w_1w_2w_5D_{13}D_{14}D_{35},~~~~~~~~~~~~~~~~~~~~~~~~~~~~~~~~~~~~~~
\ee
\be
S_{14}=x_2x_3^2x_5z_1z_2z_4z_5\,y_4^2y_5w_1w_2w_5D_{13}^2D_{24},~~~~~~~~~~~~~~~~~~~~~~~~~~~~~~~~~~~~~~~~
\ee
\be
\bal
S_{17-19}=\,&\,x_2x_3x_5z_1z_2z_4z_5\,y_4y_5w_1w_2D_{13}D_{34}\\
&\times(-\,x_3D_{15}D_{24}+x_3\,y_4w_2D_{15}+x_3\,y_5w_1D_{24}+x_2D_{15}D_{34}+x_5D_{13}D_{24}),
\eal
\ee
\be
\bal
~~~~~~~~~~~~~~S_{20-24}=\,&\,x_2x_3x_5z_1z_2z_4\,y_3^2y_4w_1w_2w_5D_{15}D_{45}\\
&\times\(-\,D_{13}D_{24}+y_4w_2D_{13}+\frac{y_4}{y_3}\,x_3z_2D_{13}+x_2z_4D_{13}+y_3w_1D_{24}+x_3z_1D_{24}\),
\eal
\ee
\be
S_{30}=s_2\,x_2x_3x_5^2z_1z_2z_4z_5\,y_3y_4y_5w_1^2w_2D_{13}D_{24}D_{34},~~~~~~~~~~~~~~~~~~~~~~~~~~~~~~~~\,
\ee
\be
S_{31}=-\,x_2x_3^2x_5z_1z_2z_4z_5\,y_3y_4y_5w_1w_2w_5D_{13}D_{14}D_{24}~~~~~~~~~~~~~~~~~~~~~~~~~~~~~~\,
\ee
for the additional surviving contribution.
The difference between the deformed $S$ on the $7\!+\!2$ cuts and the proper numerator is then
\be
\bal
&\,S\(0\,,\,0\,,z_2,w_1,x_3\!+\!\frac{y_3w_2}{z_2},\,0\,,z_2,w_2,x_3,y_3,\,0\,,\,0\,,\,0\,,y_4,z_4,\,0\,,x_5,y_5,z_5,w_5\)\\
&-\(x_3+\frac{y_3w_2}{z_2}\)^2x_3z_4\,y_4w_1\,(x_3z_2+y_3w_1)\,M\\
=\,&\,x_3x_5z_4z_5\,y_3y_4y_5w_1w_2\,(x_3z_2+y_3w_1)(x_3z_2+y_3w_2)
\[(x_3z_2+y_3w_2)\(\frac{z_4}{z_2}-1\)+\,y_4w_2\]\\
&\times[(1+s_2)x_3x_5z_2z_4\,w_1+(1+s_3)w_5(x_3z_4(x_3z_2+y_3w_2)+y_3y_4w_1w_2)],
\eal
\ee
to make this difference vanish we must take $s_2\!=\!s_3\!=\!-1$, so via $1\!+\!2s_3\!+\!s_4\!=\!0$
we also obtain $\!s_4\!=\!+1$, all of which agree with \cite{Bern:2006ew}.
We see that determining $s_2$ is a byproduct of determining $s_3$.

It is worth noticing the complexity of 5-loop topologies which have a purely internal loop:
the simple case of $T_{16}$ with 8 symmetric external cuts is clearly rather rare, as merely relaxing one cut
results in five positivity conditions that do not trivially decouple. In general, the more external cuts
a topology has, the easier its calculation might be. We will see how dramatic this qualitative criterion looks
from the case of $T_{34}$, which merely has two external cuts less than $T_{16}$ but becomes extremely complicated,
even compared to the case of $T_{32}$ which is already very nontrivial.

\subsection{Determination of $s_5$}

\begin{figure}
\begin{center}
\includegraphics[width=0.161\textwidth]{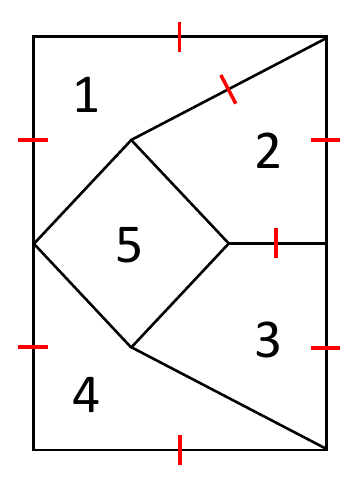}
\caption{A particular diagram of $T_{34}$ at 5-loop with 6 external and 2 internal cuts.} \label{fig-14}
\end{center}
\end{figure}

To determine $s_5$, the coefficient of $T_{34}$, turns out to be the most difficult case at 5-loop.
We again consider a particular diagram given in figure \ref{fig-14}, in which all 6 available external cuts are imposed,
now let's again impose internal cuts $D_{12}\!=\!D_{23}\!=\!0$ upon $x_1\!=\!y_1\!=\!z_2\!=\!z_3\!=\!w_4\!=\!x_4\!=\!0$.
Even though this diagram has only one external cut less than the one in figure \ref{fig-13}, it is very different from
the latter. In fact, the structure and complexity of the simplified positivity conditions are very sensitive to
the choice of cuts.

\begin{figure}
\begin{center}
\includegraphics[width=0.415\textwidth]{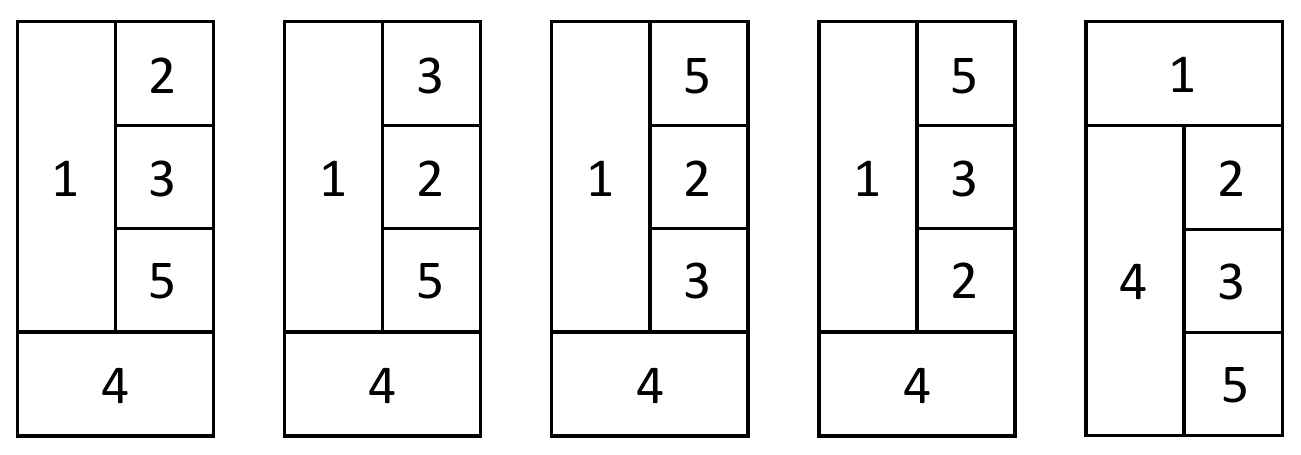}
\caption{The $(4)\!+\!1$ multiplicity of $T_5$.} \label{fig-15}
\end{center}
\end{figure}

Explicitly, for the two internal cuts we can impose
\be
w_2=w_3=w_1+\frac{x_2z_1}{y_2}\equiv\hat{w}_3, \labell{eq-15}
\ee
so the ten $D$'s can be simplified as
\be
D_{12}=D_{23}=0,~~D_{14}=y_4w_1
\ee
which are either zero or manifestly positive, as well as
\be
\bal
&D_{13}=z_1\(x_3-x_2\frac{y_3}{y_2}\)\equiv z_1x'_3,\\
&D_{15}=x_5z_1+y_5w_1-x_5z_5-y_5w_5,\\
&D_{45}=x_5z_4+w_5y_4-x_5z_5-y_5w_5,\\
&D_{24}=(x_2z_1+y_2w_1)\(\frac{z_4}{z_1+w_1y_2/x_2}+\frac{y_4}{y_2}-1\),\\
&D_{34}=(x_2z_1+y_2w_1)\,\frac{y_3}{y_2}\(\frac{x_3}{y_3}\frac{z_4}{z_1x_2/y_2+w_1}+\frac{y_4}{y_3}-1\),\\
&D_{25}=(y_5-y_2)\(w_1+z_1\frac{x_2}{y_2}-w_5\)+z_5(x_2-x_5),\\
&D_{35}=(y_5-y_3)\(w_1+z_1\frac{x_2}{y_2}-w_5\)+z_5(x_3-x_5),
\eal
\ee
where $x'_3$ is defined to trivialize $D_{13}\!>\!0$, and the rest six conditions can be analyzed more clearly
after the following reorganization:
\be
\bal
\!\!\!\!\!\!
&w_1\!+\!z_1\frac{x_5}{y_5}\!>\!w_5\!+\!z_5\frac{x_5}{y_5},~~
(y_5\!-\!y_2)\(\!w_1\!+\!z_1\frac{x_2}{y_2}\!-\!w_5\!\)\!+\!z_5(x_2\!-\!x_5)\!>\!0,~~
(y_5\!-\!y_3)\(\!w_1\!+\!z_1\frac{x_2}{y_2}\!-\!w_5\!\)\!+\!z_5(x_3\!-\!x_5)\!>\!0,\\
\!\!\!\!\!\!
&\frac{z_4}{z_5\!+\!y_5w_5/x_5}\!+\!\frac{y_4}{y_5\!+\!x_5z_5/w_5}\!>\!1,~~
\frac{z_4}{z_1\!+\!w_1y_2/x_2}\!+\!\frac{y_4}{y_2}\!>\!1,~~
\frac{z_4}{k(z_1\!+\!w_1y_2/x_2)}\!+\!\frac{y_4}{y_3}\!>\!1,
\eal
\ee
where $k\!=\!y_3x_2/(y_2x_3)\!<\!1$ due to $D_{13}\!>\!0$. In the first line we focus on $w_1,z_1$ and in the second
$z_4,y_4$, as the second line's discussion depends on how $w_1,z_1$ vary in the first line,
and its technical details are briefly given in appendix \ref{app2}. Below we just present the resulting
$d\log$ form after analyzing all possible situations of variables $y_2,y_3,y_5,x_5,x'_3,w_5,z_1,w_1,y_4,z_4$:
\be
\bal
&\frac{1}{y_2y_3y_5x_5x'_3w_5z_1w_1y_4z_4\,D_{15}D_{25}D_{35}D_{45}D_{24}D_{34}}\,\frac{\hat{w}_3}{D_{23}}\,
\frac{y_2(M_1y_2D_{34})+y_3M_2}{y_2^4}\\
\equiv\,&\frac{R}{y_2y_3y_5x_5x_3w_5z_1w_1y_4z_4}, \labell{eq-17}
\eal
\ee
where the expressions of $M_1$ and $M_2$ simplified by \textsc{Mathematica} can be referred in appendix \ref{app2},
and $R$ is the desired dimensionless ratio, which is explicitly given by
\be
\bal
R&=\frac{x_3}{x'_3}\,\frac{\hat{w}_3}{D_{15}D_{25}D_{35}D_{45}D_{24}D_{34}D_{23}}\,\frac{y_2(M_1y_2D_{34})+y_3M_2}{y_2^4}\\
&=\frac{x_3z_1\hat{w}_3}{D_{13}D_{15}D_{25}D_{35}D_{45}D_{24}D_{34}D_{23}}\,\frac{y_2(M_1y_2D_{34})+y_3M_2}{y_2^4}.
\eal
\ee
To get the overall dimensionless ratio, we also need
\be
w_2\(\frac{1}{w_2}-\frac{1}{w_2-\hat{w}_3}\)=\frac{y_2\hat{w}_3}{D_{12}},
\ee
where $\hat{w}_3$ is defined in \eqref{eq-15}, and since the positivity of $D_{14}$ is trivial, we finally obtain
\be
\frac{y_2\hat{w}_3}{D_{12}}\,\frac{D_{14}}{D_{14}}\,R=\frac{y_2\hat{w}_3D_{14}}{D_{12}D_{14}}\,
\frac{x_3z_1\hat{w}_3}{D_{13}D_{15}D_{25}D_{35}D_{45}D_{24}D_{34}D_{23}}\,\frac{y_2(M_1y_2D_{34})+y_3M_2}{y_2^4},
\ee
therefore the proper numerator is
\be
N=\hat{w}_3^2\,D_{14}\,x_3z_1\,\frac{y_2(M_1y_2D_{34})+y_3M_2}{y_2^3}
=\(w_1+\frac{x_2z_1}{y_2}\)^2y_4w_1\,x_3z_1\,\frac{y_2(M_1y_2D_{34})+y_3M_2}{y_2^3}.
\ee
On the other hand, diagrams of all topologies, orientations and configurations of loop numbers at 5-loop
that survive these $6\!+\!2$ cuts are summarized below:
\be
\begin{small}
\!\!\!\!\!\!\!\!\!\!\!\!\!\!\!\!
\begin{array}{cccccccccccccccccccccccccc}
T_1\!\! & T_3\! & T_5 & T_6\! & T_7\! & T_8 & T_9\!\! & T_{10}\! & T_{11}\! & T_{13}\! & T_{14}\! & T_{15}\! & T_{16}\!
& T_{17}\!\! & T_{18} & T_{19} & T_{20} & T_{21} & T_{22} & T_{23} & T_{24}\!
& T_{25}\! & T_{30}\! & T_{31}\! & T_{32}\! & T_{34} \\
\hline
{} & {} & {} & {} & {} & {} & {} & {} & {} & {} & {} & 4\! & 1\!
& {} & {} & {} & 4 & 2 & 4 & 4 & 4\! & {} & {} & {} & 2\! & {} \\
1\!\! & 2\! & (4)\!+\!1 & (3)\!+\!1\! & 2\! & (3) & (4)\!+\!3\!\! & 1\! & 1\! & 2\! & 2\! & {} & {}
& 2\!\! & (2)\!+\!3 & (2)\!+\!3 & (3)\!+\!1 & (3) & (3)\!+\!3 & (4)\!+\!4 & (4)\!+\!4\! & 2\! & 2\! & 1\! & {} & 1
\end{array}
\end{small}
\ee
where the first line denotes a subset of diagrams among \eqref{eq-10} which are identical to those given in \eqref{eq-16},
and the second line the additional surviving contribution. Now for some $T_i$'s, a particular orientation can contribute
more than one configuration of loop numbers, as the numbers in parentheses above denote this kind of multiplicity.
An explicit example is $(4)\!+\!1$ for $T_5$ corresponding to the diagrams given in figure \ref{fig-15},
of which the first four with different number configurations share the same orientation.

The sum of their proper numerators is
\be
\bal
&\,S\,(x_1,y_1,z_1,w_1,x_2,y_2,z_2,w_2,x_3,y_3,z_3,w_3,x_4,y_4,z_4,w_4,x_5,y_5,z_5,w_5)\\
=\,&\,x_2x_3x_5z_1z_4z_5\,y_3y_4y_5w_1w_2w_5\,(S_{15-16}
\!+\!S_{20}\!+\!S_{21}\!+\!S_{22}\!+\!S_{23}\!+\!S_{24}\!+\!S_{32}\!+\!S_{34})\\
&+\!S_1\!+\!S_3\!+\!S_5\!+\!S_6\!+\!S_7\!+\!S_8\!+\!S_9\!+\!S_{10}\!+\!S_{11}\!+\!S_{13}\!+\!S_{14}
\!+\!S_{17-19}\!+\!S'_{20-24}\!+\!S_{25}\!+\!S_{30}\!+\!S_{31}.
\eal
\ee
Recall that $S_{15-16},S_{20},S_{21},S_{22},S_{23},S_{24},S_{32}$ are already given in \eqref{eq-16}, while
\be
S_{34}=s_5\,y_2w_3D_{13}D_{14}D_{24}
\ee
is the extra term in the second line above, and each piece in the third line is given by
\be
S_1=y_2y_3y_4y_5w_1w_2w_3w_5D_{13}D_{14}D_{15}D_{24}D_{25}D_{34},
~~~~~~~~~~~~~~~~~~~~~~~~~~~~~~~~~~~~~~~~~~~~~~~~~~~~~~~~~~~~
\ee
\be
S_3=x_3x_5z_4z_5\,y_2y_3y_4y_5w_1w_2^3D_{13}D_{14}D_{15}D_{34}
+x_3x_5z_1z_5\,y_2^3y_4w_1w_2w_3w_5D_{13}D_{14}D_{34}D_{45},~~~~~~~~~~
\ee
\be
\bal
S_5=\,&\,x_2x_3x_5z_1^3\,y_4^2w_1w_2w_3w_5(y_3y_5D_{24}D_{25}D_{34}+y_2y_5D_{24}D_{34}D_{35}
+y_2y_3D_{24}D_{35}D_{45}+y_2y_3D_{25}D_{34}D_{45})\\
&+x_2x_3x_5z_4^3\,y_2y_3y_4y_5w_1^2w_2w_3D_{13}D_{15}D_{25},
\eal
\ee
\be
\bal
\!\!\!\!\!\!\!\!\!
S_6=\,&\,z_1\,y_4w_1w_2w_3w_5D_{14}D_{24}\!\(x_2\,y_3^2y_5D_{15}D_{25}D_{34}+x_3\,y_2^2y_5D_{15}D_{34}D_{35}
+x_5\,y_2^2y_3D_{13}D_{35}D_{45}\)\\
&+x_5z_4\,y_2y_3y_4y_5w_1w_2w_3^2D_{13}D_{14}D_{15}D_{24}D_{25},
\eal
\ee
\be
S_7=z_5\,y_2y_3y_4y_5w_1w_2w_3w_5D_{13}D_{14}D_{24}(x_2D_{13}D_{45}+x_3D_{15}D_{24}),
~~~~~~~~~~~~~~~~~~~~~~~~~~~~~~~~~~~~~~~~
\ee
\be
\!\!\!\!\!\!\!\!\!\!\!\!
S_8=x_5z_1z_4\,y_4w_1w_2D_{14}\!\(x_3\,y_2^2y_3w_2w_5D_{13}D_{35}D_{45}+x_2\,y_3^2y_5w_3^2D_{15}D_{24}D_{25}
+x_3\,y_2^2y_5w_2w_3D_{15}D_{34}D_{35}\),
\ee
\be
\bal
\!\!\!\!\!\!\!\!\!\!\!\!\!
S_9=\,&z_1^2\,y_4w_1w_2w_3w_5D_{14}\!\(x_2x_3\,y_3y_5^2D_{24}D_{25}D_{34}\!+\!x_2x_3\,y_2y_5^2D_{24}D_{34}D_{35}
\!+\!x_2x_5\,y_2y_3^2D_{24}D_{35}D_{45}\!+\!x_3x_5\,y_2^2y_3D_{25}D_{34}D_{45}\)\\
&\!+\!x_2^2z_1z_5\,y_3^2y_4y_5w_1w_2w_3w_5D_{13}D_{14}D_{24}D_{45}
\!+\!x_3z_4\,y_2y_3y_4y_5w_1w_2^2D_{13}D_{14}D_{15}(x_3z_5\,w_5D_{24}\!+\!x_5z_4\,w_3D_{25}),
\eal
\ee
\be
S_{10}=x_2x_3z_5^2\,y_2y_3y_4y_5w_1w_2w_3w_5D_{13}D_{14}^2D_{24},
~~~~~~~~~~~~~~~~~~~~~~~~~~~~~~~~~~~~~~~~~~~~~~~~~~~~~~~~~~~~~~~~~~\,
\ee
\be
S_{11}=x_2x_3x_5z_4z_5^2\,y_2y_3y_4y_5w_1^3w_2D_{13}D_{24}D_{34},
~~~~~~~~~~~~~~~~~~~~~~~~~~~~~~~~~~~~~~~~~~~~~~~~~~~~~~~~~~~~~~~~~~~\,
\ee
\be
S_{13}=x_2x_3x_5z_4^2z_5\,y_2y_3y_4y_5w_1^2w_2^2D_{13}D_{15}D_{34}
+x_2x_3x_5z_1^2z_5\,y_2^2y_4^2w_1w_2w_3w_5D_{13}D_{34}D_{45},~~~~~~~~~~~~~~\,
\ee
\be
S_{14}=x_2x_3x_5z_5\,y_3y_4y_5w_1w_2w_5D_{13}D_{24}\!\(z_1^2\,y_4w_3D_{24}+z_4^2\,y_2w_1D_{13}\),
~~~~~~~~~~~~~~~~~~~~~~~~~~~~~~~~~~~~~~~
\ee
\be
\bal
\!\!\!\!S_{17-19}=\,&\,x_2x_3z_4z_5\,y_2y_3y_4y_5w_1w_2w_5D_{13}D_{14}(-\,w_1D_{24}D_{35}+w_1x_2z_4D_{35}
+w_1x_3z_5D_{24}+w_2D_{14}D_{35}+w_5D_{13}D_{24})\\
&+x_2x_3z_1z_5\,y_3y_4y_5w_1w_2w_3w_5D_{14}D_{24}(-\,y_4D_{13}D_{25}+x_3z_1y_4D_{25}
+x_2z_5\,y_4D_{13}+y_3D_{14}D_{25}+y_5D_{13}D_{24})\\
&+x_2x_3z_1z_5\,y_2y_4y_5w_1w_2w_3w_5D_{14}D_{34}D_{35}(x_2z_1\,y_4+y_2D_{14}),
\eal
\ee
\be
\bal
\!\!\!\!\!S'_{20-24}=\,&\(x_2x_3x_5z_1z_4z_5\,y_2y_4y_5w_1^2w_3D_{24}D_{34}D_{35}
\!+\!x_2^2x_3z_1z_4z_5\,y_2y_4y_5w_1w_3w_5D_{14}D_{34}D_{35}\right.\\
&\left.\!\!\!+x_2x_3x_5z_1z_4z_5\,y_2^2y_4w_1w_3w_5D_{13}D_{34}D_{45}\)(x_2z_1\!+\!y_2w_1)\\
&\!\!\!+\!x_2x_3x_5z_1^2z_4\,y_2y_3y_4w_1w_2w_5D_{35}D_{45}\(\!-D_{13}D_{24}\!+\!x_2z_4D_{13}\!+\!y_4w_2D_{13}
\!+\!\frac{z_4}{z_1}\,y_2w_1D_{13}\!+\!x_3z_1D_{24}\!+\!y_3w_1D_{24}\!\)\\
&\!\!\!+\!x_2x_3x_5z_1^2z_4\,y_2y_3y_4w_1w_3w_5D_{25}D_{34}D_{45}(x_2z_1\!+\!y_2w_1)\\
&\!\!\!\!\!\!\!\!\!\!\!\!\!\!\!\!\!\!\!\!\!\!\!\!\!\!\!\!\!\!\!\!\!\!
+\!x_2x_3x_5z_1^2z_4\,y_3y_4y_5w_1w_2w_3D_{24}D_{25}\(\!-D_{15}D_{34}\!-\!\frac{z_4}{z_1}D_{13}D_{15}
\!+\!x_3z_4D_{15}\!+\!y_4w_3D_{15}\!+\!\frac{z_4}{z_1}\,y_3w_1D_{15}\!+\!x_5z_1D_{34}\!+\!y_5w_1D_{34}\!\)\\
&\!\!\!+\!x_2x_3x_5z_1z_4^2\,y_3y_4y_5w_1w_2w_3D_{13}D_{15}D_{25}(x_2z_4\!+\!y_4w_2)\\
&\!\!\!+\!x_2x_3x_5z_1^2z_4\,y_2y_4y_5w_1w_2w_3D_{34}D_{35}\(\!-D_{15}D_{24}\!+\!x_2z_4D_{15}\!+\!y_4w_2D_{15}
\!+\!\frac{z_4}{z_1}\,y_2w_1D_{15}\!+\!x_5z_1D_{24}\!+\!y_5w_1D_{24}\!\),
\eal
\ee
\be
S_{25}=-\,x_2x_3z_5\,y_2y_3y_4y_5w_1w_2w_3w_5D_{13}D_{14}D_{24}(z_4D_{15}+z_1D_{45}),
~~~~~~~~~~~~~~~~~~~~~~~~~~~~~~~~~~~~~~~~~~\,
\ee
\be
S_{30}=s_2\,x_2x_3z_1z_4z_5\,y_3y_4y_5w_1w_2w_5D_{13}D_{14}D_{24}(x_2\,y_5w_3+x_3\,y_2w_5),
~~~~~~~~~~~~~~~~~~~~~~~~~~~~~~~~~~~~~~~
\ee
\be
S_{31}=-\,x_2x_3x_5z_1z_4z_5\,y_2y_3y_4y_5w_1^2w_2w_5D_{13}D_{24}D_{34}.
~~~~~~~~~~~~~~~~~~~~~~~~~~~~~~~~~~~~~~~~~~~~~~~~~~~~~~~~~~~\,
\ee
The difference between the deformed $S$ on the $6\!+\!2$ cuts and the proper numerator is then
\be
\bal
&\,S\(0\,,\,0\,,z_1,w_1,x_2,y_2,\,0\,,w_1\!+\!\frac{x_2z_1}{y_2},x_3,y_3,\,0\,,w_1\!+\!\frac{x_2z_1}{y_2},
\,0\,,y_4,z_4,\,0\,,x_5,y_5,z_5,w_5\)\\
&-\(w_1+\frac{x_2z_1}{y_2}\)^2y_4w_1\,x_3z_1\,\frac{y_2(M_1y_2D_{34})+y_3M_2}{y_2^3}\\
=\,&\,x_2x_3x_5z_1^2z_4z_5\,y_3y_4^2y_5w_1^2w_5\,(x_3y_2-x_2y_3)\(w_1+\frac{x_2z_1}{y_2}\)^2
\[x_2z_4+(y_4-y_2)\(w_1+\frac{x_2z_1}{y_2}\)\](s_5-1),
\eal
\ee
to make this difference vanish we must take $s_5\!=\!+1$, which agrees with \cite{Bern:2007ct}.

This completes the determination of $s_1,s_2,s_3,s_4,s_5$ for all five non-rung-rule topologies at 5-loop.

\newpage
\section{Beyond 5-loop Order?}

It is clear that for the 4- and 5-loop 4-particle amplituhedra we are no longer using the Mondrian diagrammatics,
instead we use the purely amplituhedronic way to obtain the $d\log$ forms from positivity conditions simplified by
external and internal cuts, which are similar to the traditional unitarity cuts. As discussed in the end
of \cite{An:2017tbf}, it is appealing to generalize the Mondrian diagrammatics to include the non-Mondrian complexity.
In \cite{Cachazo:2008dx} there is some kind of evidence about how the Mondrian DCI topologies can be related to
non-Mondrian ones, and it would be interesting to prove those rules which determine the coefficients of non-rung-rule
topologies from the amplituhedronic perspective. All the effort on discovering new rules and patterns finally aims to
help us go beyond the current understanding of the 5-loop case, such as to explain the coefficient $+2$ of a
special 6-loop DCI topology in \cite{Bourjaily:2011hi} since we believe a simple integer coefficient must have a
simple origin. The brute-force calculation merely using positivity conditions might be significantly simplified
by clever new observations, as we have witnessed in the Mondrian diagrammatics at 3-loop and the positive cuts
at 4- and 5-loop. After extracting sufficient deeper features of positivity conditions, it is even possible to conceive
a purely combinatoric description of the amplituhedron.

Still, the standard geometric way has a lot to be excavated beyond the current primitive level. When we use positive cuts
to determine the coefficient of a particular DCI topology, this looks like ``projecting'' the entire amplituhedron
onto a subspace that contains a subset of all boundaries, we then would like to get more intuition of its geometric
interpretation. And why the DCI topologies must be planar, as a basis
in what sense they are complete, how this completeness is related to the triangulation of amplituhedron,
as well as what role dual conformal invariance plays
in the geometric picture, are very vague so far while we believe clarification of these questions will be a
significant progress. When searching for various novel formalisms and connections to mathematics to better aid the practical
calculation of physical integrands at sufficiently higher loop orders,
we will also pay attention to some aspects discussed in \cite{Eden:2012tu,Drummond:2007aua,Nguyen:2007ya} which may
provide unexpected inspirations. For example, it is interesting to explore how the off-shell finiteness finds its basis
in the amplituhedronic setting. And starting at 8-loop \cite{Bourjaily:2015bpz,Bourjaily:2016evz},
novelties such as fractional coefficients and non-$d\log$ contributions also call for amplituhedronic
explanations, if the amplituhedron manages to pass all the lower loop tests.

Besides the outlook, it is also helpful to give some remarks on the technical aspects. To simplify the determination
of coefficients as much as possible, we must maximally utilize the crucial difference in pole structure of DCI topologies,
namely, we will impose sufficient cuts to isolate the particular diagram under consideration while minimizing its
accompanying surviving diagrams of different topologies. Note that in our convention, diagrams with the same denominator
but different numerators such that they cannot be related to each other by dihedral symmetry, are considered as different
DCI topologies, such as $T_{15}$ and $T_{16}$ in figure \ref{fig-10}. If finally it is inevitable to deal with these
accompanying diagrams, we can still use cuts to separate them, so that their coefficients must
satisfy independent sub-equalities in the overall equality required by positivity conditions.

Also, as we have seen from various examples, the calculation of 4-particle loop integrands from positivity conditions
with or without cuts, is magically effective: as long as the final answer is free of spurious poles, it is correct and
physical. Besides the possible geometric interpretation using DCI topologies, this mystery should have
a more self-contained mathematical reason, which can in return refine the laborious and foamy cancelation of spurious poles.
And the process of combining the so-called $d\log$ forms, in fact, indicates properties more general than
logarithmic singularities or differential forms, as it only depends on the universal fact that the integrand is a
rational function in which physical propagators appear as simple poles. The conjectured positivity conditions further
serve as some kind of ``residue theorems'' to provide an effective prescription for constructing the integrand.
Such observations may imply that the $d\log$ forms function beyond their definitions, which may hopefully unleash the
possibility to account for the non-$d\log$ novelty from the amplituhedronic perspective at 8-loop and higher.

Finally, it has been appealing to extend the techniques for 4-particle amplituhedron to handle more
external particles and various configurations of helicities. Attempts include the recent development using sign flips
\cite{Arkani-Hamed:2017vfh,Prlina:2017azl}, and the discovery of the key role of 4-particle loop integrand from which
the integrand of more particles can be extracted \cite{Heslop:2018zut}. It is worth noticing that, positivity of the pure
loop sector and that of the supersymmetric sector encoding helicities use quite different mathematical prescriptions.
This difference somehow obstructs an effective unified framework, while from the perspective of positivity, the 4-particle
amplituhedron with pure loop sector only (and the 4-particle sign-flip constraints are trivial) is the simplest object,
in particular, it is even simpler than the pure tree amplituhedron.

\newpage
\appendix
\section{Details of the $d\log$ Form for Determining $s_2,s_3,s_4$}
\label{app1}

Below we derive the $d\log$ form for determining $s_2,s_3,s_4$, with respect to positivity conditions
\be
\bal
&\frac{z_2}{z_5+y_5w_5/x_5}+\frac{w_1}{w_5+x_5z_5/y_5}>1,\\
&\frac{x_3}{x_5+y_5w_5/z_5}+\frac{y_3}{y_5+x_5z_5/w_5}>1,~~~(z_5-z_2)\(x_3+y_3\frac{w_2}{z_2}-x_5\)+y_5(w_2-w_5)>0,\\
&\frac{z_4}{z_5+y_5w_5/x_5}+\frac{y_4}{y_5+x_5z_5/w_5}>1,~~~\frac{z_4}{z_2}+\frac{y_4}{y_3+x_3z_2/w_2}>1. \labell{eq-11}
\eal
\ee
For later convenience, we define quantities
\be
\bal
&n_3=x_3+y_3\frac{w_5}{z_5}-x_5-\frac{y_5w_5}{z_5},~~
n_5=x_3+y_3\frac{w_2}{z_2}-x_5-y_5\,\frac{w_5-w_2}{z_5-z_2},\\
&p_3=y_5+\frac{x_5z_5}{w_5},~~p_5=\frac{z_2}{w_2}\(x_5+y_5\,\frac{w_5-w_2}{z_5-z_2}\),~~
p_{35}=y_5\,\frac{z_2}{z_2-z_5},\\
&n_{24}=x_3-\frac{w_2}{z_2}\(y_5+\frac{x_5z_5}{w_5}-y_3\) \labell{eq-13}
\eal
\ee
for the discussion involving $y_3,x_3$, as well as
\be
\bal
&a_2=z_5+\frac{y_5w_5}{x_5},~~b_2=y_5+\frac{x_5z_5}{w_5},~~
a_4=z_2,~~b_4=y_3+\frac{x_3z_2}{w_2},~~z_4^\ast=\frac{b_4-b_2}{b_4/a_4-b_2/a_2},\\
&n_2=z_4\frac{b_2}{a_2}+y_4-b_2,~~n_4=z_4\frac{b_4}{a_4}+y_4-b_4,\\
&A=\(\frac{1}{z_4}-\frac{1}{z_4-z_4^\ast}\)\frac{1}{n_4}
+\(\frac{1}{z_4-z_4^\ast}-\frac{1}{z_4-a_2}\)\frac{1}{n_2}+\frac{1}{z_4-a_2}\frac{1}{y_4},~~~
B=\frac{1}{z_4y_4}\frac{n_2+b_2}{n_2},\\
&F=\frac{1}{z_4y_4}\frac{n_4+b_4}{n_4},~~~
G=\(\frac{1}{z_4}-\frac{1}{z_4-z_4^\ast}\)\frac{1}{n_2}
+\(\frac{1}{z_4-z_4^\ast}-\frac{1}{z_4-a_4}\)\frac{1}{n_4}+\frac{1}{z_4-a_4}\frac{1}{y_4} \labell{eq-14}
\eal
\ee
for the discussion involving $z_4,y_4$. We will also use identities
\be
\bal
&\frac{w_5}{z_5}-\frac{w_5-w_2}{z_5-z_2}=\frac{z_2}{z_2-z_5}\(\frac{w_5}{z_5}-\frac{w_2}{z_2}\),\\
&p_3-p_5=\frac{z_2z_5}{w_2w_5}\frac{x_5}{z_2-z_5}\(\frac{w_5}{z_5}-\frac{w_2}{z_2}\)\(z_5+\frac{y_5w_5}{x_5}-z_2\).
\labell{eq-12}
\eal
\ee
Now let's analyze all possible situations of variables $z_2,w_1,w_2,y_3,x_3,z_4,y_4$, by first separating
situations $z_2\!<\!z_5$, $z_5\!<\!z_2\!<\!z_5\!+\!y_5w_5/w_5$ and $z_2\!>\!z_5\!+\!y_5w_5/w_5$.

\subsection{$z_2\!<\!z_5$}

For $z_2\!<\!z_5$, the 1st line of \eqref{eq-11} in terms of $w_1$ is nontrivial.
The 2nd condition in its 2nd line becomes
\be
x_3+y_3\frac{w_2}{z_2}>x_5+y_5\,\frac{w_5-w_2}{z_5-z_2},
\ee
and for comparison we can rewrite the 1st condition in the same line as
\be
x_3+y_3\frac{w_5}{z_5}>x_5+y_5\frac{w_5}{z_5},
\ee
using the 1st identity in \eqref{eq-12}, for $w_2\!<\!w_5z_2/z_5$ we find
\be
w_2<w_5\frac{z_2}{z_5}\Longrightarrow\frac{w_5}{z_5}<\frac{w_5-w_2}{z_5-z_2}.
\ee
For these two conditions in the 2nd line of \eqref{eq-11}, in terms of $n_3$ and $n_5$ defined in \eqref{eq-13},
we have a clear picture in the $y_3$-$x_3$ plane: the $x_3$-intercept of $n_3\!=\!0$ is less than that of $n_5\!=\!0$,
while its slope is greater than that of $n_5\!=\!0$, therefore $n_5\!>\!0$ already implies $n_3\!>\!0$ in the 1st quadrant.

\begin{figure}
\begin{center}
\includegraphics[width=0.361\textwidth]{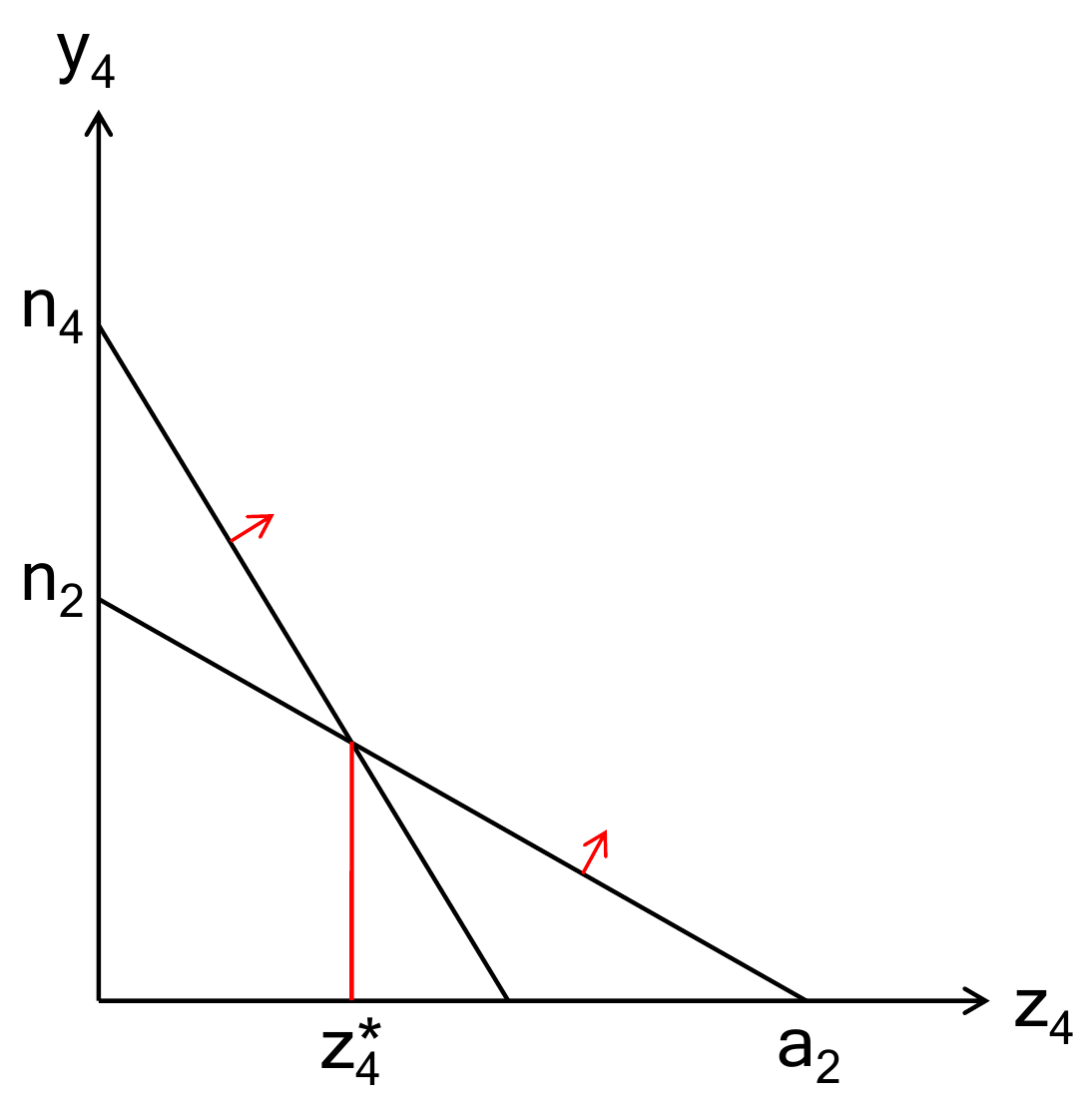}
\caption{Geometric picture of the $d\log$ form $A$.} \label{fig-16}
\end{center}
\end{figure}

For the two conditions in the 3rd line of \eqref{eq-11}, in terms of $n_2$ and $n_4$ defined in \eqref{eq-14},
since $z_2\!<\!z_5\!<\!z_5\!+\!y_5w_5/w_5$ and
\be
y_3+x_3\frac{z_2}{w_2}>y_3+x_3\frac{z_5}{w_5}>y_5+x_5\frac{z_5}{w_5},
\ee
in the $z_4$-$y_4$ plane the $y_4$-intercept of $n_4\!=\!0$ is greater than that of $n_2\!=\!0$
while its $z_4$-intercept is less than that of $n_2\!=\!0$, so they intercept at $z_4\!=\!z_4^\ast$ in the 1st quadrant.
Its $d\log$ form is given by $A$, where $z_4^\ast$ and $A$ are defined in \eqref{eq-14},
and the corresponding geometric picture is given in figure \ref{fig-16}.

\begin{figure}
\begin{center}
\includegraphics[width=0.367\textwidth]{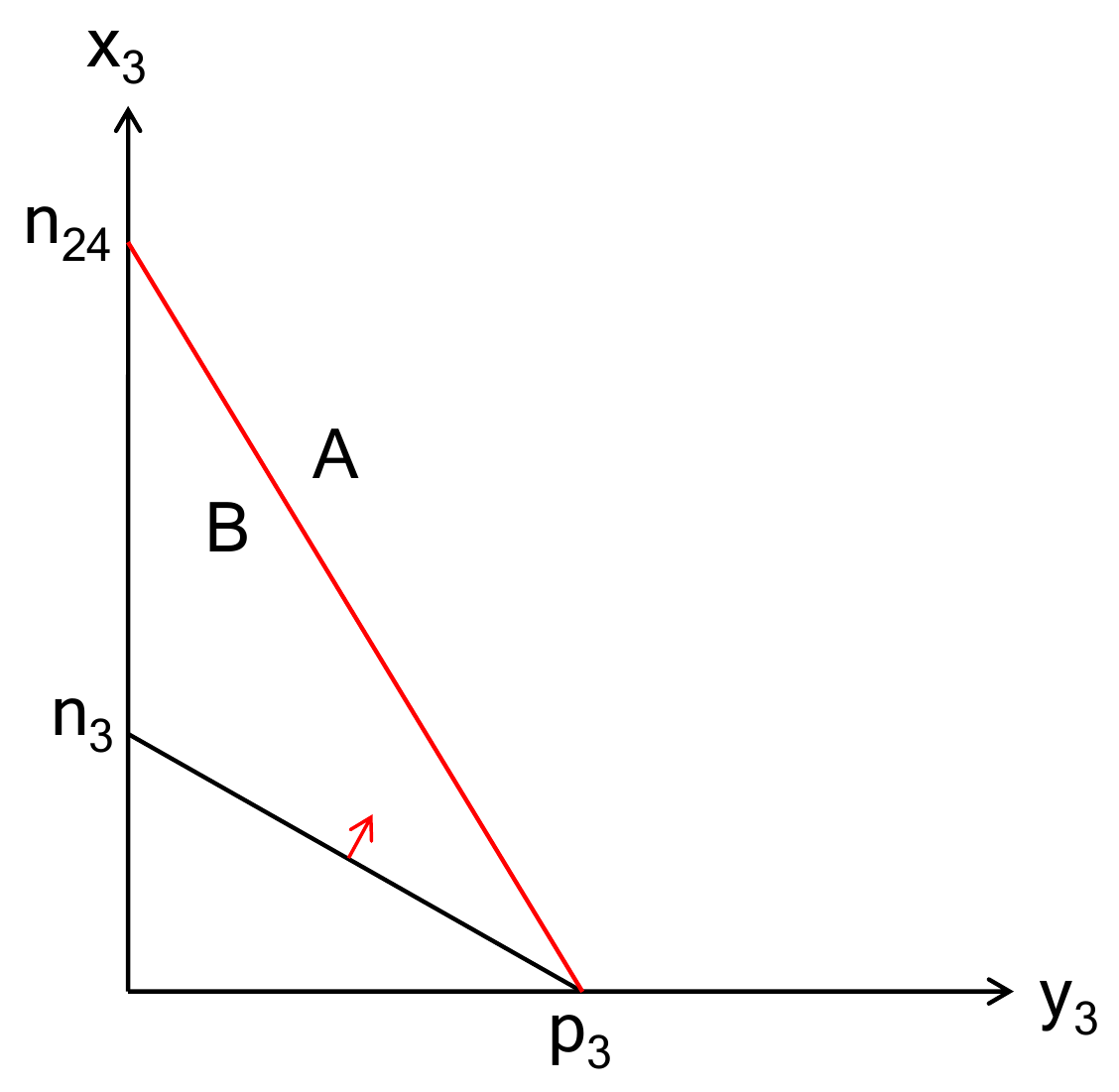}
\caption{Bifurcation of $y_3\!+\!x_3z_2/w_2\!\lessgtr\!y_5\!+\!x_5z_5/w_5$ in the $y_3$-$x_3$ plane.} \label{fig-17}
\end{center}
\end{figure}

Now for $w_2\!>\!w_5z_2/z_5$, similarly we have
\be
w_2>w_5\frac{z_2}{z_5}\Longrightarrow\frac{w_5}{z_5}>\frac{w_5-w_2}{z_5-z_2},
\ee
therefore $n_3\!>\!0$ already implies $n_5\!>\!0$. Since
\be
y_3+x_3\frac{z_2}{w_2}<y_3+x_3\frac{z_5}{w_5},
\ee
we need $n_{24}$ defined in \eqref{eq-13} for comparing $y_3\!+\!x_3z_2/w_2$ and $y_5\!+\!x_5z_5/w_5$.
If $y_3\!+\!x_3z_2/w_2\!<\!y_5\!+\!x_5z_5/w_5$, $n_2\!>\!0$ already implies $n_4\!>\!0$ in the $z_4$-$y_4$ plane,
$A$ will be replaced by $B$ defined in \eqref{eq-14}, which involves $n_2$ only.
This bifurcation divides the region of $n_3\!>\!0$ in the $y_3$-$x_3$ plane
as shown in figure \ref{fig-17}, in which $p_3$ defined in \eqref{eq-13} is the $y_3$-intercept of both $n_3\!=\!0$
and $n_{24}\!=\!0$.

In summary, the $d\log$ form for $z_2\!<\!z_5$ is given by (omitting the part of $z_2,w_1$ for the moment)
\be
\bal
S_1=&\(\frac{1}{w_2}-\frac{1}{w_2-w_5z_2/z_5}\)\frac{1}{y_3x_3}\frac{x_3+y_3w_2/z_2}{n_5}A\\
&+\frac{1}{w_2-w_5z_2/z_5}\[\(\frac{1}{y_3}-\frac{1}{y_3-p_3}\)
\(\(\frac{1}{n_3}-\frac{1}{n_{24}}\)B+\frac{1}{n_{24}}A\)+\frac{1}{y_3-p_3}\frac{1}{x_3}A\].
\eal
\ee

\subsection{$z_5\!<\!z_2\!<\!z_5\!+\!y_5w_5/w_5$}

\begin{figure}
\begin{center}
\includegraphics[width=0.368\textwidth]{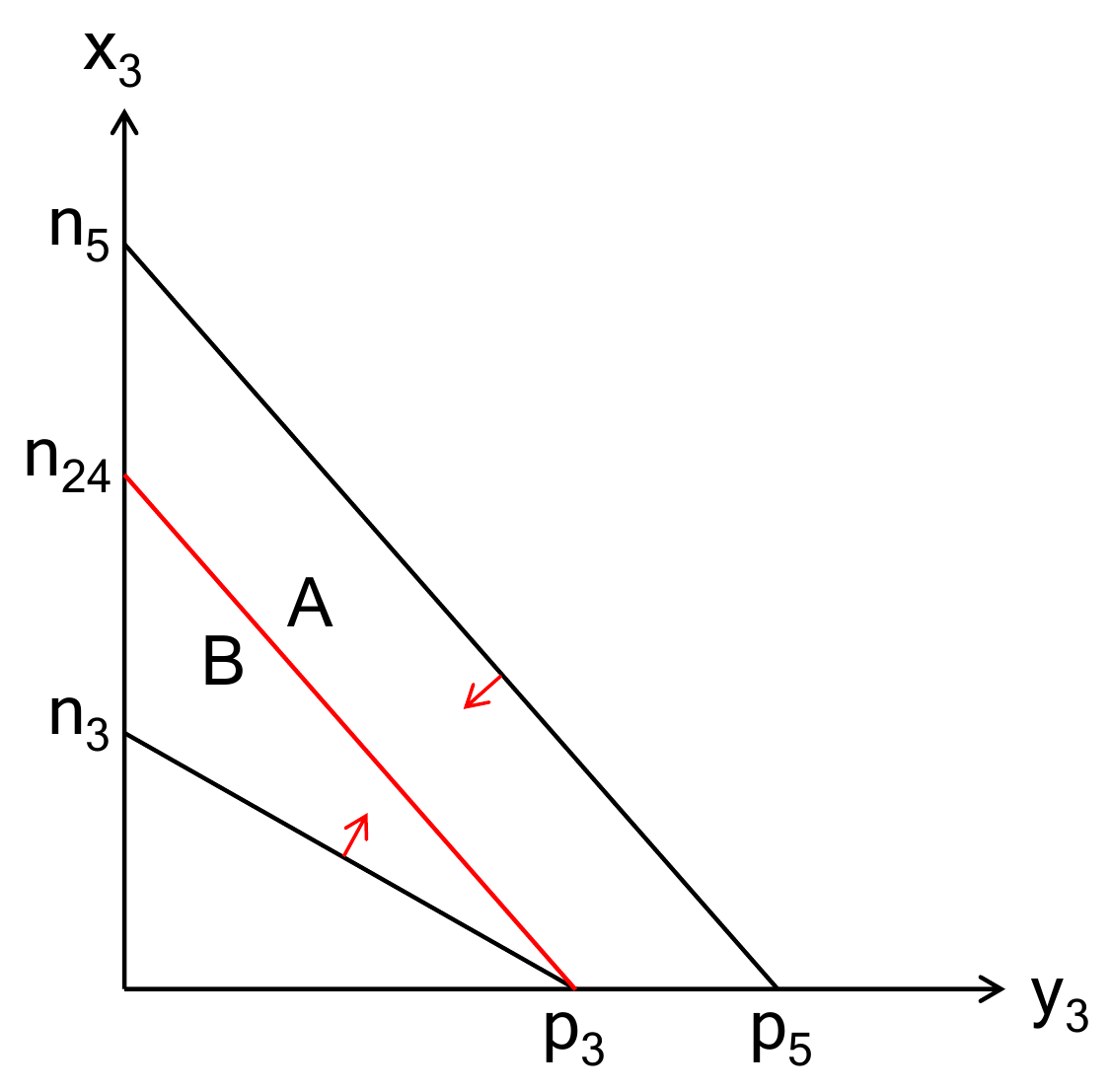}
\caption{The only contributing part of $z_5\!<\!z_2\!<\!z_5\!+\!y_5w_5/w_5$,
for which $w_2\!>\!w_5z_2/z_5$.} \label{fig-18}
\end{center}
\end{figure}

For $z_5\!<\!z_2\!<\!z_5\!+\!y_5w_5/w_5$, the 1st line of \eqref{eq-11} remains nontrivial. Its 2nd line becomes
\be
x_3+y_3\frac{w_5}{z_5}>x_5+y_5\frac{w_5}{z_5},~~x_3+y_3\frac{w_2}{z_2}<x_5+y_5\,\frac{w_2-w_5}{z_2-z_5},
\ee
using both identities in \eqref{eq-12} we find (below $p_5$ defined in \eqref{eq-13} is the $y_3$-intercept of $n_5\!=\!0$)
\be
\bal
w_2\lessgtr w_5\frac{z_2}{z_5}&\Longrightarrow\frac{w_5}{z_5}\gtrless\frac{w_2-w_5}{z_2-z_5}\\
&\Longrightarrow p_3\gtrless p_5.
\eal
\ee
If $w_2\!<\!w_5z_2/z_5$, both the $x_3$- and $y_3$-intercept of $n_3\!=\!0$ are greater than that of $n_5\!=\!0$,
so regions of $n_3\!>\!0$ and $n_5\!<\!0$ have no overlap. Therefore only the $w_2\!>\!w_5z_2/z_5$ part contributes,
for which both the $x_3$- and $y_3$-intercept of $n_3\!=\!0$ are less than that of $n_5\!=\!0$
as shown in figure \ref{fig-18}. In this case, we again need $n_{24}$ to divide the region, as the slope of $n_{24}\!=\!0$
is greater than that of $n_3\!=\!0$ ($n_{24}\!=\!0$ is parallel to $n_5\!=\!0$).

In summary, the $d\log$ form for $z_5\!<\!z_2\!<\!z_5\!+\!y_5w_5/w_5$ is given by
\be
\!\!\!\!\!\!\!\!\!\!\!\!\!\!
S_2=\frac{1}{w_2-w_5z_2/z_5}\[\(\frac{1}{y_3}-\frac{1}{y_3-p_3}\)
\(\(\frac{1}{n_3}-\frac{1}{n_{24}}\)B+\(\frac{1}{n_{24}}-\frac{1}{n_5}\)A\)
+\(\frac{1}{y_3-p_3}-\frac{1}{y_3-p_5}\)\(\frac{1}{x_3}-\frac{1}{n_5}\)A\].
\ee

\subsection{$z_2\!>\!z_5\!+\!y_5w_5/w_5$}

\begin{figure}
\begin{center}
\includegraphics[width=0.362\textwidth]{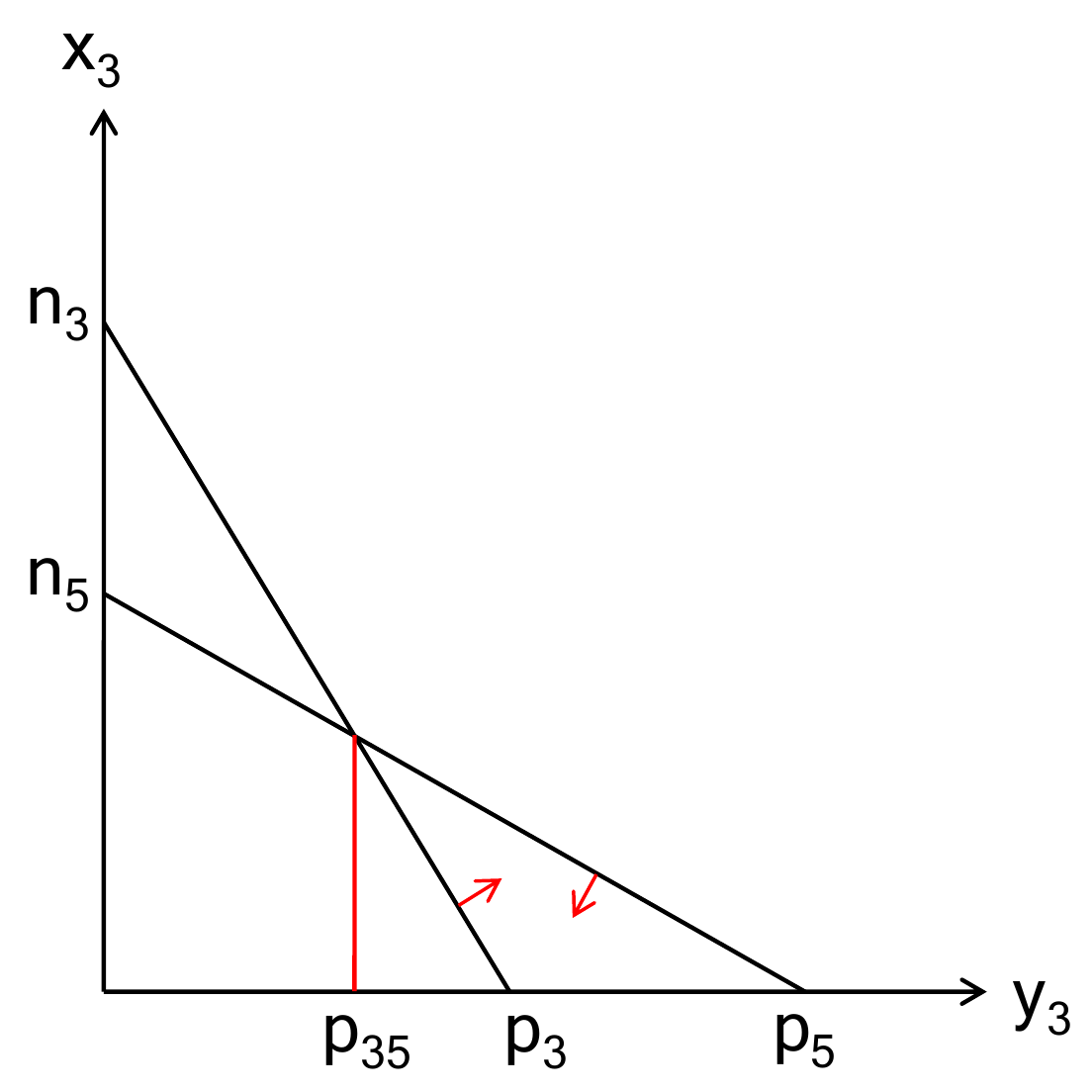}
\caption{$n_3\!=\!0$ and $n_5\!=\!0$ intercept when $w_2\!<\!w_5z_2/z_5$.} \label{fig-19}
\end{center}
\end{figure}

\begin{figure}
\begin{center}
\includegraphics[width=0.364\textwidth]{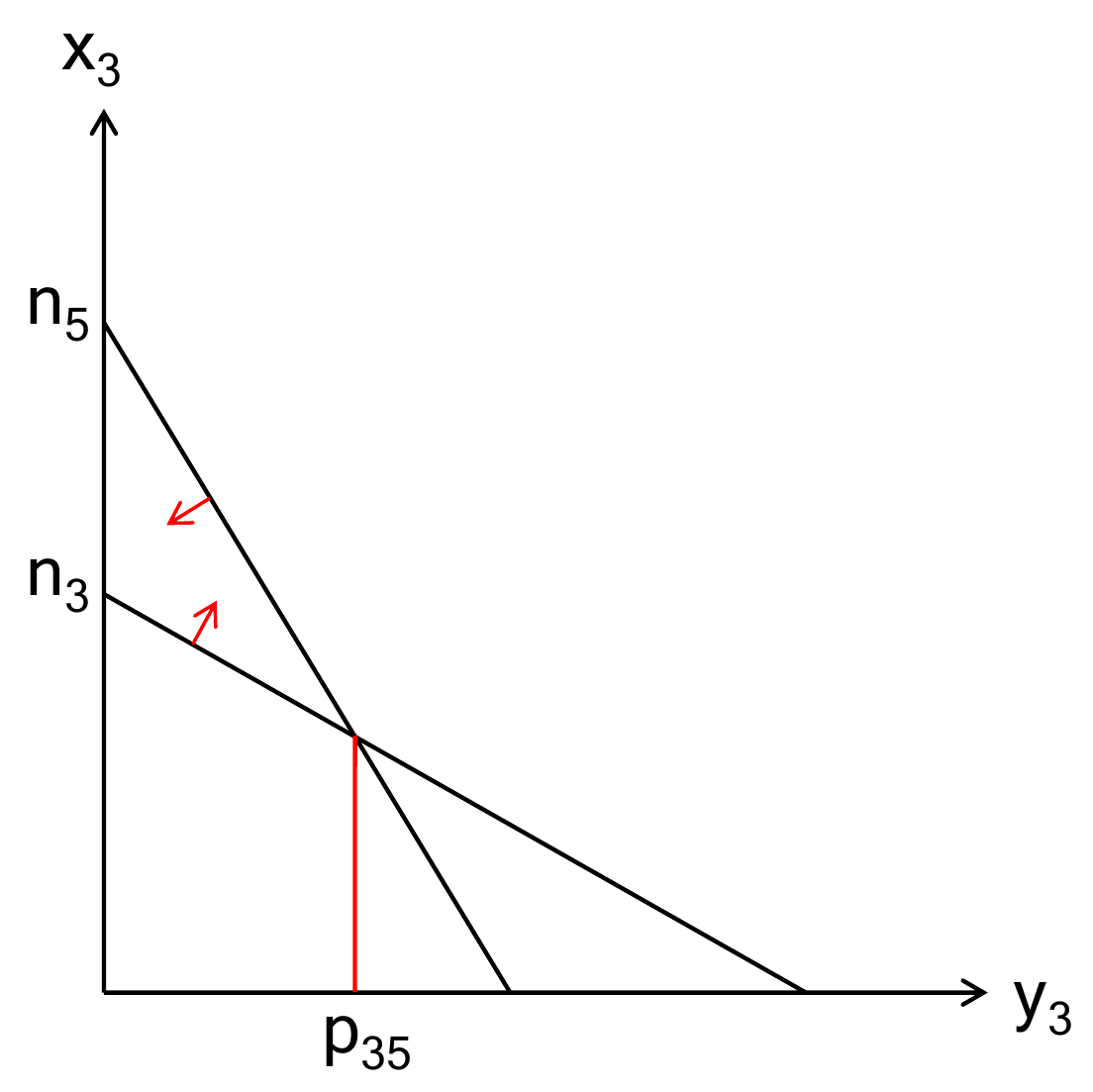}
\caption{$n_3\!=\!0$ and $n_5\!=\!0$ intercept when $w_2\!>\!w_5z_2/z_5$.} \label{fig-20}
\end{center}
\end{figure}

For $z_2\!>\!z_5\!+\!y_5w_5/w_5$, the 1st line of \eqref{eq-11} now becomes trivial. Its 2nd line remains the same as
that for $z_5\!<\!z_2\!<\!z_5\!+\!y_5w_5/w_5$, but there is a slight difference in the 2nd identity in \eqref{eq-12} as
\be
\bal
w_2\lessgtr w_5\frac{z_2}{z_5}&\Longrightarrow\frac{w_5}{z_5}\gtrless\frac{w_2-w_5}{z_2-z_5}\\
&\Longrightarrow p_3\lessgtr p_5,
\eal
\ee
so that $n_3\!=\!0$ and $n_5\!=\!0$ always intercept, and its geometric pictures are given in figures \ref{fig-19}
and \ref{fig-20} with respect to $w_2\!\lessgtr\!w_5z_2/z_5$. For $w_2\!<\!w_5z_2/z_5$ we again have
\be
y_3+x_3\frac{z_2}{w_2}>y_3+x_3\frac{z_5}{w_5}>y_5+x_5\frac{z_5}{w_5},
\ee
and since $z_2\!>\!z_5\!+\!y_5w_5/w_5$, $n_4\!>\!0$ already implies $n_2\!>\!0$ in the $z_4$-$y_4$ plane.
Its $d\log$ form is given by $F$ defined in \eqref{eq-14}, which involves $n_4$ only. For $w_2\!>\!w_5z_2/z_5$,
since $n_{24}\!=\!0$ intercepts $n_3\!=\!0$ at $p_3$ with $p_3\!>\!p_5$
and $n_{24}\!=\!0$ is parallel to $n_5\!=\!0$, $n_5\!<\!0$ already implies $n_{24}\!<\!0$, which means
\be
y_3+x_3\frac{z_2}{w_2}<y_5+x_5\frac{z_5}{w_5},
\ee
and hence $F$ will be replaced by $G$ defined in \eqref{eq-14}, as it can be obtained from $A$ by
switching $n_2,a_2,b_2\!\leftrightarrow\!n_4,a_4,b_4$.

In summary, the $d\log$ form for $z_2\!>\!z_5\!+\!y_5w_5/w_5$ is given by
\be
\bal
S_3=&\(\frac{1}{w_2}-\frac{1}{w_2-w_5z_2/z_5}\)\[\(\frac{1}{y_3-p_{35}}-\frac{1}{y_3-p_3}\)
\(\frac{1}{n_3}-\frac{1}{n_5}\)+\(\frac{1}{y_3-p_3}-\frac{1}{y_3-p_5}\)\(\frac{1}{x_3}-\frac{1}{n_5}\)\]F\\
&+\frac{1}{w_2-w_5z_2/z_5}\(\frac{1}{y_3}-\frac{1}{y_3-p_{35}}\)\(\frac{1}{n_3}-\frac{1}{n_5}\)G.
\eal
\ee
\\ \\
Collecting $S_1,S_2,S_3$, the overall $d\log$ form is then
\be
\bal
&\[\(\frac{1}{z_2}-\frac{1}{z_2-z_5}\)S_1+\(\frac{1}{z_2-z_5}-\frac{1}{z_2-z_5-y_5w_5/x_5}\)S_2\]
\frac{1}{x_5z_2/y_5+w_1-x_5z_5/y_5-w_5}\\
&+\frac{1}{z_2-z_5-y_5w_5/x_5}\frac{1}{w_1}\,S_3=\frac{M}{z_2^3w_1w_2y_3x_3z_4y_4\,D_{15}D_{35}D_{25}D_{45}D_{24}},
\eal
\ee
where $M$ is the numerator simplified by \textsc{Mathematica} as given in the expression below \eqref{eq-19}.

\section{Details of the $d\log$ Form for Determining $s_5$}
\label{app2}

Below we present the $d\log$ form for determining $s_5$ with a brief description of its derivation,
with respect to positivity conditions
\be
\bal
\!\!\!\!\!\!
&w_1\!+\!z_1\frac{x_5}{y_5}\!>\!w_5\!+\!z_5\frac{x_5}{y_5},~~
(y_5\!-\!y_2)\(\!w_1\!+\!z_1\frac{x_2}{y_2}\!-\!w_5\!\)\!+\!z_5(x_2\!-\!x_5)\!>\!0,~~
(y_5\!-\!y_3)\(\!w_1\!+\!z_1\frac{x_2}{y_2}\!-\!w_5\!\)\!+\!z_5(x_3\!-\!x_5)\!>\!0,\\
\!\!\!\!\!\!
&\frac{z_4}{z_5\!+\!y_5w_5/x_5}\!+\!\frac{y_4}{y_5\!+\!x_5z_5/w_5}\!>\!1,~~
\frac{z_4}{z_1\!+\!w_1y_2/x_2}\!+\!\frac{y_4}{y_2}\!>\!1,~~
\frac{z_4}{k(z_1\!+\!w_1y_2/x_2)}\!+\!\frac{y_4}{y_3}\!>\!1, \labell{eq-18}
\eal
\ee
where $k\!=\!y_3x_2/(y_2x_3)\!<\!1$. Recall that we focus on $w_1,z_1$ in the first line and $z_4,y_4$ in the second,
so that the discussions can be done within two planes: the $z_1$-$w_1$ and the $y_4$-$z_4$ plane. For a clear picture,
we can rewrite the 2nd and 3rd conditions in the 1st line as
\be
\bal
w_1+z_1\frac{x_2}{y_2}&>w_5+z_5\,\frac{x_5-x_2}{y_5-y_2}~~\textrm{for}~y_2<y_5\\
&<w_5+z_5\,\frac{x_2-x_5}{y_2-y_5}~~\textrm{for}~y_2>y_5,\\
\eal
\ee
\be
\bal
w_1+z_1\frac{x_2}{y_2}&>w_5+z_5\,\frac{x_5-x_3}{y_5-y_3}~~\textrm{for}~y_3<y_5\\
&<w_5+z_5\,\frac{x_3-x_5}{y_3-y_5}~~\textrm{for}~y_3>y_5.\\
\eal
\ee
We also have noticed that since $k\!<\!1$, if $y_3\!<\!y_2$ the 2nd condition in the 2nd line already implies
the 3rd, which explains the factor $D_{34}$ in the numerator of \eqref{eq-17}.
There is another tricky issue depending on the relation between
$y_2$ and $y_3$ as well, namely before we impose $w_2\!=\!w_3$ for setting $D_{23}\!=\!0$, we have
\be
D_{23}=(y_3-y_2)(w_2-w_3),
\ee
so there is a bifurcation of $y_3\!\lessgtr\!y_2$ in the relevant dimensionless ratio
\be
\frac{y_2}{y_2-y_3}\frac{w_3}{w_3-w_2}\,R_1+\frac{y_3}{y_3-y_2}\frac{w_2}{w_2-w_3}\,R_2
\to\frac{\hat{w}_3}{D_{23}}\,(y_2R_1+y_3R_2)
\ee
after imposing $w_2\!=\!w_3\!=\!\hat{w}_3$, where $R_1$ and $R_2$ are proportional to $M_1$ and $M_2$ in \eqref{eq-17}
respectively which are the numerators simplified by \textsc{Mathematica} as given in the expressions below.

As indicated above, it is better to separately consider situations $y_3\!<\!y_2\!<\!y_5$, $y_3\!<\!y_5\!<\!y_2$,
$y_5\!<\!y_3\!<\!y_2$, $y_2\!<\!y_3\!<\!y_5$, $y_2\!<\!y_5\!<\!y_3$ and $y_5\!<\!y_2\!<\!y_3$ first, then depending on
each case we may need to discuss various situations involving $x_5,x'_3,w_5$ as well.
For example, to compare $x_5/y_5$ and $x_2/y_2$ involves $x_5$. And in the identity
which will be frequently used in the relevant discussions
\be
\frac{x_5-x_2}{y_5-y_2}-\frac{x_5-x_3}{y_5-y_3}=\frac{y_2-y_3}{(y_5-y_2)(y_5-y_3)}
\(x_5+x'_3\,\frac{y_5-y_2}{y_2-y_3}-x_2\frac{y_5}{y_2}\),
\ee
both $x_5$ and $x'_3$ are involved. Finally in the 2nd line of \eqref{eq-18}, to compare $y_5\!+\!x_5z_5/w_5$,
$y_2$ and $y_3$ may also involve $w_5$ given a fixed order of $y_2,y_3,y_5$.\\ \\

$M_1 = w_1^4 y_2^3 y_4 (y_3 - y_5) y_5^2 (w_5 (y_2 - y_4) - x_5 z_4) +
   w_1^3 y_2^2 y_5 (-2 w_5^2 y_2 (y_2 - y_4) y_4 (y_3 - y_5) y_5 +
      x_5 z_4 (x_3 y_2 y_4 y_5 z_5 -
         x_2 (y_3 - y_5) (3 y_4 y_5 z_1 + y_2 y_5 z_4 - y_2^2 z_5 + y_2 y_4 z_5) +
         x_5 y_2 y_4 (y_5 (z_1 - 2 z_5) + y_3 (-z_1 + z_5))) +
      w_5 y_4 (x_3 y_2 (-y_2 + y_4) y_5 z_5 -
         x_2 (y_3 - y_5) (3 y_4 y_5 z_1 + y_2 (y_5 (-3 z_1 + z_4) + y_4 z_5)) +
         x_5 y_2 (y_5 (y_4 z_1 - 2 y_5 z_4 - 2 y_4 z_5) +
            y_3 (-y_4 z_1 + 2 y_5 z_4 + y_4 z_5) +
            y_2 (y_3 z_1 - y_5 z_1 - y_3 z_5 + 2 y_5 z_5)))) -
   x_2 x_5 z_1 (w_5^3 y_2^2 y_3 (y_2 - y_5) (-y_4 + y_5) (y_4 z_1 + y_2 z_4) +
      x_2 x_5 y_5 z_1 z_4 (y_4 z_1 +
         y_2 (z_4 - z_5)) (x_2 (y_3 - y_5) z_1 + (-x_3 + x_5) y_2 z_5) +
      w_5^2 y_2 (x_2 z_1 (y_4 y_5 (y_4 y_5 + y_3 (-2 y_4 + y_5)) z_1 +
            y_2^2 y_3 (y_4 - y_5) z_4 +
            y_2 (-y_5^2 (y_4 z_1 - y_4 z_4 + y_5 z_4) +
               y_3 (y_4^2 z_1 - 2 y_4 y_5 z_4 + 2 y_5^2 z_4))) -
         y_2 (y_4 z_1 + y_2 z_4) (x_3 y_2 (y_4 - y_5) z_5 -
            x_5 y_3 (y_5 z_4 + y_4 z_5 - 2 y_5 z_5 + y_2 (-z_4 + z_5)))) +
      w_5 (-x_2^2 (y_3 - y_5) y_5 z_1^2 (-y_4^2 z_1 +
            y_2 (y_4 (z_1 - z_4) + y_5 z_4)) +
         x_5 y_2^2 (-x_3 y_2 + x_5 y_3) (y_4 z_1 + y_2 z_4) (z_4 - z_5) z_5 +
         x_2 y_2 z_1 (x_3 y_5 (-y_4^2 z_1 + y_2 (y_4 z_1 - y_4 z_4 + y_5 z_4)) z_5 +
            x_5 (y_2^2 y_3 z_4 (z_4 - z_5) +
               y_4 y_5 z_1 (y_5 z_4 + y_4 z_5 + y_3 (-2 z_4 + z_5)) +
               y_2 (y_3 (y_4 z_1 - 2 y_5 z_4) (z_4 - z_5) +
                  y_5 (y_5 z_4 (z_4 - 2 z_5) + y_4 (-z_1 + z_4) z_5)))))) +
   w_1 (w_5^3 y_2^2 y_4 (x_5 y_2 y_3 (y_2 - y_5) (y_4 - y_5) z_1 -
         x_2 (y_3 - y_5) y_5^2 (y_4 z_1 + y_2 (-z_1 + z_4))) +
      x_2 x_5 y_5 z_4 (-x_2^2 (y_3 - y_5) z_1 (y_4 z_1 (y_5 z_1 + y_2 z_5) +
            y_2 (y_5 z_1 z_4 + y_2 (-z_1 + z_4) z_5)) - (x_3 -
            x_5) x_5 y_2^2 z_5 (y_4 z_1 (-2 z_1 + z_5) +
            y_2 (z_4 z_5 + z_1 (-z_4 + z_5))) +
         x_2 y_2 (x_3 z_5 (y_4 z_1 (y_5 z_1 + y_2 z_5) +
               y_2 (y_5 z_1 z_4 + y_2 (-z_1 + z_4) z_5)) +
            x_5 (y_4 z_1 (y_5 z_1 (3 z_1 - 2 z_5) - y_2 z_5^2) +
               y_2 (y_2 (z_1 - z_4) z_5^2 +
                  2 y_5 z_1 (z_1 z_4 - z_1 z_5 - z_4 z_5)) +
               y_3 z_1 (y_4 z_1 (-3 z_1 + z_5) +
                  y_2 (-2 z_1 z_4 + 2 z_1 z_5 + z_4 z_5))))) -
      w_5 (x_5^2 y_2^3 (-x_3 y_2 + x_5 y_3) y_4 z_1 (z_4 - z_5) z_5 +
         x_2^3 y_4 (y_3 - y_5) y_5 z_1 (y_4 y_5 z_1^2 + y_2^2 z_4 z_5 +
            y_2 z_1 (y_5 (-z_1 + z_4) + y_4 z_5)) -
         x_2^2 y_2 y_5 (x_3 y_4 z_5 (y_4 y_5 z_1^2 + y_2^2 z_4 z_5 +
               y_2 z_1 (y_5 (-z_1 + z_4) + y_4 z_5)) -
            x_5 (y_2^2 z_4 z_5 (y_3 (z_1 - z_4) + y_5 (-z_1 + z_4) + y_4 z_5) +
               y_4 z_1^2 (y_3 (3 y_4 z_1 - 2 y_5 z_4 - y_4 z_5) +
                  y_5 (-3 y_4 z_1 + 2 y_5 z_4 + 2 y_4 z_5)) +
               y_2 z_1 (2 y_5^2 z_4 (z_1 + z_4) + y_4^2 z_5^2 +
                  y_4 y_5 (3 z_1^2 + 3 z_4 z_5 - 2 z_1 (z_4 + z_5)) +
                  y_3 (-2 y_5 z_4 (z_1 + z_4) +
                    y_4 (-3 z_1^2 + 2 z_1 z_4 + z_1 z_5 - 2 z_4 z_5))))) +
         x_2 x_5 y_2^2 (x_3 y_5 z_5 (y_4 z_1 (y_5 z_4 + y_4 (-2 z_1 + z_5)) +
               y_2 (y_5 z_4 (z_1 + z_4) +
                  y_4 (2 z_1^2 + z_4 z_5 - z_1 (z_4 + z_5)))) +
            x_5 (y_2^2 y_3 z_1 z_4 (z_4 - z_5) +
               y_4 y_5 z_1 (2 y_5 z_4 (z_1 - z_5) + y_4 (2 z_1 - z_5) z_5 +
                  y_3 (-4 z_1 z_4 + 2 z_1 z_5 + z_4 z_5)) +
               y_2 (y_3 (2 y_4 z_1^2 (z_4 - z_5) +
                    y_5 z_4 (-2 z_1 z_4 + 2 z_1 z_5 + z_4 z_5)) +
                  y_5 (y_5 z_4 (z_1 (z_4 - 2 z_5) - 2 z_4 z_5) +
                    y_4 z_5 (-2 z_1^2 - z_4 z_5 + z_1 (z_4 + z_5))))))) -
      w_5^2 y_2 (-x_2^2 y_4 (y_3 - y_5) y_5 (2 y_4 y_5 z_1^2 + y_2^2 z_4 z_5 +
            y_2 z_1 (-2 y_5 z_1 + 2 y_5 z_4 + y_4 z_5)) +
         x_5 y_2^2 y_4 z_1 (x_3 y_2 (-y_4 + y_5) z_5 +
            x_5 y_3 (y_5 z_4 + y_4 z_5 - 2 y_5 z_5 + y_2 (-z_4 + z_5))) +
         x_2 y_2 (x_3 y_4 y_5^2 (y_4 z_1 + y_2 (-z_1 + z_4)) z_5 +
            x_5 (y_2^2 y_3 (y_4 - y_5) z_1 z_4 +
               y_4 y_5 z_1 (y_3 (-4 y_4 z_1 + 2 y_5 z_1 + y_5 z_4 + y_4 z_5) -
                  y_5 (-2 y_4 z_1 + y_5 z_4 + 2 y_4 z_5)) +
               y_2 (y_3 (2 y_4^2 z_1^2 + y_5^2 z_4 (2 z_1 + z_4) +
                    y_4 y_5 (z_4 z_5 - z_1 (2 z_4 + z_5))) -
                  y_5^2 (y_5 z_4 (z_1 + z_4) +
                    y_4 (2 z_1^2 + 2 z_4 z_5 - z_1 (z_4 + 2 z_5)))))))) +
   w_1^2 y_2 (w_5^3 y_2^2 (y_2 - y_4) y_4 (y_3 - y_5) y_5^2 -
      w_5^2 y_2 y_4 (y_5 (x_3 y_2 (-y_2 + y_4) y_5 z_5 -
            x_2 (y_3 - y_5) (4 y_4 y_5 z_1 +
               y_2 (-4 y_5 z_1 + 2 y_5 z_4 + y_4 z_5))) +
         x_5 y_2 (y_2 (-y_5^2 (z_1 - 2 z_5) + y_3 (y_4 z_1 - y_5 z_5)) +
            y_5 (y_5 (-y_5 z_4 + y_4 (z_1 - 2 z_5)) +
               y_3 (y_5 (z_1 + z_4) + y_4 (-2 z_1 + z_5))))) +
      x_5 y_5 z_4 ((x_3 - x_5) x_5 y_2^2 y_4 (z_1 - z_5) z_5 -
         x_2^2 (y_3 - y_5) (y_4 z_1 (3 y_5 z_1 + 2 y_2 z_5) +
            y_2 (2 y_5 z_1 z_4 + y_2 (-2 z_1 + z_4) z_5)) +
         x_2 y_2 (x_3 z_5 (2 y_4 y_5 z_1 + y_2 y_5 z_4 - y_2^2 z_5 + y_2 y_4 z_5) +
            x_5 (y_4 (y_5 z_1 (3 z_1 - 4 z_5) - y_2 z_5^2) +
               y_2 (y_2 z_5^2 + y_5 (z_1 z_4 - z_1 z_5 - 2 z_4 z_5)) +
               y_3 (y_4 z_1 (-3 z_1 + 2 z_5) +
                  y_2 (-z_1 z_4 + z_1 z_5 + z_4 z_5))))) -
      w_5 (x_2^2 y_4 (y_3 - y_5) y_5 (3 y_4 y_5 z_1^2 + y_2^2 z_4 z_5 +
            y_2 z_1 (-3 y_5 z_1 + 2 y_5 z_4 + 2 y_4 z_5)) +
         x_5 y_2^2 y_4 (x_3 y_5 z_5 (y_5 z_4 + y_2 (z_1 - z_5) + y_4 (-z_1 + z_5)) +
             x_5 (y_2 (y_3 z_1 (z_4 - z_5) + y_5 z_5 (-z_1 + z_5)) +
               y_5 (y_5 z_4 (z_1 - 2 z_5) + y_4 (z_1 - z_5) z_5 +
                  y_3 (z_4 z_5 + z_1 (-2 z_4 + z_5))))) -
         x_2 y_2 y_5 (x_3 y_4 z_5 (2 y_4 y_5 z_1 +
               y_2 (y_5 (-2 z_1 + z_4) + y_4 z_5)) -
            x_5 (y_2^2 (y_3 - y_5) z_4 z_5 +
               y_4 z_1 (y_3 (3 y_4 z_1 - 4 y_5 z_4 - 2 y_4 z_5) +
                  y_5 (-3 y_4 z_1 + 4 y_5 z_4 + 4 y_4 z_5)) +
               y_2 (y_5^2 z_4 (z_1 + 2 z_4) + y_4^2 z_5^2 +
                  y_4 y_5 (3 z_1^2 + 3 z_4 z_5 - z_1 (z_4 + 4 z_5)) -
                  y_3 (y_5 z_4 (z_1 + 2 z_4) +
                    y_4 (3 z_1^2 + 2 z_4 z_5 - z_1 (z_4 + 2 z_5))))))));$\\

$M_2 = w_1^5 y_2^4 (y_2 - y_4) y_4 (y_2 - y_5) y_5^2 (w_5 (-y_3 + y_4) + x_5 z_4) +
   w_1^4 y_2^3 y_5 (2 w_5^2 y_2 (y_2 - y_4) (y_3 - y_4) y_4 (y_2 - y_5) y_5 +
      w_5 y_4 (x_3 y_2 (y_2 - y_4) y_5 (-y_5 z_4 - y_4 z_5 + y_2 (z_4 + z_5)) +
         x_5 y_2 (y_2 -
            y_4) (y_5 (-y_4 z_1 + 2 y_5 z_4 + y_3 (z_1 - 2 z_5) + 2 y_4 z_5) +
            y_2 (y_4 z_1 - 2 y_5 z_4 - y_4 z_5 + y_3 (-z_1 + z_5))) +
         x_2 (4 y_4 (-y_3 + y_4) y_5^2 z_1 +
            y_2^2 (y_4 y_5 (4 z_1 - z_4 - z_5) +
               y_3 (-4 y_5 z_1 + y_5 z_4 + y_4 z_5)) +
            y_2 (-y_3 (-4 y_4 y_5 z_1 + y_5^2 (-4 z_1 + z_4) + y_4^2 z_5) +
               y_4 y_5 (y_5 (-4 z_1 + z_4) + y_4 (-4 z_1 + z_5))))) +
      x_5 z_4 (x_3 y_2 y_4 y_5 (-y_2 z_4 + y_5 z_4 - y_3 z_5 + y_4 z_5) +
         x_5 y_2 (y_2 - y_4) y_4 (y_2 z_1 - y_5 z_1 - y_2 z_5 + 2 y_5 z_5) +
         x_2 (4 y_4^2 y_5^2 z_1 +
            y_2^2 (y_3 (y_5 z_4 - y_3 z_5) +
               y_4 (4 y_5 z_1 - y_5 z_4 + y_3 z_5 - y_5 z_5)) +
            y_2 (y_3 y_5 (-y_5 z_4 + y_3 z_5) +
               y_4 (y_5^2 (-4 z_1 + z_4) + y_3^2 z_5 - y_3 y_5 z_5) +
               y_4^2 (-y_3 z_5 + y_5 (-4 z_1 + z_5)))))) +
   x_2 x_5 z_1 (-w_5^3 y_2^3 (y_3 - y_5) (-y_4 +
         y_5) (x_3 y_2 z_4 (y_4 z_1 + y_2 z_4) +
         x_2 z_1 (y_4^2 z_1 - y_2 (y_4 z_1 + y_3 z_4 - y_4 z_4))) -
      x_2 x_5 y_5 z_1 z_4 (-x_3 (x_3 - x_5) y_2^2 z_4 (y_4 z_1 +
            y_2 (z_4 - z_5)) z_5 +
         x_2^2 z_1 (-y_2^2 (y_4 (z_1 - z_4) + y_3 (z_4 - z_5)) (z_1 - z_5) +
            y_2 (y_4^2 z_1 (z_1 - z_5) + y_4 (z_1 - z_4) (y_5 z_1 - y_3 z_5) +
               y_3 (z_4 - z_5) (y_5 z_1 - y_3 z_5)) +
            y_4 z_1 (y_3 (-y_3 + y_5) z_5 + y_4 (-y_5 z_1 + y_3 z_5))) +
         x_2 y_2 (x_3 (y_4 z_1 + y_2 (z_4 - z_5)) (-y_5 z_1 z_4 +
               y_2 z_4 (z_1 - z_5) + (-y_4 z_1 + y_3 (z_1 + z_4)) z_5) +
            x_5 z_1 z_5 (y_4 (y_4 z_1 - y_3 z_5) +
               y_2 (y_4 (-z_1 + z_4) + y_3 (-z_4 + z_5))))) -
      w_5 (-x_3 (x_3 - x_5) x_5 y_2^4 z_4 (y_4 z_1 + y_2 z_4) (z_4 - z_5) z_5 +
         x_2 y_2^2 z_1 (x_3^2 y_5 z_4 (-y_4^2 z_1 +
               y_2 (y_4 (z_1 - z_4) + y_5 z_4)) z_5 -
            x_5^2 y_2 (-y_4^2 z_1 + y_2 (y_4 (z_1 - z_4) + y_3 z_4)) (z_4 -
               z_5) z_5 +
            x_3 x_5 (y_4 y_5 z_1 z_4 (y_5 z_4 + y_4 z_5) +
               y_2 (y_5^2 z_4^2 (z_4 - 2 z_5) + y_3 y_4 z_1 z_4 (z_4 - z_5) +
                  y_4 y_5 z_4^2 (-2 z_1 + z_5) + y_4^2 z_1 z_5 (-z_4 + z_5)) +
               y_2^2 (z_4 - z_5) (-2 y_5 z_4^2 + y_4 (z_1 - z_4) z_5 +
                  y_3 z_4 (z_4 + z_5)))) +
         x_2^3 y_5 z_1^2 (y_2^2 (y_3 (y_4 z_1 (z_1 - z_4) + y_5 z_4 (z_1 - z_5)) -
               y_4^2 (z_1 - z_4) (z_1 - z_5)) +
            y_4 z_1 (-y_4^2 y_5 z_1 + y_3 (y_4 y_5 z_1 + y_5^2 z_4 + y_4^2 z_5)) +
            y_2 (y_4^2 z_1 (y_5 (z_1 - z_4) + y_4 (z_1 - z_5)) +
               y_3^2 y_5 z_4 z_5 -
               y_3 (y_4 y_5 z_1^2 + y_5^2 z_1 z_4 +
                  y_4^2 (z_1^2 + z_1 z_5 - z_4 z_5)))) +
         x_2^2 y_2 z_1 (x_5 z_1 (y_2^2 (-y_3 (y_4 z_1 - y_4 z_4 - 2 y_5 z_4) (z_4 -
                    z_5) + y_4 y_5 (z_1 - z_4) (2 z_4 - z_5) +
                  y_3^2 z_4 (-z_4 + z_5)) +
               y_4 y_5 (y_4^2 z_1 z_5 - 2 y_3 y_5 z_4 z_5 +
                  y_4 z_1 (y_5 z_4 - y_3 z_5)) +
               y_2 (y_3 (-y_5^2 z_4 (z_4 - 2 z_5) + y_4^2 z_1 (z_4 - z_5) +
                    y_4 y_5 z_1 z_5) +
                  y_4 y_5 z_4 (y_5 (-z_1 + z_4) + y_4 (-2 z_1 + z_5)))) +
            x_3 y_5 (-y_4^2 z_1 (y_5 z_1 z_4 + y_4 z_1 z_5 - y_3 z_4 z_5) -
               y_2^2 (y_5 z_4^2 (z_1 - z_5) +
                  y_4 (z_1 - z_4) (-z_4 z_5 + z_1 (z_4 + z_5))) +
               y_2 (y_5 z_4 (y_5 z_1 z_4 - y_3 (z_1 + z_4) z_5) +
                  y_4 z_4 (y_3 (-z_1 + z_4) z_5 + y_5 z_1 (z_1 - z_4 + z_5)) +
                  y_4^2 z_1 (-2 z_4 z_5 + z_1 (z_4 + 2 z_5)))))) +
      w_5^2 y_2 (x_2^2 z_1^2 (y_4 y_5^2 (y_3 y_4 z_1 - y_4^2 z_1 + y_3 y_5 z_4) +
            y_2^2 (-y_4 (2 y_4 - y_5) y_5 (z_1 - z_4) + y_3^2 (y_4 - y_5) z_4 +
               y_3 (y_4^2 (z_1 - z_4) - y_4 y_5 z_4 + 2 y_5^2 z_4)) -
            y_2 (y_4^2 y_5 (-2 y_4 z_1 + y_5 z_4) +
               y_3 (y_4^3 z_1 + y_4 y_5^2 z_1 + y_5^3 z_4))) -
         x_3 y_2^3 z_4 (y_4 z_1 + y_2 z_4) (x_3 (-y_4 + y_5) z_5 +
            x_5 (y_5 z_4 + y_4 z_5 - 2 y_5 z_5 + y_3 (-z_4 + z_5))) -
         x_2 y_2 z_1 (-x_5 y_2 (-y_4^2 z_1 +
               y_2 (y_4 (z_1 - z_4) + y_3 z_4)) (y_5 (z_4 - 2 z_5) + y_4 z_5 +
               y_3 (-z_4 + z_5)) +
            x_3 (y_4^2 y_5^2 z_1 z_4 +
               y_2 (y_3 y_4 (y_4 - y_5) z_1 z_4 + y_4 y_5^2 z_4^2 - y_5^3 z_4^2 -
                  y_4^3 z_1 z_5 + y_4^2 y_5 z_1 (-2 z_4 + z_5)) +
               y_2^2 (y_4^2 (z_1 - z_4) z_5 +
                  y_5 z_4 (2 y_5 z_4 - y_3 (z_4 + z_5)) +
                  y_4 (y_3 z_4 (z_4 + z_5) +
                    y_5 (z_1 (z_4 - z_5) + z_4 (-2 z_4 + z_5)))))))) +
   w_1^3 y_2^2 (w_5^3 y_2^2 (y_2 - y_4) y_4 (-y_3 + y_4) (y_2 - y_5) y_5^2 +
      w_5^2 y_2 y_4 (x_5 y_2 (y_2 -
            y_4) (-y_5^2 (-y_4 z_1 + y_5 z_4 + y_3 (z_1 - 2 z_5) + 2 y_4 z_5) +
            y_2 (y_3 (y_4 z_1 - y_5 z_5) +
               y_5 (y_5 (z_1 + z_4) + y_4 (-2 z_1 + z_5)))) -
         y_5 (x_3 y_2 (y_2 - y_4) y_5 (-2 y_5 z_4 - y_4 z_5 + y_2 (2 z_4 + z_5)) +
            x_2 (6 y_4 (-y_3 + y_4) y_5^2 z_1 +
               y_2^2 (y_4 y_5 (6 z_1 - 2 z_4 - z_5) +
                  y_3 (-6 y_5 z_1 + 2 y_5 z_4 + y_4 z_5)) +
               y_2 (y_4 y_5 (-6 y_4 z_1 - 6 y_5 z_1 + 2 y_5 z_4 + y_4 z_5) +
                  y_3 (6 y_4 y_5 z_1 + 6 y_5^2 z_1 - 2 y_5^2 z_4 -
                    y_4^2 z_5))))) +
      x_5 y_5 z_4 (y_2^2 y_4 (x_3^2 y_5 z_4 z_5 + x_5^2 (y_2 - y_4) (z_1 - z_5) z_5 +
             x_3 x_5 (y_5 z_4 (z_1 - 2 z_5) - (y_3 - y_4) (z_1 - z_5) z_5 +
               y_2 z_4 (-z_1 + z_5))) +
         x_2^2 (6 y_4^2 y_5^2 z_1^2 -
            3 y_2 z_1 (y_3 y_5 (y_5 z_4 - y_3 z_5) +
               y_4^2 (2 y_5 z_1 + y_3 z_5 - y_5 z_5) +
               y_4 (y_5^2 (2 z_1 - z_4) - y_3^2 z_5 + y_3 y_5 z_5)) +
            y_2^2 (y_3 (y_5 z_4 (3 z_1 - z_5) + y_3 (-3 z_1 + z_4) z_5) +
               y_4 (y_3 (3 z_1 - z_4) z_5 +
                  y_5 (6 z_1^2 + z_4 z_5 - 3 z_1 (z_4 + z_5))))) +
         x_2 y_2 (x_3 (3 y_4 y_5 z_1 (y_5 z_4 + (-y_3 + y_4) z_5) +
               y_2^2 (-y_5 z_4^2 + z_5 (-y_4 z_5 + y_3 (z_4 + z_5))) +
               y_2 (y_4^2 z_5^2 + y_5 z_4 (y_5 z_4 - 2 y_3 z_5) -
                  y_4 (y_5 z_4 (3 z_1 - 2 z_5) + y_3 z_5 (z_4 + z_5)))) +
            x_5 (y_4 z_1 (y_3 (y_3 - y_5) z_5 +
                  y_4 (4 y_5 z_1 - y_3 z_5 - 6 y_5 z_5)) +
               y_2^2 (y_3 (z_1 (z_4 - z_5) - z_4 z_5) +
                  y_4 (4 z_1^2 + z_5 (z_4 + z_5) - z_1 (z_4 + 4 z_5))) +
               y_2 (-y_4^2 (-2 z_1 + z_5)^2 +
                  y_4 (y_3 z_5 (z_1 + z_5) +
                    y_5 (-4 z_1^2 + z_1 z_4 + 6 z_1 z_5 - 2 z_4 z_5)) +
                  y_3 (-y_3 z_5^2 + y_5 (-z_1 z_4 + z_1 z_5 + 2 z_4 z_5)))))) +
      w_5 (x_2^2 y_4 y_5 (6 y_4 (-y_3 + y_4) y_5^2 z_1^2 +
            3 y_2 z_1 (-y_3 (-2 y_4 y_5 z_1 + y_5^2 (-2 z_1 + z_4) + y_4^2 z_5) +
                y_4 y_5 (y_5 (-2 z_1 + z_4) + y_4 (-2 z_1 + z_5))) +
            y_2^2 (y_3^2 z_4 z_5 +
               y_4 y_5 (6 z_1^2 + z_4 z_5 - 3 z_1 (z_4 + z_5)) -
               y_3 (y_4 (-3 z_1 + z_4) z_5 +
                  y_5 (6 z_1^2 - 3 z_1 z_4 + z_4 z_5)))) +
         y_2^2 y_4 (x_3^2 (-y_2 + y_4) y_5^2 z_4 z_5 +
            x_5^2 (y_2 -
               y_4) (y_5 (y_5 z_4 (z_1 - 2 z_5) - (y_3 - y_4) (z_1 - z_5) z_5) +
               y_2 (y_3 z_1 (z_4 - z_5) + y_5 (-2 z_1 z_4 + z_1 z_5 + z_4 z_5))) +
             x_3 x_5 y_5 (y_4 y_5 z_4 (z_1 - 3 z_5) + y_4^2 (z_1 - z_5) z_5 +
               y_2^2 (z_1 - z_5) (z_4 + z_5) + y_5 z_4 (-2 y_5 z_4 + y_3 z_5) +
               y_2 (-y_4 (z_1 - z_5) (z_4 + 2 z_5) +
                  y_5 z_4 (-z_1 + 2 (z_4 + z_5))))) -
         x_2 y_2 y_5 (x_3 y_4 (-3 y_4 y_5 z_1 (y_5 z_4 + y_4 z_5) +
               y_2^2 (y_4 z_5^2 - y_5 (3 z_1 - z_4) (z_4 + z_5)) +
               y_2 (y_5^2 (3 z_1 - z_4) z_4 - y_4^2 z_5^2 +
                  y_4 (3 y_5 z_1 z_4 + 6 y_5 z_1 z_5 + y_3 z_4 z_5 -
                    2 y_5 z_4 z_5))) +
            x_5 (y_4 z_1 (2 y_4 y_5 (-2 y_4 z_1 + 3 y_5 z_4 + 3 y_4 z_5) +
                  y_3 (4 y_4 y_5 z_1 + y_5^2 z_4 + y_4^2 z_5 - 6 y_4 y_5 z_5)) +
               y_2 (y_3^2 (y_4 + y_5) z_4 z_5 -
                  y_3 (y_5^2 z_4 (z_1 + 2 z_4) +
                    y_4^2 (4 z_1^2 - 2 z_1 z_5 + z_4 z_5) +
                    y_4 y_5 (4 z_1^2 - 6 z_1 z_5 + 3 z_4 z_5)) +
                  y_4 (2 y_5^2 z_4 (-3 z_1 + z_4) + y_4^2 (-2 z_1 + z_5)^2 +
                    y_4 y_5 (4 z_1^2 - 7 z_1 z_4 - 6 z_1 z_5 + 3 z_4 z_5))) -
               y_2^2 (y_3^2 z_4 z_5 -
                  y_3 (y_5 z_4 (z_1 + 2 z_4) +
                    y_4 (4 z_1^2 + 2 z_4 z_5 - z_1 (z_4 + 3 z_5))) +
                  y_4 (y_5 z_4 (-6 z_1 + 2 z_4 + z_5) +
                    y_4 (4 z_1^2 + z_5 (z_4 + z_5) -
                    z_1 (z_4 + 4 z_5)))))))) +
   w_1^2 y_2 (w_5^3 y_2^2 y_4 (x_5 y_2^2 (y_2 - y_4) (y_4 - y_5) (-y_3 +
            y_5) z_1 + (y_2 - y_5) y_5^2 (x_3 y_2 (y_2 - y_4) z_4 +
            x_2 (y_3 - y_4) (2 y_4 z_1 + y_2 (-2 z_1 + z_4)))) +
      x_5 y_5 z_4 (x_3 (x_3 - x_5) x_5 y_2^3 y_4 z_4 (z_1 - z_5) z_5 +
         x_2^3 z_1 (4 y_4^2 y_5^2 z_1^2 +
            y_2 z_1 (3 y_3 y_5 (-y_5 z_4 + y_3 z_5) +
               y_4^2 (-4 y_5 z_1 - 3 y_3 z_5 + 3 y_5 z_5) +
               y_4 (y_5^2 (-4 z_1 + 3 z_4) + 3 y_3^2 z_5 - 3 y_3 y_5 z_5)) +
            y_2^2 (y_3 (y_5 z_4 (3 z_1 - 2 z_5) + y_3 (-3 z_1 + 2 z_4) z_5) +
               y_4 (y_3 (3 z_1 - 2 z_4) z_5 +
                  y_5 (4 z_1^2 + 2 z_4 z_5 - 3 z_1 (z_4 + z_5))))) +
         x_2 y_2^2 (x_3^2 z_4 z_5 (2 y_4 y_5 z_1 + y_2 y_5 z_4 - y_2^2 z_5 +
               y_2 y_4 z_5) -
            x_3 x_5 (y_2^2 z_4 (z_1 (z_4 - z_5) - z_4 z_5) +
               y_4 z_1 (z_5 (-3 y_4 z_1 + y_3 (3 z_1 + z_4 - 2 z_5) +
                    2 y_4 z_5) + y_5 (-3 z_1 z_4 + 4 z_4 z_5)) +
               y_2 (y_4 (3 z_1^2 z_4 + 2 z_4 z_5^2 +
                    z_1 z_5 (-4 z_4 + z_5)) + (y_5 z_4 - y_3 z_5) (2 z_4 z_5 +
                    z_1 (-z_4 + z_5)))) +
            x_5^2 z_5 (y_4 z_1 (-3 y_4 z_1 + y_3 z_5 + 2 y_4 z_5) +
               y_2 (y_3 (z_1 (z_4 - z_5) - z_4 z_5) +
                  y_4 (3 z_1^2 + z_4 z_5 - z_1 (z_4 + 2 z_5))))) +
         x_2^2 y_2 (x_3 (3 y_4 y_5 z_1^2 (y_5 z_4 + (-y_3 + y_4) z_5) +
               y_2^2 (y_5 z_4^2 (-2 z_1 + z_5) + (2 z_1 - z_4) z_5 (-y_4 z_5 +
                    y_3 (z_4 + z_5))) +
               y_2 z_1 (2 y_4^2 z_5^2 + 2 y_5 z_4 (y_5 z_4 - 2 y_3 z_5) -
                  y_4 (y_5 z_4 (3 z_1 - 4 z_5) + 2 y_3 z_5 (z_4 + z_5)))) +
            x_5 (3 y_4 z_1^2 (y_3 (y_3 - y_5) z_5 +
                  y_4 (2 y_5 z_1 - y_3 z_5 - 2 y_5 z_5)) +
               y_2^2 (y_4 (2 z_1 - z_4) (3 z_1^2 - 3 z_1 z_5 + z_5^2) +
                  y_3 (3 z_1^2 (z_4 - z_5) + z_4 z_5^2 +
                    z_1 z_5 (-3 z_4 + z_5))) +
               y_2 z_1 (-2 y_4^2 (3 z_1^2 - 3 z_1 z_5 + z_5^2) +
                  y_4 (y_3 z_5 (3 z_1 - z_4 + 2 z_5) +
                    y_5 (-6 z_1^2 + 3 z_1 z_4 + 6 z_1 z_5 - 4 z_4 z_5)) +
                  y_3 (y_3 (z_4 - 3 z_5) z_5 +
                    y_5 (-3 z_1 z_4 + 3 z_1 z_5 + 4 z_4 z_5)))))) -
      w_5^2 y_2 (x_2^2 y_4 y_5 (6 y_4 (-y_3 + y_4) y_5^2 z_1^2 +
            2 y_2 z_1 (y_3 (3 y_4 y_5 z_1 + y_5^2 (3 z_1 - 2 z_4) - y_4^2 z_5) +
               y_4 y_5 (-3 y_5 z_1 + 2 y_5 z_4 + y_4 (-3 z_1 + z_5))) +
            y_2^2 (y_3^2 z_4 z_5 +
               y_4 y_5 (6 z_1^2 + z_4 z_5 - 2 z_1 (2 z_4 + z_5)) -
               y_3 (y_4 (-2 z_1 + z_4) z_5 +
                  y_5 (6 z_1^2 - 4 z_1 z_4 + z_4 z_5)))) +
         y_2^2 y_4 (x_3^2 (-y_2 + y_4) y_5^2 z_4 z_5 -
            x_5^2 y_2 (y_2 - y_4) z_1 (y_5 (z_4 - 2 z_5) + y_4 z_5 +
               y_3 (-z_4 + z_5)) +
            x_3 x_5 (y_5^2 z_4 (-y_5 z_4 + y_4 (z_1 - 2 z_5)) +
               y_2^2 (y_4 z_1 z_5 + y_5 (z_1 (z_4 - z_5) - z_4 z_5)) +
               y_2 (y_3 (y_4 - y_5) z_1 z_4 - y_4^2 z_1 z_5 +
                  y_5^2 z_4 (z_4 + 2 z_5) +
                  y_4 y_5 (-2 z_1 z_4 + z_1 z_5 + z_4 z_5)))) +
         x_2 y_2 (x_5 (y_4 y_5^2 z_1 (y_4 (3 y_4 z_1 - 2 y_5 z_4 - 4 y_4 z_5) -
                  y_3 (3 y_4 z_1 + y_5 z_4 - 4 y_4 z_5)) +
               y_2 (y_4 y_5 (y_5^2 (2 z_1 - z_4) z_4 +
                    2 y_4^2 z_1 (-3 z_1 + z_5) +
                    y_4 y_5 (3 z_1 z_4 + 4 z_1 z_5 - 2 z_4 z_5)) +
                  y_3 (3 y_4^3 z_1^2 + y_5^3 z_4 (z_1 + z_4) -
                    2 y_4^2 y_5 z_1 z_5 +
                    y_4 y_5^2 (3 z_1^2 - 4 z_1 z_5 + 2 z_4 z_5))) +
               y_2^2 (y_3^2 (-y_4 + y_5) z_1 z_4 +
                  y_3 (y_4^2 z_1 (-3 z_1 + z_4) - y_5^2 z_4 (2 z_1 + z_4) +
                    y_4 y_5 (z_1 z_4 + 2 z_1 z_5 - z_4 z_5)) +
                  y_4 y_5 (y_5 (-3 z_1^2 - z_1 z_4 + z_4^2) +
                    y_4 (6 z_1^2 + z_4 z_5 - 2 z_1 (z_4 + z_5))))) +
            x_3 y_4 y_5 (y_2^2 y_5 (2 z_1 - z_4) (2 z_4 + z_5) +
               2 y_4 y_5 z_1 (2 y_5 z_4 + y_4 z_5) -
               y_2 (2 y_5^2 (2 z_1 - z_4) z_4 +
                  y_4 (y_3 z_4 z_5 + y_5 (-2 z_4 z_5 + 4 z_1 (z_4 + z_5))))))) +
       w_5 (x_2^3 y_4 y_5 z_1 (4 y_4 (-y_3 + y_4) y_5^2 z_1^2 +
            y_2 z_1 (y_4 y_5 (-4 y_4 z_1 - 4 y_5 z_1 + 3 y_5 z_4 + 3 y_4 z_5) +
               y_3 (4 y_4 y_5 z_1 + y_5^2 (4 z_1 - 3 z_4) - 3 y_4^2 z_5)) +
            y_2^2 (2 y_3^2 z_4 z_5 +
               y_4 y_5 (4 z_1^2 + 2 z_4 z_5 - 3 z_1 (z_4 + z_5)) +
               y_3 (y_4 (3 z_1 - 2 z_4) z_5 +
                  y_5 (-4 z_1^2 + 3 z_1 z_4 - 2 z_4 z_5)))) +
         x_5 y_2^3 y_4 (x_5^2 y_2 (y_2 - y_4) z_1 (z_4 - z_5) z_5 -
            x_3^2 y_5 z_4 z_5 (y_5 z_4 + y_2 (z_1 - z_5) + y_4 (-z_1 + z_5)) -
            x_3 x_5 (y_2^2 z_1 (z_4 - z_5) z_5 +
               y_5 z_4 (y_5 z_4 (z_1 - 2 z_5) + y_4 (z_1 - z_5) z_5) +
               y_2 (y_3 z_1 z_4 (z_4 - z_5) + y_4 z_1 z_5 (-z_4 + z_5) +
                  y_5 z_4 (-2 z_1 z_4 + z_5 (z_4 + z_5))))) +
         x_2^2 y_2 y_5 (x_3 y_4 (3 y_4 y_5 z_1^2 (y_5 z_4 + y_4 z_5) +
               y_2 z_1 (y_5^2 z_4 (-3 z_1 + 2 z_4) + 2 y_4^2 z_5^2 -
                  y_4 (3 y_5 z_1 z_4 + 6 y_5 z_1 z_5 + 2 y_3 z_4 z_5 -
                    4 y_5 z_4 z_5)) +
               y_2^2 (-z_5 (y_4 (2 z_1 - z_4) z_5 + y_3 z_4 (z_4 + z_5)) +
                  y_5 (z_4^2 z_5 + 3 z_1^2 (z_4 + z_5) -
                    2 z_1 z_4 (z_4 + z_5)))) +
            x_5 (-3 y_4 z_1^2 (y_3 (y_5^2 z_4 + 2 y_4 y_5 (z_1 - z_5) +
                    y_4^2 z_5) + 2 y_4 y_5 (y_5 z_4 + y_4 (-z_1 + z_5))) +
               y_2 z_1 (-y_3^2 (2 y_4 + 3 y_5) z_4 z_5 +
                  y_3 (y_5^2 z_4 (3 z_1 + 4 z_4) + y_4^2 (6 z_1^2 + z_4 z_5) +
                    6 y_4 y_5 (z_1^2 - z_1 z_5 + z_4 z_5)) +
                  y_4 (2 y_5^2 (3 z_1 - 2 z_4) z_4 +
                    3 y_4 y_5 (-2 z_1^2 + 3 z_1 z_4 + 2 z_1 z_5 - 2 z_4 z_5) -
                    2 y_4^2 (3 z_1^2 - 3 z_1 z_5 + z_5^2))) +
               y_2^2 (y_3^2 (2 z_1 - z_4) z_4 z_5 +
                  y_4 (y_5 z_4 (-6 z_1^2 + 4 z_1 z_4 + 2 z_1 z_5 - z_4 z_5) +
                    y_4 (2 z_1 - z_4) (3 z_1^2 - 3 z_1 z_5 + z_5^2)) +
                  y_3 (y_5 z_4 (-3 z_1^2 + z_4 z_5 + z_1 (-4 z_4 + z_5)) +
                    y_4 (-6 z_1^3 - 4 z_1 z_4 z_5 + 3 z_1^2 (z_4 + z_5) +
                    z_4 z_5 (z_4 + z_5)))))) +
         x_2 y_2^2 (x_3^2 y_4 y_5 z_4 z_5 (2 y_4 y_5 z_1 +
               y_2 (y_5 (-2 z_1 + z_4) + y_4 z_5)) +
            x_3 x_5 y_5 (y_4 z_1 (y_4^2 (3 z_1 - 2 z_5) z_5 +
                  2 y_5 z_4 (-2 y_5 z_4 + y_3 z_5) +
                  y_4 z_4 (3 y_5 z_1 - y_3 z_5 - 6 y_5 z_5)) +
               y_2 (-y_5 z_4 (z_1 + 2 z_4) (y_5 z_4 - y_3 z_5) +
                  y_4^2 (-2 z_4 z_5^2 + 4 z_1 z_5 (z_4 + z_5) -
                    3 z_1^2 (z_4 + 2 z_5)) +
                  y_4 z_4 (y_3 (z_1 + z_4) z_5 +
                    y_5 (-3 z_1^2 + 5 z_1 z_4 + 3 z_1 z_5 - 4 z_4 z_5))) +
               y_2^2 (z_4^2 (y_5 (z_1 + 2 z_4) - y_3 z_5) +
                  y_4 (3 z_1^2 (z_4 + z_5) + z_4 z_5 (z_4 + z_5) -
                    z_1 (z_4^2 + 4 z_4 z_5 + 2 z_5^2)))) +
            x_5^2 (y_4 y_5 z_1 (2 y_3 y_5 z_4 z_5 + y_4^2 z_5 (-3 z_1 + 2 z_5) +
                  y_4 (-3 y_5 z_1 z_4 + 3 y_3 z_1 z_5 + 4 y_5 z_4 z_5 -
                    2 y_3 z_5^2)) +
               y_2^2 (y_3^2 z_1 z_4 (z_4 - z_5) +
                  y_4 y_5 (-z_4^2 z_5 + z_1 z_4 (2 z_4 + z_5) +
                    z_1^2 (-6 z_4 + 3 z_5)) +
                  y_3 (y_4 z_1 (3 z_1 - z_4) (z_4 - z_5) +
                    y_5 z_4 (-2 z_1 z_4 + 2 z_1 z_5 + z_4 z_5))) -
               y_2 (y_3 (3 y_4^2 z_1^2 (z_4 - z_5) +
                    y_4 y_5 z_5 (3 z_1^2 - 2 z_1 z_5 + z_4 z_5) +
                    y_5^2 z_4 (-z_1 z_4 + 2 z_1 z_5 + 2 z_4 z_5)) +
                  y_4 y_5 (y_4 (-6 z_1^2 z_4 - z_4 z_5^2 +
                    z_1 z_5 (3 z_4 + 2 z_5)) +
                    y_5 z_4 (-3 z_1^2 - 2 z_4 z_5 + z_1 (z_4 + 4 z_5)))))))) -
    w_1 (w_5^3 y_2^2 (x_3 x_5 y_2^3 y_4 (y_3 - y_5) (-y_4 + y_5) z_1 z_4 -
         x_2^2 (y_3 - y_4) y_4 (y_2 - y_5) y_5^2 z_1 (y_4 z_1 + y_2 (-z_1 + z_4)) +
         x_2 y_2 (x_3 y_4 (y_2 - y_5) y_5^2 z_4 (y_4 z_1 + y_2 (-z_1 + z_4)) +
            x_5 y_2 (y_3 - y_5) (y_4 - y_5) z_1 (-2 y_4^2 z_1 +
               y_2 (2 y_4 z_1 + y_3 z_4 - y_4 z_4)))) +
      x_2 x_5 y_5 z_4 (x_3 (x_3 - x_5) x_5 y_2^3 z_4 z_5 (y_4 z_1 (-2 z_1 + z_5) +
            y_2 (z_4 z_5 + z_1 (-z_4 + z_5))) +
         x_2^3 z_1^2 (-y_4^2 y_5^2 z_1^2 +
            y_2 z_1 (y_3 y_5 (y_5 z_4 - y_3 z_5) +
               y_4^2 (y_5 (z_1 - z_5) + y_3 z_5) +
               y_4 (y_5^2 (z_1 - z_4) - y_3^2 z_5 + y_3 y_5 z_5)) +
            y_2^2 (-y_4 (z_1 - z_4) (y_5 (z_1 - z_5) + y_3 z_5) +
               y_3 (y_3 (z_1 - z_4) z_5 + y_5 z_4 (-z_1 + z_5)))) +
         x_2 y_2^2 (-x_3^2 z_4 z_5 (y_4 z_1 (y_5 z_1 + y_2 z_5) +
               y_2 (y_5 z_1 z_4 + y_2 (-z_1 + z_4) z_5)) +
            x_5^2 z_1 z_5 (y_4 z_1 (3 y_4 z_1 - 2 y_3 z_5 - y_4 z_5) +
               y_2 (y_4 (-3 z_1^2 + 2 z_1 z_4 + z_1 z_5 - z_4 z_5) +
                  y_3 (-2 z_1 z_4 + 2 z_1 z_5 + z_4 z_5))) +
            x_3 x_5 (y_2^2 z_4 (2 z_1^2 (z_4 - z_5) + z_4 z_5^2 +
                  z_1 z_5 (-2 z_4 + z_5)) +
               y_4 z_1^2 (y_5 (-3 z_1 z_4 + 2 z_4 z_5) +
                  z_5 (y_3 (3 z_1 + 2 z_4 - z_5) + y_4 (-3 z_1 + z_5))) +
               y_2 z_1 (y_3 z_5 (2 z_1 z_4 + z_4^2 - 2 z_1 z_5 - 3 z_4 z_5) +
                  2 y_5 z_4 (-z_1 z_4 + z_1 z_5 + z_4 z_5) +
                  y_4 (3 z_1^2 z_4 + 2 z_4 z_5^2 +
                    z_1 z_5 (-5 z_4 + 2 z_5))))) -
         x_2^2 y_2 z_1 (x_3 (y_4 y_5 z_1^2 (y_5 z_4 + (-y_3 + y_4) z_5) +
               y_2^2 (y_5 z_4^2 (-z_1 + z_5) + (z_1 - z_4) z_5 (-y_4 z_5 +
                    y_3 (z_4 + z_5))) -
               y_2 z_1 (-y_4^2 z_5^2 + y_5 z_4 (-y_5 z_4 + 2 y_3 z_5) +
                  y_4 (y_5 z_4 (z_1 - 2 z_5) + y_3 z_5 (z_4 + z_5)))) +
            x_5 (y_4 z_1^2 (3 y_3 (y_3 - y_5) z_5 +
                  y_4 (4 y_5 z_1 - 3 y_3 z_5 - 2 y_5 z_5)) +
               y_2^2 (y_3 (3 z_1^2 (z_4 - z_5) + z_4 z_5^2 +
                    z_1 z_5 (-3 z_4 + 2 z_5)) +
                  y_4 (4 z_1^3 - z_4 z_5^2 + z_1 z_5 (3 z_4 + z_5) -
                    z_1^2 (3 z_4 + 4 z_5))) +
               y_2 z_1 (-y_4^2 (-2 z_1 + z_5)^2 +
                  y_4 (y_3 z_5 (3 z_1 - 2 z_4 + z_5) +
                    y_5 (-4 z_1^2 + 3 z_1 z_4 + 2 z_1 z_5 - 2 z_4 z_5)) +
                  y_3 (y_3 (2 z_4 - 3 z_5) z_5 +
                    y_5 (-3 z_1 z_4 + 3 z_1 z_5 + 2 z_4 z_5)))))) +
      w_5^2 y_2 (x_3 x_5 y_2^4 y_4 z_1 z_4 (x_3 (-y_4 + y_5) z_5 +
            x_5 (y_5 z_4 + y_4 z_5 - 2 y_5 z_5 + y_3 (-z_4 + z_5))) +
         x_2^3 y_4 y_5 z_1 (2 y_4 (-y_3 + y_4) y_5^2 z_1^2 +
            y_2 z_1 (y_3 (2 y_4 y_5 z_1 + 2 y_5^2 (z_1 - z_4) - y_4^2 z_5) +
               y_4 y_5 (2 y_5 (-z_1 + z_4) + y_4 (-2 z_1 + z_5))) +
            y_2^2 (y_4 y_5 (z_1 - z_4) (2 z_1 - z_5) + y_3^2 z_4 z_5 -
               y_3 (y_4 (-z_1 + z_4) z_5 +
                  y_5 (2 z_1^2 - 2 z_1 z_4 + z_4 z_5)))) +
         x_2 y_2^2 (x_3^2 y_4 y_5^2 z_4 (y_4 z_1 + y_2 (-z_1 + z_4)) z_5 -
            x_5^2 y_2 z_1 (-2 y_4^2 z_1 +
               y_2 (2 y_4 z_1 + y_3 z_4 - y_4 z_4)) (y_5 (z_4 - 2 z_5) + y_4 z_5 +
                y_3 (-z_4 + z_5)) +
            x_3 x_5 (y_4 y_5^2 z_1 z_4 (-y_5 z_4 + 2 y_4 (z_1 - z_5)) +
               y_2 (2 y_3 y_4 (y_4 - y_5) z_1^2 z_4 - y_5^3 z_4^2 (z_1 + z_4) -
                  2 y_4^3 z_1^2 z_5 +
                  y_4^2 y_5 z_1 (-4 z_1 z_4 + 2 z_1 z_5 + z_4 z_5) +
                  2 y_4 y_5^2 z_4 (-z_4 z_5 + z_1 (z_4 + z_5))) +
               y_2^2 (y_4^2 z_1 (2 z_1 - z_4) z_5 +
                  y_5 z_4 (y_5 z_4 (2 z_1 + z_4) - y_3 z_1 (z_4 + z_5)) +
                  y_4 (y_3 z_1 z_4 (z_4 + z_5) +
                    y_5 (-2 z_1 z_4^2 + 2 z_1^2 (z_4 - z_5) + z_4^2 z_5))))) +
          x_2^2 y_2 (x_3 y_4 y_5 (y_4 y_5 z_1^2 (2 y_5 z_4 + y_4 z_5) +
               y_2^2 (-y_3 z_4^2 z_5 +
                  y_5 (z_4^2 z_5 + z_1^2 (2 z_4 + z_5) -
                    z_1 z_4 (2 z_4 + z_5))) -
               y_2 z_1 (2 y_5^2 (z_1 - z_4) z_4 +
                  y_4 (y_3 z_4 z_5 + 2 y_5 (-z_4 z_5 + z_1 (z_4 + z_5))))) +
            x_5 z_1 (y_4 y_5^2 z_1 (y_4 (3 y_4 z_1 - y_5 z_4 - 2 y_4 z_5) +
                  y_3 (-3 y_4 z_1 - 2 y_5 z_4 + 2 y_4 z_5)) +
               y_2 (y_4 y_5 (y_5^2 (z_1 - z_4) z_4 + y_4^2 z_1 (-6 z_1 + z_5) +
                    y_4 y_5 (3 z_1 z_4 + 2 z_1 z_5 - 2 z_4 z_5)) +
                  y_3 (3 y_4^3 z_1^2 + y_5^3 z_4 (2 z_1 + z_4) -
                    y_4^2 y_5 z_1 z_5 +
                    y_4 y_5^2 (3 z_1^2 - 2 z_1 z_5 + 2 z_4 z_5))) +
               y_2^2 (2 y_3^2 (-y_4 + y_5) z_1 z_4 -
                  y_3 (y_4^2 z_1 (3 z_1 - 2 z_4) + y_5^2 z_4 (4 z_1 + z_4) +
                    y_4 y_5 (z_4 z_5 - z_1 (2 z_4 + z_5))) +
                  y_4 y_5 (y_5 (-3 z_1^2 + z_1 z_4 + z_4^2) +
                    y_4 (6 z_1^2 + z_4 z_5 - z_1 (4 z_4 + z_5))))))) +
      w_5 (-x_3 (x_3 - x_5) x_5^2 y_2^5 y_4 z_1 z_4 (z_4 - z_5) z_5 +
         x_2^4 y_4 y_5 z_1^2 ((y_3 - y_4) y_4 y_5^2 z_1^2 +
            y_2 z_1 (y_4 y_5 (y_5 (z_1 - z_4) + y_4 (z_1 - z_5)) +
               y_3 (-y_4 y_5 z_1 + y_5^2 (-z_1 + z_4) + y_4^2 z_5)) +
            y_2^2 (-y_4 y_5 (z_1 - z_4) (z_1 - z_5) - y_3^2 z_4 z_5 +
               y_3 (y_4 (-z_1 + z_4) z_5 + y_5 (z_1^2 - z_1 z_4 + z_4 z_5)))) +
         x_2 x_5 y_2^3 (-x_5^2 y_2 z_1 (-2 y_4^2 z_1 +
               y_2 (2 y_4 z_1 + y_3 z_4 - y_4 z_4)) (z_4 - z_5) z_5 +
            x_3^2 y_5 z_4 z_5 (y_4 z_1 (y_5 z_4 + y_4 (-2 z_1 + z_5)) +
               y_2 (y_5 z_4 (z_1 + z_4) +
                  y_4 (2 z_1^2 + z_4 z_5 - z_1 (z_4 + z_5)))) +
            x_3 x_5 (y_4 y_5 z_1 z_4 (2 y_5 z_4 (z_1 - z_5) +
                  y_4 (2 z_1 - z_5) z_5) +
               y_2^2 (y_4 z_1 (2 z_1 - z_4) (z_4 - z_5) z_5 +
                  y_5 z_4^2 (-2 z_1 z_4 + 2 z_1 z_5 + z_4 z_5) +
                  y_3 z_1 z_4 (z_4^2 - z_5^2)) +
               y_2 (2 y_3 y_4 z_1^2 z_4 (z_4 - z_5) +
                  2 y_4^2 z_1^2 z_5 (-z_4 + z_5) +
                  y_5^2 z_4^2 (z_1 (z_4 - 2 z_5) - 2 z_4 z_5) +
                  y_4 y_5 z_4 (-4 z_1^2 z_4 - z_4 z_5^2 +
                    z_1 z_5 (2 z_4 + z_5))))) +
         x_2^2 y_2^2 (-x_3^2 y_4 y_5 z_4 z_5 (y_4 y_5 z_1^2 + y_2^2 z_4 z_5 +
               y_2 z_1 (y_5 (-z_1 + z_4) + y_4 z_5)) +
            x_3 x_5 y_5 (y_4 z_1^2 (y_4^2 z_5 (-3 z_1 + z_5) +
                  y_5 z_4 (2 y_5 z_4 - y_3 z_5) +
                  y_4 z_4 (-3 y_5 z_1 + 2 y_3 z_5 + 3 y_5 z_5)) +
               y_2 z_1 (y_5 z_4 (2 y_5 z_4 (z_1 + z_4) -
                    y_3 (2 z_1 + 3 z_4) z_5) +
                  y_4^2 (2 z_4 z_5^2 + 3 z_1^2 (z_4 + 2 z_5) -
                    z_1 z_5 (5 z_4 + 2 z_5)) +
                  y_4 z_4 (-2 y_3 z_1 z_5 +
                    y_5 (3 z_1^2 - 4 z_1 z_4 + 4 z_4 z_5))) +
               y_2^2 (z_4^2 (-y_5 (z_1 + z_4) (2 z_1 - z_5) +
                    y_3 (z_1 - z_4) z_5) +
                  y_4 (z_4^2 z_5^2 - 3 z_1^3 (z_4 + z_5) -
                    z_1 z_4 z_5 (2 z_4 + z_5) +
                    z_1^2 (2 z_4^2 + 5 z_4 z_5 + z_5^2)))) +
            x_5^2 z_1 (y_4 y_5 z_1 (-4 y_3 y_5 z_4 z_5 + y_4^2 (3 z_1 - z_5) z_5 +
                  y_4 (y_5 z_4 (3 z_1 - 2 z_5) + y_3 z_5 (-3 z_1 + z_5))) +
               y_2^2 (2 y_3^2 z_1 z_4 (-z_4 + z_5) +
                  y_4 y_5 (z_1^2 (6 z_4 - 3 z_5) + z_4^2 z_5 +
                    z_1 z_4 (-4 z_4 + z_5)) -
                  y_3 (y_4 z_1 (3 z_1 - 2 z_4) (z_4 - z_5) +
                    y_5 z_4 (-4 z_1 z_4 + 4 z_1 z_5 + z_4 z_5))) +
               y_2 (y_3 (3 y_4^2 z_1^2 (z_4 - z_5) +
                    y_4 y_5 z_5 (3 z_1^2 - z_1 z_5 + z_4 z_5) +
                    2 y_5^2 z_4 (-z_1 z_4 + 2 z_1 z_5 + z_4 z_5)) +
                  y_4 y_5 (y_5 z_4 (-3 z_1^2 - 2 z_4 z_5 + 2 z_1 (z_4 + z_5)) +
                    y_4 (-6 z_1^2 z_4 - z_4 z_5^2 +
                    z_1 z_5 (3 z_4 + z_5)))))) +
         x_2^3 y_2 y_5 z_1 (x_3 y_4 (-y_4 y_5 z_1^2 (y_5 z_4 + y_4 z_5) +
               y_2 z_1 (y_5^2 (z_1 - z_4) z_4 - y_4^2 z_5^2 +
                  y_4 (y_5 z_1 z_4 + 2 y_5 z_1 z_5 + y_3 z_4 z_5 -
                    2 y_5 z_4 z_5)) +
               y_2^2 (z_5 (y_4 (z_1 - z_4) z_5 + y_3 z_4 (z_4 + z_5)) -
                  y_5 (z_4^2 z_5 + z_1^2 (z_4 + z_5) - z_1 z_4 (z_4 + z_5)))) +
            x_5 (y_4 z_1^2 (2 y_4 y_5 (-2 y_4 z_1 + y_5 z_4 + y_4 z_5) +
                  y_3 (4 y_4 y_5 z_1 + 3 y_5^2 z_4 + 3 y_4^2 z_5 -
                    2 y_4 y_5 z_5)) +
               y_2 z_1 (y_3^2 (y_4 + 3 y_5) z_4 z_5 -
                  y_3 (y_5^2 z_4 (3 z_1 + 2 z_4) +
                    y_4^2 (4 z_1^2 + 2 z_1 z_5 - z_4 z_5) +
                    y_4 y_5 (4 z_1^2 - 2 z_1 z_5 + 3 z_4 z_5)) +
                  y_4 (2 y_5^2 z_4 (-z_1 + z_4) + y_4^2 (-2 z_1 + z_5)^2 +
                    y_4 y_5 (4 z_1^2 - 5 z_1 z_4 - 2 z_1 z_5 + 3 z_4 z_5))) -
               y_2^2 (y_3^2 (z_1 - z_4) z_4 z_5 +
                  y_3 (y_5 z_4 (-3 z_1^2 - 2 z_1 z_4 + 2 z_1 z_5 + z_4 z_5) +
                    y_4 (-4 z_1^3 - 2 z_1 z_4 z_5 + z_4 z_5 (z_4 + z_5) +
                    z_1^2 (3 z_4 + z_5))) +
                  y_4 (-y_5 (z_1 - z_4) z_4 (2 z_1 - z_5) +
                    y_4 (4 z_1^3 - z_4 z_5^2 + z_1 z_5 (3 z_4 + z_5) -
                    z_1^2 (3 z_4 + 4 z_5))))))));$


\end{document}